\documentclass[a4paper,11pt]{article}
\pdfoutput=1 % if your are submitting a pdflatex (i.e. if you have
\usepackage{jcappub} % for details on the use of the package, please
\usepackage[T1]{fontenc} % if needed
\usepackage{calc}
\usepackage{rotating}
\usepackage[english]{babel}
\usepackage{graphicx}
\usepackage{subfig}
\usepackage{float}
\usepackage{amsmath}
\usepackage{amssymb}
\usepackage{amsthm}
\usepackage{latexsym}
\usepackage{dcolumn}
\usepackage{geometry}
\usepackage{color}
\usepackage{hyperref}
\usepackage{mciteplus}

\def\gtwid{\mathrel{\raise.3ex\hbox{$>$\kern-.75em\lower1ex\hbox{$\sim
$}}}}
\def\vio{\mathrel{\hbox{$E$\kern-.60em\hbox{$/
$}}}}
\newcommand{\newc}{\newcommand*}

\long\def\begincomment#1\endcomment{%
        \begingroup\sf\baselineskip12pt#1\endgroup}

\newc{\etal}{\textrm{et al.}} 
\newc{\eg}{\textrm{e.g.}} 
\newc{\ie}{\textrm{i.e.}}
\newc{\etc}{\textrm{etc}}
\newc\vs{\textrm{vs.}}
\newc{\cl}{\rm {C.L.}}
\newc{\ev}{\ensuremath{\,\mathrm{eV}}}
\newc{\kev}{\ensuremath{\,\mathrm{keV}}}
\newc{\mev}{\ensuremath{\,\mathrm{MeV}}}
\newc{\gev}{\ensuremath{\,\mathrm{GeV}}}
\newc{\tev}{\ensuremath{\,\mathrm{TeV}}}
\newc{\MeV}{\mev} 
\newc{\TeV}{\tev}
\newc{\invpb}{\ensuremath{/\text{pb}}}
\newc{\invfb}{\ensuremath{\,\text{fb}^{-1}}}
\newc\nb{\ensuremath{\,\mathrm{nb}}} \newc\pb{\ensuremath{\,\mathrm{pb}}} \newc\fb{\ensuremath{\,\mathrm{fb}}}
\newc\pc{\ensuremath{\,\mathrm{pc}}}
\newc\kpc{\ensuremath{\,\mathrm{kpc}}}
\newc\mpc{\ensuremath{\,\mathrm{Mpc}}}
\newc\ps{\ensuremath{\,\mathrm{ps}}} 
\newc\cmeter{\ensuremath{\,\mathrm{cm}}} 
\newc\meter{\ensuremath{\,\mathrm{m}}} 
\newc\kmeter{\ensuremath{\,\mathrm{km}}}
\newc\second{\ensuremath{\,\mathrm{s}}}
\newc\msecond{\ensuremath{\,\mathrm{ms}}}
\newc\nsecond{\ensuremath{\,\mathrm{ns}}}
\newc\psecond{\ensuremath{\,\mathrm{ps}}}
\newc{\chisqmin}{\ensuremath{\chi^2_{\mathrm{min}}}}
\newc{\Delchisq}{\ensuremath{\Delta\chi^2}}
\newc{\chisq}{\ensuremath{\chi^2}}
\newc{\like}{\ensuremath{\mathcal{L}}}
\newc\lsim{\ensuremath{\mathrel{\rlap{\lower4pt\hbox{\hskip1pt$\sim$}}\raise1pt\hbox{$<$}}}}
\newc\gsim{\ensuremath{\mathrel{\rlap{\lower4pt\hbox{\hskip1pt$\sim$}}\raise1pt\hbox{$>$}}}}
\newc{\VEV}[1]{\ensuremath{\langle #1 \rangle}}
\newc{\dl}{\ensuremath{\stackrel{\leftarrow}{D}}}
\newc{\dr}{\ensuremath{\stackrel{\rightarrow}{D}}}
\newc{\bcenter}{\begin{center}}    \newc{\ecenter}{\end{center}}
\newc{\bfl}{\begin{flushleft}}    \newc{\efl}{\end{flushleft}}
\newc{\bfr}{\begin{flushright}}    \newc{\efr}{\end{flushright}}

\newc{\bi}{\begin{itemize}}
\newc{\ei}{\end{itemize}}
\newc{\bed}{\begin{description}}
\newc{\eed}{\end{description}}
\newc{\ben}{\begin{enumerate}}
\newc{\een}{\end{enumerate}}

\newc{\be}{\begin{equation}}
\newc{\ee}{\end{equation}}
\newc{\bea}{\begin{eqnarray}}
\newc{\eea}{\end{eqnarray}}
\newc{\ra}{\rightarrow}
\newc{\alphas}{\ensuremath{\alpha_s}}
\newc{\alphatwo}{\ensuremath{\alpha_2}}
\newc{\alphaone}{\ensuremath{\alpha_1}}
\newc{\alphai}[1]{\ensuremath{\alpha_{#1}}}
\newc{\alphaem}{\ensuremath{\alpha_{\mathrm{em}}}}
\newc{\alphaeff}{\ensuremath{\alpha_{\mathrm{eff}}}}
\newc{\sineff}{\ensuremath{\sin \theta_{\mathrm{eff}}}}
\newc{\sinsqeff}{\ensuremath{\sin^2 \theta_{\mathrm{eff}}}}
\newc{\dalphahad}{\ensuremath{\Delta \alpha_{\mathrm{had}}}}
\newc{\yt}{\ensuremath{h_t}} \newc{\yb}{\ensuremath{h_b}} \newc{\ytau}{\ensuremath{h_{\tau}}}
\newc\mz{\ensuremath{M_Z}} 
\newc\mw{\ensuremath{m_W}}
\newc\mZ{\mz}        \newc\mW{\mw}
\newc\mhsm{\ensuremath{ m_{H_{\mathrm{SM}}}}}
\newc{\mtop}{\ensuremath{ m_t}}               \newc{\mtpole}{\ensuremath{ M_t}}
\newc{\mbottom}{\ensuremath{ m_b}} 
\newc{\mtau}{\ensuremath{ m_{\tau}}}
\newc{\mt}{\mtpole}
\newc{\mb}{\mbottom} 
\newc{\rgg}{\ensuremath{R_{h}(\gamma\gamma)}}
\newc{\rzz}{\ensuremath{R_{h}(ZZ)}}
\newc{\rtwogg}{\ensuremath{R_{h_2}(\gamma\gamma)}}
\newc{\rtwozz}{\ensuremath{R_{h_2}(ZZ)}}
\newc{\ronegg}{\ensuremath{R_{h_1}(\gamma\gamma)}}
\newc{\ronezz}{\ensuremath{R_{h_1}(ZZ)}}
\newc{\rsiggg}{\ensuremath{R_{h_\textrm{sig}}(\gamma\gamma)}}
\newc{\rsigzz}{\ensuremath{R_{h_\textrm{sig}}(ZZ)}}
\newc{\llbar}{\ensuremath{\ell\bar{\ell}}}
\newc{\tauptaum}{\ensuremath{ \tau^+\tau^-}}
\newc{\qqbar}{\ensuremath{ q\bar{q}}} \newc{\ppbar}{\ensuremath{ p\bar{p}}}
\newc{\bbbar}{\ensuremath{ b\bar{b}}} \newc{\ttbar}{\ensuremath{ t\bar{t}}}
\newc{\ffbar}{\ensuremath{ f\bar{f}}} \newc{\tautaubar}{\ensuremath{ \tau\bar{\tau}}}
\newc{\mchi}{\ensuremath{m_{\chi}}}
\newc{\squark}{\ensuremath{\tilde{q}}}
\newc{\slepton}{\ensuremath{\tilde{l}}}
\newc{\gluino}{\ensuremath{\tilde{g}}} 
\newc{\mgluino}{\ensuremath{{m_{\gluino}}}}
\newc{\tone}{\ensuremath{{\tilde{t}_1}}}
\newc{\sthw}{\ensuremath{ \sin\theta_W}}              \newc{\cthw}{\ensuremath{\cos\theta_W}}
\newc{\tanthw}{\ensuremath{ \tan\theta_W}}              \newc{\cotthw}{\ensuremath{\cot\theta_W}}
\newc{\ssqthw}{\ensuremath{\sin^2 \theta_W}}
\newc{\msbar}{\ensuremath{\overline{MS}}} \newc{\drbar}{\ensuremath{\overline{DR}}}
\newc{\mtmtsmmsbar}{\ensuremath{ m_t(m_t)^{\msbar}_{{\mathrm{SM}}}}}
\newc{\mtmtsmdrbar}{\ensuremath{ m_t(m_t)^{\drbar}_{{\mathrm{SM}}}}}
\newc{\mtmtmssmdrbar}{\ensuremath{ m_t(m_t)^{\drbar}_{{\mathrm{SUSY}}}}}
\newc{\mbmbmsbar}{\ensuremath{ m_b(m_b)^{\msbar} }}
\newc{\mbmbsmmsbar}{\ensuremath{ m_b(m_b)^{\msbar}_{{\mathrm{SM}}}}}
\newc{\mbmzsmmsbar}{\ensuremath{ m_b(\mz)^{\msbar}_{{\mathrm{SM}}}}}
\newc{\mbmzsmdrbar}{\ensuremath{ m_b(\mz)^{\drbar}_{{\mathrm{SM}}}}}
\newc{\mbmzmssmdrbar}{\ensuremath{ m_b(\mz)^{\drbar}_{{\mathrm{SUSY}}}}}
\newc{\mtaumzsmmsbar}{\ensuremath{ m_{\tau}(\mz)^{\msbar}_{{\mathrm{SM}}}}}
\newc{\mtaumzsmdrbar}{\ensuremath{ m_{\tau}(\mz)^{\drbar}_{{\mathrm{SM}}}}}
\newc{\mtaumzmssmdrbar}{\ensuremath{ m_{\tau}(\mz)^{\drbar}_{{\mathrm{SUSY}}}}}
\newc{\alphasmzms}{\ensuremath{\alpha_s(M_Z)^{\overline{MS}}}}
\newc{\alphaimzms}[1]{\ensuremath{\alpha_{#1}(M_Z)^{\overline{MS}}}}
\newc{\alphaemmz}{\ensuremath{\alpha_{\mathrm{em}}(M_Z)^{\overline{MS}}}}
\newc{\mzero}{\ensuremath{{m_0}}}
\newc{\mhalf}{\ensuremath{ m_{1/2}}}
\newc{\tanb}{\ensuremath{\tan\beta}}
\newc{\azero}{\ensuremath{ A_0}}
\newc{\bzero}{\ensuremath{ B_0}}
\newc{\signmu}{\ensuremath{\rm{sgn}\,\mu}}
\newc{\mueff}{\ensuremath{\mu_{\rm{eff}}}}
\newc{\lam}{\ensuremath{{\lambda}}}
\newc{\kap}{\ensuremath{{\kappa}}}
\newc{\alam}{\ensuremath{{A_{\lambda}}}}
\newc{\akap}{\ensuremath{{A_{\kappa}}}}
\newc{\hs}{\ensuremath{ H_s}}      
\newc{\mhs}{\ensuremath{ m_{H_s}}} 
\newc{\mgut}{\ensuremath{ M_{\rm GUT}}}
\newc{\mplanck}{\ensuremath{ M_{\rm P}}}      \newc{\mpl}{\ensuremath{ M_{\rm Pl}}}
\newc{\msusy}{\ensuremath{ M_{\rm SUSY}}}      \newc{\ms}{\ensuremath{ M_{\rm S}}}
 \newc{\mhl}{\ensuremath{m_\hl}} 
 \newc{\mhone}{\ensuremath{m_{h_1}}} 
 \newc{\mhtwo}{\ensuremath{m_{h_2}}} 
 \newc{\mglu}{\ensuremath{m_{\tilde g}}} 
 \newc{\mul}{\ensuremath{m_{\tilde{u}_L}}} 
 \newc{\mtone}{\ensuremath{m_{\tilde{t}_1}}} 
 \newc{\ma}{\ensuremath{m_A}} 
 \newc{\maone}{\ensuremath{m_{a_1}}} 
 \newc{\matwo}{\ensuremath{m_{a_2}}}
 \newc{\hone}{\ensuremath{h_1}}
 \newc{\htwo}{\ensuremath{h_2}}
 \newc{\aone}{\ensuremath{a_1}}
 \newc{\atwo}{\ensuremath{a_2}}
 \newc{\mhu}{\ensuremath{ m_{H_u}}}       
 \newc{\mhd}{\ensuremath{ m_{H_d}}}
 \newc{\mhusq}{\ensuremath{ m_{H_u}^2}}       
 \newc{\mhdsq}{\ensuremath{ m_{H_d}^2}}
 \newc{\mhuew}{\ensuremath{ m^{\ast}_{H_u}}}       
 \newc{\mhdew}{\ensuremath{ m^{\ast}_{H_d}}}
 \newc{\mhuewsq}{\ensuremath{ m^{\ast\, 2}_{H_u}}}       
 \newc{\mhdewsq}{\ensuremath{ m^{\ast\, 2}_{H_d}}}
 \newc{\hu}{\ensuremath{ H_u}}       
 \newc{\hd}{\ensuremath{ H_d}}
 \newc{\barmhu}{\ensuremath{ \bar{m}_{H_u}}}
 \newc{\barmhd}{\ensuremath{ \bar{m}_{H_d}}}

 \newc{\mqthree}{\ensuremath{m_{\widetilde{Q}_3}^2}}
 \newc{\muthree}{\ensuremath{m_{\tilde{u}_3}^2}}
 \newc{\mdthree}{\ensuremath{m_{\tilde{d}_3}^2}}
 \newc{\mlthree}{\ensuremath{m_{\widetilde{L}_3}^2}}
 \newc{\methree}{\ensuremath{m_{\tilde{e}_3}^2}}
 \newc{\mqtwo}{\ensuremath{m_{\widetilde{Q}_2}^2}}
 \newc{\mutwo}{\ensuremath{m_{\tilde{u}_2}^2}}
 \newc{\mdtwo}{\ensuremath{m_{\tilde{d}_2}^2}}
 \newc{\mltwo}{\ensuremath{m_{\widetilde{L}_2}^2}}
 \newc{\metwo}{\ensuremath{m_{\tilde{e}_2}^2}}
 \newc{\mqone}{\ensuremath{m_{\widetilde{Q}_1}^2}}
 \newc{\muone}{\ensuremath{m_{\tilde{u}_1}^2}}
 \newc{\mdone}{\ensuremath{m_{\tilde{d}_1}^2}}
 \newc{\mlone}{\ensuremath{m_{\widetilde{L}_1}^2}}
 \newc{\meone}{\ensuremath{m_{\tilde{e}_1}^2}}
 \newc{\msmul}{\ensuremath{m_{\tilde{\mu}_L}}}
 \newc{\msmur}{\ensuremath{m_{\tilde{\mu}_R}}}
 \newc{\msneumu}{\ensuremath{m_{\tilde{\nu}_{\mu}}}}
 \newc{\mone}{\ensuremath{M_1}}
 \newc{\monesq}{\ensuremath{M_1^2}}
 \newc{\mtwo}{\ensuremath{M_2}}
 \newc{\mtwosq}{\ensuremath{M_2^2}}
 \newc{\mthree}{\ensuremath{M_3}}
 \newc{\mthreesq}{\ensuremath{M_3^2}}
 \newc{\atau}{\ensuremath{{A_{\tau}}}}
 \newc{\at}{\ensuremath{{A_{t}}}}
 \newc{\ab}{\ensuremath{{A_{b}}}}
 \newc{\atausq}{\ensuremath{{A_{\tau}^2}}}
 \newc{\atsq}{\ensuremath{{A_{t}^2}}}
 \newc{\absq}{\ensuremath{{A_{b}^2}}}
 \newc{\dmzero}{\ensuremath{\Delta{_{m_0}}}}
 \newc{\dmhalf}{\ensuremath{\Delta{_{m_{1/2}}}}}
 \newc{\dmu}{\ensuremath{\Delta{_{\mu}}}}

 \newc{\pten}{\ensuremath{\psi_{10}}}
 \newc{\ffive}{\ensuremath{\phi_{5}}}
 \newc{\hfive}{\ensuremath{h_{5}}}
 \newc{\hbfive}{\ensuremath{h_{\bar{5}}}}
 \newc{\thet}{\ensuremath{\theta_{50}}}
 \newc{\thetb}{\ensuremath{\theta_{\,\overline{50}}}}
 \newc{\ptenhat}{\ensuremath{\hat{\psi}_{10}}}
 \newc{\ffivehat}{\ensuremath{\hat{\phi}_{5}}}
 \newc{\hfivehat}{\ensuremath{\hat{h}_{5}}}
 \newc{\hbfivehat}{\ensuremath{\hat{h}_{\bar{5}}}}
 \newc{\thethat}{\ensuremath{\hat{\theta}_{50}}}
 \newc{\thetbhat}{\ensuremath{\hat{\theta}_{\,\overline{50}}}}
 \newc{\si}{\ensuremath{\Sigma}}
 \newc{\mfive}{\ensuremath{m_5^2}}
 \newc{\mten}{\ensuremath{m_{10}^2}}
 \newc{\dfive}{\ensuremath{\Delta^2_5}}
 \newc{\dbfive}{\ensuremath{\Delta^2_{\bar{5}}}}
 \newc{\dfifty}{\ensuremath{\Delta^2_{50}}}
 \newc{\dfiftyb}{\ensuremath{\Delta^2_{\,\overline{50}}}}
 \newc{\msi}{\ensuremath{m_{\Sigma}^2}}
 \newc{\lamh}{\ensuremath{\lambda_{H}}}
 \newc{\lamhb}{\ensuremath{\lambda_{\bar{H}}}}
 \newc{\ah}{\ensuremath{A_{H}}}
 \newc{\ahb}{\ensuremath{A_{\bar{H}}}}
 \newc{\lams}{\ensuremath{\lambda_{S}}}
 \newc{\as}{\ensuremath{A_{S}}}
 \newc{\lamsig}{\ensuremath{\lambda_{\si}}}
 \newc{\asig}{\ensuremath{A_{\si}}}

 \newc{\msten}{\ensuremath{m_{16}^2}}
 \newc{\mhun}{\ensuremath{m_{126}^2}}
 \newc{\mhunb}{\ensuremath{m_{\bar{126}}^2}}
 \newc{\mthun}{\ensuremath{m_{210}^2}}
 \newc{\ahun}{\ensuremath{A_{\bar{126}}}}
 \newc{\yhun}{\ensuremath{Y_{\bar{126}}}}
 \newc{\aten}{\ensuremath{A_{10}}}
 \newc{\yten}{\ensuremath{Y_{10}}}
 \newc{\alone}{\ensuremath{A_{\lambda_1}}}
 \newc{\altwo}{\ensuremath{A_{\lambda_2}}}
 \newc{\althree}{\ensuremath{A_{\lambda_3}}}
 \newc{\althreeb}{\ensuremath{A_{\bar{\lambda_3}}}}
 \newc{\lone}{\ensuremath{\lambda_1}}
 \newc{\ltwo}{\ensuremath{\lambda_2}}
 \newc{\lthree}{\ensuremath{\lambda_3}}
 \newc{\lthreeb}{\ensuremath{\bar{\lambda_3}}}

\newc{\sigsip}{\ensuremath{\sigma^{\rm SI}_{p}}}	\newc{\sigsin}{\ensuremath{\sigma^{\rm SI}_{n}}}
\newc{\sigsdp}{\ensuremath{\sigma^{\rm SD}_{p}}}	\newc{\sigsdn}{\ensuremath{\sigma^{\rm SD}_{n}}}
\newc{\sigsi}{\ensuremath{\sigma^{\rm SI}}}	\newc{\sigsd}{\ensuremath{\sigma^{\rm SD}}}
\newc{\sigv}{\ensuremath{\sigma v}}
\newc{\abund}{\ensuremath{ \Omega h^2}}
\newc{\omegadm}{\ensuremath{ \Omega_{{\rm DM}}}}     \newc{\abunddm}{\ensuremath{ \Omega_{{\rm DM}} h^2}} 
\newc{\omegam}{\ensuremath{ \Omega_{{\rm m}}}}       \newc{\abundm}{\ensuremath{ \Omega_{{\rm m}} h^2}}
\newc{\omegab}{\ensuremath{ \Omega_{{\rm b}}}}	\newc{\abundb}{\ensuremath{ \Omega_{{\rm b}} h^2}}
\newc{\omegatot}{\ensuremath{ \Omega_{{\rm TOT}}}}
\newc{\omegacdm}{\ensuremath{ \Omega_{{\rm CDM}}}}   \newc{\abundcdm}{\ensuremath{ \Omega_{{\rm CDM}} h^2}}
\newc{\omegalambda}{\ensuremath{ \Omega_{\Lambda}}} \newc{\abundlambda}{\ensuremath{ \Omega_{\Lambda} h^2}}
\newc{\omegarad}{\ensuremath{ \Omega_{{\rm rad}}}}  \newc{\abundrad}{\ensuremath{ \Omega_{{\rm rad}} h^2}}
\newc{\rhocrit}{\ensuremath{ \rho_{\rm crit}}}
\newc{\rhochi}{\ensuremath{ \rho_{\chi}}}
\newc{\abunchi}{\ensuremath{\Omega_\chi h^2}}
\newc{\abundlsp}{\ensuremath{\Omega_{\rm LSP}h^2}}
\newc{\amu}{\ensuremath{ a_{\mu}}}        \newc{\amususy}{\ensuremath{ a_{\mu}^{\mathrm{SUSY}}}}
\newc{\amuexpt}{\ensuremath{ a_{\mu}^{\mathrm{expt}}}}        \newc{\amusm}{\ensuremath{ a_{\mu}^{\mathrm{SM}}}}
\newc\deltaamu{\ensuremath{\Delta a_{\mu}}} \newc{\deltaamususy}{\ensuremath{\delta a_{\mu}^{\mathrm{SUSY}}}}
\newc\gmtwo{\ensuremath{ (g-2)_{\mu}}} 
\newc{\deltagmtwomususy}{\ensuremath{\delta\left(g-2\right)_{\mu}^{\mathrm{SUSY}}}}
\newc{\deltagmtwomu}{\ensuremath{\delta\left(g-2\right)_{\mu}}}
\newc\BR{\ensuremath{\rm BR}}
\newc\bsgamma{\ensuremath{ b\rightarrow s \gamma }}
\newc\bxsgamma{\ensuremath{\overline{B}\rightarrow X_{s}\gamma}}
\newc\brbsgamma{\ensuremath{\BR\left(\bsgamma\right)}}
\newc\brbxsgamma{\ensuremath{\BR\left(\bxsgamma\right)}}
\newc\bsmumu{\ensuremath{B_s\to\mu^+\mu^-}}
\newc\brbsmumu{\ensuremath{\BR\left(B_s\to\mu^+\mu^-\right)}}
\newc\bdmmumu{\ensuremath{\overline{B}_d\to\mu^+\mu^-}}
\newc\bbbarmix{\ensuremath{\overline{B}_s\mbox{-}B_s}}      % B_s mixing
\newc\delmbs{\ensuremath{\Delta M_{B_s}}}
\newc{\butaunu}{\ensuremath{B_u \rightarrow \tau \nu}}
\newc{\brbutaunu}{\ensuremath{\BR\left(B_u \rightarrow \tau \nu\right)}}
\newcommand*{\reftable}[1]{Table~\ref{#1}}         
       
\newcommand*{\reffig}[1]{Fig.~\ref{#1}}
 
        \newcommand*{\refeq}[1]{Eq.~(\ref{#1})}
     \newcommand*{\refsec}[1]{Sec.~\ref{#1}}

\let\oldcite\cite
\renewcommand*{\cite}{~\oldcite}

\newcommand*{\hl}{\ensuremath{h}}

\restylefloat{figure}

\title{\boldmath Reconstructing WIMP properties through an interplay of
  signal measurements in direct detection, Fermi-LAT, and CTA searches
  for dark matter}

\author[a,b]{Leszek Roszkowski,}
\author[a]{Enrico Maria Sessolo,}
\author[a]{\\Sebastian Trojanowski}
\author[a]{and Andrew J.~Williams}

\affiliation[a]{National Centre for Nuclear Research,\\Ho{\. z}a 69, 00-681 Warsaw, Poland}
\affiliation[b]{Department of Physics and Astronomy, University of Sheffield,\\Sheffield S3 7RH, United Kingdom}

\emailAdd{L.Roszkowski@sheffield.ac.uk}
\emailAdd{enrico.sessolo@ncbj.gov.pl}
\emailAdd{sebastian.trojanowski@ncbj.gov.pl}
\emailAdd{andrew.williams.2009@live.rhul.ac.uk}

\abstract{We examine the projected ability to reconstruct the mass, scattering,
and annihilation cross section of dark matter in the new generation of
large underground detectors, XENON-1T, SuperCDMS, and DarkSide-G2, in
combination with diffuse gamma radiation from expected 15~years of
data from Fermi-LAT observation of 46~local spiral dwarf galaxies and
projected CTA sensitivity to a signal from the Galactic Center. To
this end we consider several benchmark points spanning a wide range of
WIMP mass, different annihilation final states, and large enough event
rates to warrant detection in one or more experiments. As previously
shown, below some 100\gev\ only direct detection experiments will in
principle be able to reconstruct WIMP mass well. This may, in case a
signal at Fermi-LAT is also detected, additionally help restricting
\sigv\ and the allowed decay branching rates. In the intermediate range
between some 100\gev\ and up a few hundred GeV, direct and indirect
detection experiments can be used in complementarity to ameliorate the
respective determinations, which in individual experiments can at best
be rather poor, thus making the WIMP reconstruction in this mass range
very challenging.  At large WIMP mass, $\sim1\tev$, CTA will have the 
ability to reconstruct mass, annihilation cross section, and the
allowed decay branching rates to very good precision for the
$\tau^+\tau^-$ or purely leptonic final state, good for the $W^+W^-$
case, and rather poor for $b\bar{b}$. A substantial
improvement can potentially be achieved by reducing the systematic
uncertainties, increasing exposure, or by an additional measurement at
Fermi-LAT that would help reconstruct the annihilation cross section
and the allowed branching fractions to different final states.}
 
\begin{document}
\maketitle
\flushbottom

%%%%%%%%%%%%%%%%%%%%%%%%%%%%%%%%%%%%%%%%%%%%%%%%%%%%%%%%%%%%%%%%%%%%%%%%%%%%%%%%%%%%%
\section{\label{sec:intro}Introduction}
%%%%%%%%%%%%%%%%%%%%%%%%%%%%%%%%%%%%%%%%%%%%%%%%%%%%%%%%%%%%%%%%%%%%%%%%%%%%%%%%%%%%%

%Global changes:
%\begin{itemize}
%\item 
%masses $\ra$  mass
%
%\item 
%cross section or cross-section?
%\item 
%Fermi-LAT or Fermi LAT or FermiLAT?
%
%\item 
%likelihood $\ra$ likelihood function 
%\item 
%uncertainty on $\ra$  uncertainty in or uncertainty of
%\item 
%\item 
%
%\item 
%\item 
%
%\item 
%
%
%\end{itemize}

In recent years several experiments have progressively put the paradigm
of the weakly interactive massive particle (WIMP) as a cold dark
matter (DM) to an increasing test.
%lr *** Weakly interactive massive particles (WIMPs) are to this date among
%lr *** the best candidates for the dark matter (DM) in the Universe. They
%lr *** arise naturally in many extensions of the Standard Model, they were in
%lr *** thermal equilibrium with it in the early Universe, and they can easily
%lr *** be brought to yield the measured value of the relic abundance, thus
%lr *** being prototypical thermal relics.  For these reasons, in recent years
%lr *** the effort to test the WIMP paradigm has been progressively stepped up
%lr *** and increasingly stringent limits on the WIMP scattering and
%lr *** annihilation cross section have been set in different experiments,
%lr *** albeit an unmistakeable WIMP detection signal has yet to be announced.
Currently the strongest bounds on the spin-independent WIMP-proton
scattering cross section, \sigsip, have been achieved in
the Xenon-based underground detector
LUX\cite{Akerib:2013tjd,Akerib:2015rjg} which improved the previous
best limit of XENON100\cite{Aprile:2012nq}. For WIMP mass in the
range $10-100\gev$ the limit is about $\sigsip\lesssim
10^{-45}\textrm{ cm}^2$ at the 90\%~C.L., and it becomes weaker for
larger mass.  While underground detectors have also provided bounds
on the spin-dependent cross
section\cite{Amole:2015pla,Amole:2016pye,Akerib:2016lao}, the
strongest limits on the scattering to the proton, $\sigsdp\lesssim
10^{-40}\textrm{ cm}^2$, have been determined indirectly in neutrino
telescopes\cite{Adrian-Martinez:2013ayv,PhysRevLett.114.141301,Aartsen:2016exj}.
As for measurements of the present-day DM annihilation cross section
times WIMP relative velocity, \sigv, which, following the
convention used in the literature, we will refer to as simply the
annihilation cross section, the strength of the upper bounds also
depends on the WIMP mass.  For $\mchi< 1\tev$ the most
stringent limit comes from the Fermi-LAT Collaboration in the 6-year
data from observation of 15 dwarf spheroidal satellite galaxies of the
Milky Way (dSphs)\cite{Ackermann:2015zua} while for
$\mchi\gsim 1\tev$ from observation of the Galactic Center (GC) with
the air Cherenkov telescope H.E.S.S.\cite{Abramowski:2011hc}.

With experiment probing deeper and deeper into plausible regions of
WIMP cross sections, it is interesting to investigate to what extent
a detection of a genuine WIMP signal in one or more experiments will
allow one to actually reconstruct the WIMP mass and other properties. This issue
will be addressed in the present paper.

The reconstruction abilities in direct detection underground
experiments have been extensively studied\cite{Green:2008rd,Strigari:2009zb,Pato:2010zk,Pato:2012fw,Chou:2010qt,Shan:2011jz,Shan:2011ct,
Kavanagh:2012nr,Kavanagh:2013wba,Peter:2013aha,Newstead:2013pea},
leading to the conclusion that a determination of DM particle
properties from the signal should be achievable only in the relatively
low WIMP mass range $\mchi\lsim100\gev$.  
As regards the reconstruction of the DM properties from indirect detection experiments,
studies were performed for Fermi-LAT, in putative data from the GC\cite{Bernal:2010ip,Bernal:2011pz}, 
showing good accuracy in reconstructed mass below $\sim 200\gev$ for large signals. 
Reconstruction of the DM properties from a spectral analysis of a putative strong signal in dSphs was 
performed in\cite{PalomaresRuiz:2010pn}. Moreover, in\cite{Calore:2014xka} the WIMP mass and annihilation cross section 
were also reconstructed from the real Fermi-LAT Galactic Center Excess data, 
under the assumption that it originates from DM annihilation.  

As regards a complementary
use of direct and indirect detection searches for WIMP reconstruction,
an early study was performed in\cite{Bernal:2008zk} where an interplay
between expected XENON100 sensitivity and projected Fermi-LAT data from the GC was investigated at
$\mchi\lsim50\gev$. More recent studies of the interplay of the projected sensitivity of XENON-1T\cite{Aprile:2015uzo}
and IceCube on reconstruction of the DM properties can be found in\cite{Arina:2013jya} and\cite{Kavanagh:2014rya} (see 
also\cite{Mena:2007ty,Shan:2011ng,Pato:2011de,Cerdeno:2013gqa} for further studies on related topics). 

We believe that, with much new available data and
forthcoming new experiments, it is now an opportune time to update and
enhance such analyses. For definiteness, in this paper we will
focus on direct detection measurements of \sigsip\ -- and will neglect
\sigsdp\ as typically giving a subdominant signal rate -- and on
realistic projected measurements at Fermi-LAT and the Cherenkov Telescope Array (CTA)\cite{Acharya:2013sxa} 
of the diffuse gamma radiation. We will not consider neutrino flux data nor antimatter (positron, antiproton,
antideutron) data, the latter of which are prone to large
astrophysical uncertainties. Finally, we will not in this paper assume
any positive information from the LHC or any future collider, like ILC
or FCC. Indeed, detecting in a collider a WIMP-like particle would
not necessarily imply that it actually comprises DM as such a particle
could be unstable on a cosmological time scale.

From the theoretical point of view,  increasing attention has recently
been given to WIMPs with mass at the 1\tev\ scale, which are a
prediction of several models. For example, in models with GUT-unified
supersymmetry, the most statistically significant region is an almost pure higgsino with $\mchi\sim1\tev$
uniformly over wide ranges of the parameter space (see,
e.g.,\cite{Roszkowski:2014wqa} and references therein). 
It emerges in global analyses as a robust solution most naturally implied by
primarily the 125\gev\ mass value of the Higgs
boson and the correct relic abundance.  The $\sim1\tev$ higgsino features
the present-day \sigv\ around the ``canonical'' thermal expectation
value of $\sim 10^{-26}\textrm{ cm}^3/\textrm{s}$, but is not
uncommon to find regions of the parameter space where \sigv\ can be as
high as $2\times 10^{-25}\textrm{ cm}^3/\textrm{s}$.  
At the same time, because of its very large higgsino component, \sigsip\ is
typically in the range of $10^{-46}$--$10^{-45}\textrm{ cm}^2$.  Other
models\cite{Hall:2011jd,Arvanitaki:2012ps,Hall:2012zp}, also based
on supersymmetry, can feature as the DM particle a 2--3\tev\
wino, which features promising indirect detection prospects (see, e.g.,\cite{Hryczuk:2014hpa}).

The intrinsic characteristics of this type of heavy candidates give
reasons for optimism when it comes to the DM detection prospects in
the immediate future.  Tonne-scale underground detectors like LUX,
XENON-1T\cite{Aprile:2015uzo} or LZ\cite{Malling:2011va}, based on Xe as a target, or even
larger detectors based on Ar, like the proposed DarkSide-G2\cite{Aalseth:2015}
should just about cover the above-mentioned expected range for
\sigsip.  At the same time, it has been shown in several
analyses\cite{Doro:2012xx,Wood:2013taa,Pierre:2014tra,Silverwood:2014yza,Roszkowski:2014iqa,Lefranc:2015pza,Carr:2015hta},
including one by some of us, that CTA is bound to
reach unprecedented sensitivity to \sigv\ for WIMPs with a mass in the TeV regime.

On the other hand, there is also a chance that a discovery will point
to a WIMP with mass much less than a TeV. Indeed, this is what the
possible signal excess at the GC that emerged in Fermi-LAT data a few
years ago\cite{Goodenough:2009gk,Hooper:2010mq} might be hinting 
at.\footnote{However, a non-DM explanation has recently been proposed\cite{Bartels:2015aea}.} 
In that case, some recent studies have shown that the signal can be
fitted by WIMPs whose \sigv\ is around the canonical thermal value and
the mass in the range $20-40\gev$ (see, e.g.,\cite{Calore:2014xka}) or $100-200\gev$ (e.g.,\cite{Agrawal:2014oha,Calore:2014nla,Caron:2015wda,Gherghetta:2015ysa,Bertone:2015tza}),
depending on the annihilation final state.  For a WIMP with these
characteristics, then, the most likely venue for discovery is at the
moment given by Fermi-LAT. The Collaboration will continue analyzing
data from dSphs, as more of these objects are discovered.  In this paper we consider projected data 
in a hypothetical 15 year 46 dSphs analysis,
which should improve the discovery reach in the low WIMP mass region.
 
There are some insurmountable barriers plus several sources of uncertainty that can interfere with the
ability to reconstruct the correct WIMP mass, scattering, or
annihilation cross section. As pertains to direct detection,
particularly troubling is the fact that, for $\mchi\gsim100\gev$, event spectra become basically independent of \mchi. 
Additionally there are uncertainties in the density and velocity
distribution of the DM in the galactic halo\cite{Vogelsberger:2008qb,Ling:2009eh,2010A&A...509A..25W,Catena:2011kv,2012sf2a.conf..121S}, 
and uncertainties in the detectors' response.  For indirect detections based on gamma
rays from the GC, in addition to the above-mentioned uncertainties associated to
the halo and its profile\cite{1990ApJ...356..359H,Navarro:1995iw,Burkert:1995yz,Einasto:2009zd,Graham:2005xx} 
there are also uncertainties in the cosmic-ray flux and in
how the molecular clouds affect secondary gamma-ray emission. For
dSphs, uncertainties in the $J$-factors can play an important role. 

Assuming a genuine signal is detected in one or more experiment,
direct or in the diffuse gamma ray mode, and given a realistic
assessment of the current and future uncertainties, the questions we want to address in this paper are the following: 
\begin{itemize}
\item How well can the regions of  (\mchi, \sigsip) and (\mchi, \sigv)
  be determined by direct detection and gamma ray experiments,
  respectively, as well as by a complementary use of both to help disentangle possible degeneracies?
\item How well can WIMP mass reconstruction be achieved in the gamma-ray
  experiments Fermi-LAT and CTA?
\item How well can Fermi-LAT and CTA do in WIMP annihilation final state reconstruction?
\item What are the challenges and possible obstacles to obtain improved results? 

%lr *** \item Are there regions of the (\mchi, \sigv) and (\mchi, \sigsip) parameter space where complementary use of different experiments 
%lr *** can help disentangle possible degeneracies?
%lr *** \item How good is CTA for mass reconstruction?
%lr *** \item How good is CTA for final state reconstruction?
%lr *** \item What should be improved to obtain better results? 
\end{itemize}

%Additionally, we will also investigate the impact of the astrophysical
%uncertainties in the case when one already knows the final state, were
%because it was measured somewhere else, or because of theoretical
%inference.

Our strategy will be as follows. We will assume some ``true'' DM WIMP
by specifying its mass and other properties relevant to the type
of measurement under consideration. For instance, in direct detection we will specify
\mchi\ and \sigsip.  We will then generate a MC sample of
possible other configurations which could mimic the signal from the
assumed ``true'' DM WIMP. We emphasize that in this paper we will not be
assuming any particle physics model or scenario. For this reason we
will therefore not be imposing any limits from colliders nor from the
relic abundance of dark matter.

The paper is organized as follows. In \refsec{sec:like} we 
review the formalism for generating a mock signal in several 
selected direct and indirect detection experiments. We consider different 
nuclear targets for direct detection, and for indirect detection we consider 
gamma rays from dSphs and the GC. We conclude the section by presenting the 
explicit form of the likelihood functions. In \refsec{sec:bench}
we describe the selected benchmark points that will generate the signals to fit 
and the scanning methodology. We present the reconstruction results
in \refsec{sec:results}, and we conclude in \refsec{sec:summary}.

%%%%%%%%%%%%%%%%%%%%%%%%%%%%%%%%%%%%%%%%%%%%%%%%%%%%%%%%%%%%%
\section{\label{sec:like}Reconstructing a signal}
%%%%%%%%%%%%%%%%%%%%%%%%%%%%%%%%%%%%%%%%%%%%%%%%%%%%%%%%%%%%%%

We proceed to define the mock signals we will be trying to reconstruct to a $\sim2\sigma$ precision. 
As was mentioned in \refsec{sec:intro}, we focus in this paper on the
measurements of \sigsip\ in direct detection search experiments and of
\sigv\ and gamma-ray spectra in Fermi-LAT and CTA. 
%lr ***   those experiment showing
%lr *** the greatest promise for detection of a signal in ``traditional" WIMP scenarios like the 
%lr *** Minimal Supersymmetric Standard Model. 
We leave an analysis of other experimental modes to future work.  

\subsection{Signal in underground detectors}

Experiments like XENON, LUX, DarkSide\cite{Agnes:2014bvk}, SuperCDMS\cite{Brink:2005ej}, and other, which are based
on a specific fiducial volume of some target material placed in deep
underground and well shielded detectors, measure the nuclear recoil
differential rate $dR/dE_R$ of the struck nucleon.

Following several papers\cite{Green:2008rd,Strigari:2009zb,Pato:2010zk,Pato:2012fw,Chou:2010qt,Shan:2011jz,Shan:2011ct,
Kavanagh:2012nr,Kavanagh:2013wba,Peter:2013aha,Newstead:2013pea} that performed similar analyses
we parametrize the recoil energy in terms of a set of input
parameters, $\{p\}_{\textrm{DD}}=\{\mchi, \sigsip, v_0,
v_{\textrm{esc}}, \rho_0\}$, where 
% lr *** \mchi\ is the WIMP mass, \sigsip\ the spin-independent
% WIMP-proton(nucleon) scattering cross section,
$v_0$ is the circular velocity of the DM halo, $v_{\textrm{esc}}$ is the
Galactic escape velocity at the Sun's position, and $\rho_0$ is the
local DM density. Explicitly one gets
\be 
\frac{dR}{dE_R}=\frac{\sigsip}{2\mchi \mu_{\chi p}^2} A^2 F_N^2(E_R)\mathcal{G}(v_{\textrm{min}},v_{\textrm{esc}})\,,\label{DDrate}
\ee
where $\mu_{\chi p}^2$ is the reduced mass of the WIMP-nucleon system,
$A$ is the atomic mass number of the target nucleus,
$F_N^2(E_R)$ is the nuclear form factor of the target
nucleus, and $\mathcal{G}(v_{\textrm{min}},v_{\textrm{esc}})$
parametrizes the DM velocity distribution in the halo. Here we
consider it as given by an integration of a Maxwell-Boltzmann
distribution, $f(\mathbf{v},v_0)$, and neglect the uncertainties associated with deviating from it\cite{Kuhlen:2009vh,McCabe:2010zh,Green:2010gw},  
\be 
\mathcal{G}(v_{\textrm{min}},v_{\textrm{esc}})=\rho_0\int_{v_{\textrm{min}}<|\mathbf{v}|<v_{\textrm{esc}}}\frac{f(\mathbf{v},v_0)}{|\mathbf{v}|}d^3v\,.\label{astro}
\ee 
The ``minimal'' velocity, $v_{\textrm{min}}$, also depends on the recoil energy $E_R$ and atomic mass of the target 
$m_N$\,,\footnote{We assume here elastic scattering of DM particles off nuclei (for a discussion of
inelastic scattering see, e.g.,\cite{Peter:2013aha} and references therein).}
\be
v_{\textrm{min}}=\frac{1}{\sqrt{2 E_R m_N}}\left(\frac{E_R m_N}{\mu_{_{\chi N}}}\right)\,.
\ee

Because the experiments are in general well shielded by
construction, and the purity of the target material is very high, we
neglect in this paper a very small residual background which is
usually further reduced by applying one or another discrimination
method.  The effect of the residual backgrounds on WIMP property
reconstruction has been considered in the
literature\cite{Chou:2010qt,Shan:2011ct}.
Note that we also neglect in this paper the 
limitation arising from the statistical scattering of the measured event around the real $dR/dE_R$\cite{Strege:2012kv}.

One can obtain information on the input parameters
$\{p\}_{\textrm{DD}}$ by confronting the expected signal times
exposure to the (mock) measured signal.  This is done here through a
binned likelihood function that is a product of Poisson distributions
independently evaluated in $i=1,...,N_{\textrm{DD}}$ energy bins
equally spaced in logarithmic intervals.
We choose here logarithmic intervals to overcome possible bias towards the low recoil-energy bins. 
This can occur when the signal is not very strong, as the tail of the event distribution becomes 
indistinguishable from zero in the bins above a certain cutoff energy, which does not in general coincide with the nominal 
cutoff energy of the experiment.

In each bin the measured number of events,
$n_i$, is compared to the calculated signal, $\mu_i(\{p\}_{\textrm{DD}})$, given by
\be 
\mu_i=\textrm{exposure}\times\int_{E_{R,i-1}}^{E_{R,i}}\frac{dR}{dE_R} dE_R\,.\label{observation}
\ee 
The likelihood function reads
\be 
\mathcal{L}_{\textrm{DD}}=\prod_{i=1}^{N_{\textrm{DD}}}\frac{\mu_i^{n_i}e^{-\mu_i}}{n_i!}\,.\label{likeDD}
\ee

%\be 
%\mathcal{G}(v_{\textrm{min}})=\frac{\rho_0}{2v_{\textrm{esc}}}\left[\erf\left(\frac{v_{\textrm{esc}}-v_{\textrm{min}}}{v_0}\right)+
%\erf\left(\frac{v_{\textrm{esc}}+v_{\textrm{min}}}{v_0}\right)-\erf\left(\frac{v_{\textrm{esc}}-v_{\textrm{max}}}{v_0}\right)-
%\erf\left(\frac{v_{\textrm{esc}}+v_{\textrm{max}}}{v_0}\right)\right]\,,
%\ee 
%which follows from an explicit integration of a Gaussian $f(\mathbf{v})$ in \refeq{astro}~[cita].

%%%%%%%%%%%%%%%%%%%%%%%%%%%%%%%%%%%%%%%%%%%%%%%%%%%%%%%%%%%%%%%%%%%%%

\subsection{Gamma-ray signals from dwarf spheroidal galaxies and the Galactic Center}

We will now consider a WIMP discovery in gamma-ray observatories.
Because of their highly reduced backgrounds, the targets most
sensitive to observation of DM in the $10-250\gev$ range are the dSphs
from the Local Group.
To analyze a realistic scenario we consider here the sensitivity
obtained with 15~years data, from observation of 46 dSphs.  
We mimic 46 dSphs by modeling the signal in the presently discovered 23 dSphs, and successively
doubling it.
Conversely, for DM masses above 250\gev\ the highest sensitivity
is expected at CTA in observation of the GC.

The expected differential flux from WIMP annihilation depends on the input parameters
$\{p\}_{\gamma\textrm{-rays}}=\{\mchi,\sigv,f\}$, where $f$ parametrizes a set of branching ratios to 
different annihilation final states.
In the case of one dSphs, the observed flux reads 
\be
\left(\frac{d\Phi}{dE}\right)_{\textrm{dSphs}}=\frac{\sigma v}{8\pi\mchi^2}\,J\frac{dN_{\gamma}}{dE}\,,\label{dSphsflux}
\ee
where $J$ is the $J$-factor measured for all dSphs from velocity dispersion relations\cite{2009ApJ...704.1274W,2010MNRAS.406.1220W,2015MNRAS.451.2524M,Ackermann:2013yva} and 
$dN_{\gamma}/dE$ is the prompt gamma-ray spectrum from WIMP annihilation, which depends on the annihilation final state.

In the case of observation of the GC, one must additionally factor in
the large uncertainties of the halo profile determinations.  
We do so by assuming a generalized Navarro-Frenk-White (NFW) profile\cite{Navarro:1995iw}
\be 
\rho(r)=\frac{\rho_0\left(1+\frac{R_{\odot}}{r_s}\right)^{3-\gamma_{\textrm{NFW}}}}{\left(\frac{r}{R_{\odot}}\right)^{\gamma_{\textrm{NFW}}}\left(1+\frac{r}{r_s}\right)^{3-\gamma_{\textrm{NFW}}}}
\ee 
where we fix the scale radius $r_s=20\textrm{ kpc}$ and the distance of the Solar System from the GC, 
$R_{\odot}=8.5\textrm{ kpc}$.
This introduces two new free input parameters to our analysis: 
$\{p\}_{\gamma\textrm{-rays}}=\{\mchi,\sigv,f, \rho_0, \gamma_{\textrm{NFW}}\}$.
The $J$-factor for a region of the GC characterized by a solid angle $\Delta\Omega$ is 
given by integrating $\rho^2(r)$ along the line of sight in the corresponding angular region,
\be 
J_{\Delta\Omega}=\int_{\Delta\Omega} \int_{\textrm{l.o.s.}} 
\rho^2[r(\theta)]dr(\theta)d\Omega\,.
\ee 
 
A calculation of the flux from the GC must take into account the
prompt spectrum and the secondary spectrum from inverse Compton (IC)
scattering of electrons off the CMB, starlight, and IR
radiation\cite{Cirelli:2010xx,Lefranc:2015pza}, so that one gets
\be
\left(\frac{d\Phi}{dE}\right)_{\textrm{GC}}=\frac{\sigma v}{8\pi\mchi^2}\left(J_{\Delta\Omega}\frac{dN_{\gamma}}{dE}
+\frac{1}{E^2}\int_{m_e}^{\mchi}dE_s 
\bar{I}_{\textrm{IC},\Delta\Omega}(E,E_s)\frac{dN_{e^{\pm}}}{dE_s}\right)\,,\label{DMflux}
\ee
where $dN_{e^{\pm}}/dE$ stands for the electron/positron spectra from WIMP annihilation.
The quantity $\bar{I}_{\textrm{IC},\Delta\Omega}$ parametrizes the cumulative effect of the differential IC 
radiation power per unit of Galactic coordinates\cite{Cirelli:2010xx}, which we here call $I_{\textrm{IC}}(E_{\gamma},E_s,l,b)$. 
It is explicitly given by
\begin{equation}
\bar{I}_{\textrm{IC},\Delta\Omega}(E_{\gamma},E_s)=4 R_{\odot}\rho_0^2\int_{\Delta l,\Delta b} I_{\textrm{IC}}(E_{\gamma},E_s,l,b) dl db\cos b\,,
\end{equation}
where $R_{\odot}$ is the distance of the Sun from the GC
and we use for the computation the form of $I_{\textrm{IC}}(E_{\gamma},E_s,l,b)$ provided by\cite{Cirelli:2010xx}.

\subsubsection{Likelihood function for Fermi-LAT dSphs\label{sec:LAT}}

We consider a binned likelihood function for the putative data in 46 dSphs at Fermi-LAT. 

The gamma-ray flux, \refeq{dSphsflux}, is binned into $N_{\textrm{Fermi}}=17$ energy bins 
and then confronted to the putative observed residual flux. 
Each different dwarf galaxy is labeled by the index $j=1,...,46$ so that 
the expected flux in each bin is given by $\frac{d\Phi_j}{dE_i}\equiv\frac{d\Phi_j}{dE}(E_i)$\,.
 
For each dSphs the corresponding $J$-factor, $\bar{J}_j$, has been computed by the Fermi-LAT Collaboration
with a relative logarithmic uncertainty, $\sigma_{j}$\cite{Ackermann:2013yva}.
The expected flux is compared to the mock observed flux in bin $ij$, $\frac{d\bar{\Phi}_j}{dE_i}$, 
through the likelihood function    
\begin{multline}
\mathcal{L}_{\textrm{dSphs}}=\prod_{j=1}^{46}\left\{\int \frac{dJ_j}{\log(10)\bar{J}_j\sqrt{2\pi}\sigma_j}\exp\left[-\frac{(\log_{10} J_j-\log_{10}\bar{J}_j)^2}
{2\sigma_j^2}\right]\right.\times\\
\times\left.\prod_{i=1}^{N_{\textrm{Fermi}}}\frac{1}{\sqrt{2\pi}\overline{\sigma}_{ij}}\exp\left[-\frac{\left(\frac{d\Phi_j}{dE_i}-\frac{d\bar{\Phi}_j}{dE_i}\right)^2}
{2\overline{\sigma}_{ij}^2}\right]\right\}\,.\label{likeFL}
\end{multline}

The uncertainties $\overline{\sigma}_{ij}$ are estimated by generating
the total expected flux from the signal, as well as the isotropic and diffuse
background\cite{Fermi1} for each dwarf galaxy. This is translated into the total number
of expected events for the observation period by convolving with the
exposure map and observation time\cite{Fermi2}. 
The uncertainty in the total number of gamma rays in a given 
observation region and energy bin is then taken as the Poisson
uncertainty.  The error in the residual flux is calculated as
the uncertainty in the total number of gamma rays converted back into
a flux by dividing by the exposure.  
Note that the uncertainties $\overline{\sigma}_{ij}$ are thus related to
the backgrounds in each dwarf galaxy and each energy bin and therefore also
depend on $J_j$.

\subsubsection{Likelihood function for CTA\label{sec:CTA}}

\begin{figure}[t]
\centering
\includegraphics[width=0.50\textwidth]{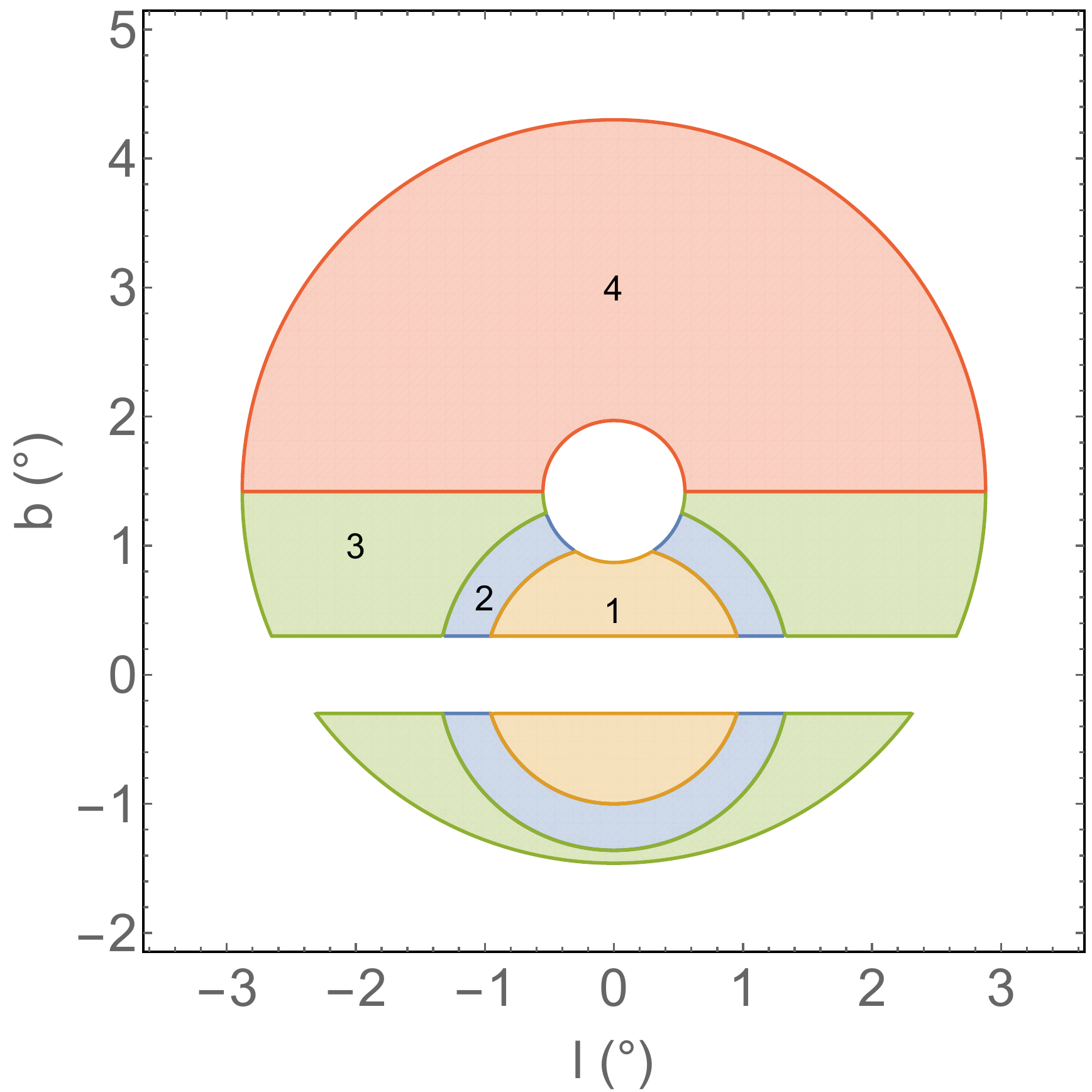}
\caption{ Our generalization of the Ring Method.  Region~1 (beige), is inside the circle of angular radius $\Delta\leq
  1.0^{\circ}$.  Region~2 (light blue) is the angular corona included
  in $1.00^{\circ}<\Delta\leq 1.36^{\circ}$. Region~3 is marked with
  light green and Region~4 with salmon pink. The ring's center is
  offset from the GC by $b_{\textrm{off}}=1.42^{\circ}$. Its inner
  radius is $r_1=0.55^{\circ}$ and its outer radius is
  $r_2=2.88^{\circ}$.  The strip of the sky at $|b|<0.3^{\circ}$
  (Galactic plane) does not belong to any of the regions.
%lr *** Our generalization of the Ring Method. In orange, Region~1, 
%lr ***   included in the circle of angular radius $\Delta\leq 1.0^{\circ}$.
%lr ***   In blue, Region~2: it is the angular corona included in
%lr ***   $1.00^{\circ}<\Delta\leq 1.36^{\circ}$.  In green Region~3 and in
%lr ***   salmon pink Region~4. The ring's center is offset from the GC by
%lr ***   $b_{\textrm{off}}=1.42^{\circ}$. The ring's inner radius is
%lr ***   $r_1=0.55^{\circ}$ and the outer radius is $r_2=2.88^{\circ}$.  The
%lr ***   strip of sky at $|b|<0.3^{\circ}$ does not belong to any of the
%lr ***   regions.  
}
\label{fig:regions}
\end{figure}

To construct the likelihood function for CTA we use a generalization
of the \textit{Ring Method} based on four spatial regions or patches
in sky.\footnote{The Ring Method is generally based on two spatial
  regions\cite{Doro:2012xx,Wood:2013taa,Pierre:2014tra,Silverwood:2014yza}.
  We will show in \refsec{sec:results} that, by adding one region to the
  traditional two, one improves significantly the background reduction, and
  the results do not then improve as much by adding
  additional patches.}  The four regions we consider are identified in
the plane of the Galactic angular coordinates $l$ and $b$, as shown in
\reffig{fig:regions}.  Region~1, in which DM annihilations are likely to play a more important role than
in other places in the sky, is defined as a circle of angular radius
$\Delta_1=1.0^{\circ}$ around the GC.  Region 2 covers the angular
corona included between $\Delta_1$ and $\Delta_2=1.36^{\circ}$.  In addition 
we define 
two regions (Regions 3 and 4) where the expected count has only a minor contribution from the DM signal. 
Regions 3 and 4 are respectively included in
the lower and upper halves of a ring centred at the offset coordinate
$b_{\textrm{off}}=1.42^{\circ}$.  The inner angular region of radius
$r_1=0.55^{\circ}$ is not considered, whereas the outer radius of the
ring including Regions~3 and~4 is $r_2=2.88^{\circ}$.  Regions~1, 2,
and 3 belong to the lower half of the ring of outer radius $r_2$ without
overlapping.  The strip of sky characterised by $|b|<0.3^{\circ}$
about the Galactic plane does not belong to any of the considered regions.

%%%%%%%%%%%%%%%%%%%%%%%%%%%%%%%%
%%%%%%%%%%%%%%%%%%%%%%%%%%%%%%%%

We bin the $\gamma$-ray spectra into $N_{\textrm{CTA}}=30$ energy
bins, $i=1,..,N_{\textrm{CTA}}$, logarithmically spaced.  Regions 1 to
4 are labeled with an index $j$.  For each bin, $ij$, the signal is
assumed to be composed of three parts: the DM signal, $\mu^{\textrm{DM}}_{ij}$,
which originates from WIMP annihilations in the halo; the isotropic
background $\mu^{\textrm{CR}}_{ij}$ from cosmic-ray (CR) showers,
which is obtained by detailed MC simulation and was provided to us by
the CTA Collaboration\cite{JCarr}; and the Galactic Diffuse Emission (GDE)
background\cite{2012ApJ...750....3A,Ackermann:2014usa,Fermi2}, 
$\mu^{\textrm{GDE}}_{ij}$, which is obtained from Fermi-LAT data for the energy bins below 500\gev, and
extrapolated with a power law for higher energies, as described, e.g.,
in\cite{Silverwood:2014yza,Lefranc:2015pza}.

Explicitly, this reads  
\be 
\mu^{\textrm{DM}}_{ij}=t_{\textrm{obs}}\,\int_{\Delta E_i}dE \frac{1}{\sqrt{2 \pi \delta(E)^2}} \int_{30\gev}^{\mchi}dE'\,
\left(\frac{d\Phi_j}{dE'}\right)_{\textrm{GC}}
A_{\text{eff}}(E')e^{-\frac{(E - E')^2}{2\delta(E)^2}}\,,\label{DMsign}
\ee
where $A_{\text{eff}}$ is the effective area of the detector,
$\delta(E)$ is the energy resolution, and $(d\Phi_j/dE)_{\textrm{GC}}$
is the gamma-ray flux from WIMP annihilation in the $j$th region of
the sky, defined in \refeq{DMflux}.  For $A_{\text{eff}}$ and
$\delta(E)$ we use recent instrument response functions
provided by the CTA Collaboration\cite{JCarr}.  
The $J$-factors, $J_j$, are calculated over each region of the sky.

In order to obtain the total count number per bin $\mu_{ij}$, which is to be confronted 
with our putative signal, $n_{ij}$, the backgrounds must be added to the DM signal,
\be 
\mu_{ij}\left(R^{\textrm{CR}}_i,R^{\textrm{GDE}}_i\right)=\mu^{\textrm{DM}}_{ij}+R^{\textrm{CR}}_i \mu^{\textrm{CR}}_{ij}+R^{\textrm{GDE}}_i \mu^{\textrm{GDE}}_{ij}\,,\label{BGunc}
\ee 
where we parametrize the uncertainty in the normalization of the CR
and GDE backgrounds with additional energy bin-dependent factors
$R^{\textrm{CR}}_i$ and $R^{\textrm{GDE}}_i$.  Throughout the paper we
assume that the measured value of the total CR flux and of the GDE at
energies $<500\gev$ indirectly imply that $R^{\textrm{CR}}_i$ and
$R^{\textrm{GDE}}_i$ are normally distributed around 1. We adopt
conservative uncertainties $\sigma_{\textrm{CR}}=10\%$ and
$\sigma_{\textrm{GDE}}=20\%$.  The CTA likelihood function is thus
given by
\begin{multline}
\mathcal{L}_{\textrm{CTA}} = \prod_{i=1}^{N_{\textrm{CTA}}}\left\{ \int dR^{\textrm{CR}}_i e^{-\frac{\left(1-R^{\textrm{CR}}_i\right)^2}{2\sigma_{\textrm{CR}}^2}} \int dR^{\textrm{GDE}}_i 
e^{-\frac{\left(1-R^{\textrm{GDE}}_i\right)^2}{2\sigma_{\textrm{GDE}}^2}}\right.\\
\left.\times\left[\prod_{j=1}^{4} 
\frac{\mu_{ij}\left(R^{\textrm{CR}}_i,R^{\textrm{GDE}}_i\right)^{n_{ij}}}{n_{ij}!} e^{-\mu_{ij}\left(R^{\textrm{CR}}_i,R^{\textrm{GDE}}_i\right)}\right]\right\}\,.\label{binlike}
\end{multline}

Note that, neglecting for a moment the spread in $\rho_0$ and $\gamma_{\textrm{NFW}}$, a minimum
number of three regions in the sky is necessary to constrain
independently the three quantities at the origin of the flux
normalization: $\sigv$, $R^{\textrm{CR}}_i$, and $R^{\textrm{GDE}}_i$.
It has been previously shown in the literature\cite{Silverwood:2014yza} that, when
deriving CTA upper bounds on the DM annihilation cross section, a
morphological analysis that divides the sky in several regions leads
to a significant increase in sensitivity with respect to the
traditional Ring Method.  We will see in \refsec{sec:results} that
for reconstructing the DM properties the greatest improvement is
obtained by extending the Ring Method by one region, even in the case 
where $\rho_0$ and $\gamma_{\textrm{NFW}}$ are treated as nuisance parameters,
and that adding even more patches
introduces a more moderate, incremental improvement.

Finally, note that we do not include in this study systematic uncertainties in the CTA detector response\cite{Silverwood:2014yza}, 
which are to some extent unknown at this point, and further systematic effects that 
can be present such as varying acceptance across the field of view or uncertainties in the effective area. 

%We find the 95\%~C.L. limit by setting $n_{ij}$ equal to the number of background-only photons 
%and calculating ${\cal L}$ by increasing the annihilation cross section from the best fit value (\sigv = 0) 
%until the difference in $-2 \ln {\cal L}$ from the best fit value is 2.71 (one-sided 95\%~C.L.).
%We use an updated\cite{JCarr} background estimate provided by the CTA Collaboration\cite{montecarlo}.

%%%%%%%%%%%%%%%%%%%%%%%%%%%%%%%%%%%%%%%%%%%%%%%%%%%%%%%%%%%%%%%%%%%%%%%%%%%%%%%%%%%%%%%%%%%%%%%%%%%%%%%%%

%%%%%%%%%%%%%%%%%%%%%%%%%%%%%%%%%%%%%%%%%%%%%%%%%%%%%%%%%%%%%%%%%%%%%%%%%%%%%%%%%%%%
\section{\label{sec:bench}Benchmark points and scanning methodology}
%%%%%%%%%%%%%%%%%%%%%%%%%%%%%%%%%%%%%%%%%%%%%%%%%%%%%%%%%%%%%%%%%%%%%%%%%%%%%%%%%

We engineer eight benchmark points to generate mock signals in the experiments described in \refsec{sec:like}. 
The benchmark point parameters are summarized in \reftable{tab:bench}.

%%%%%%%%%%%%%%%%%%%%%%%%%%%%%%%%%%%%%%%%%%%%%%%%%%%%%%%%%%%%%%%%%%
\begin{table}[bt!]
   \centering\footnotesize
   %\topcaption{Table captions are better up top} % requires the topcapt package
   \begin{tabular}{|c|c|c|c|c|c|} % Column formatting, @{} suppresses leading/trailing space
      \hline
       & \textbf{BP1} & \textbf{BP2} & \textbf{BP3} & \textbf{BP4(a, b, c, d)} & \textbf{BP5}  \\
      \hline
      \hline
       \mchi & 25\gev & 100\gev & 250\gev & 1000\gev & 1000\gev \\
       \hline
       \sigv & $8\times 10^{-27}\textrm{ cm}^3/\textrm{s}$ & $2\times 10^{-26}\textrm{ cm}^3/\textrm{s}$ & $4\times 10^{-26}\textrm{ cm}^3/\textrm{s}$ & $2\times 10^{-25}\textrm{ cm}^3/\textrm{s}$ & $3\times 10^{-26}\textrm{ cm}^3/\textrm{s}$ \\
       \hline
       \sigsip & $2\times 10^{-46}\textrm{ cm}^2$ & $3\times 10^{-46}\textrm{ cm}^2$ & $5\times 10^{-46}\textrm{ cm}^2$ & $2\times 10^{-45}\textrm{ cm}^2$ & $2\times 10^{-45}\textrm{ cm}^2$ \\
       \hline
       Final state & & & & (a) $b\bar{b}$ (b) $W^+W^-$ & \\
       (hadronic scans) & $b\bar{b}$ & $b\bar{b}$ & $b\bar{b}$ & (c) $\tau^+\tau^-$ & $W^+W^-$ \\
      \hline
       Final state & & & & & \\
       (leptonic scan) &  & & & (d) $\mu^+\mu^-$ & \\
       \hline
   \end{tabular}
   \caption{Parameters of the 8 benchmark points that ought to be reconstructed through our scanning procedure. In ``hadronic'' scans we 
   allow for 4 final-state channels such that the branching fractions sum up to 1: $f_{b\bar{b}}+f_{W^+W^-}+f_{hh}+f_{\tau^+\tau^-}=1$. 
   In the ``leptonic'' scan we 
   allow for 3 final-state channels: $f_{b\bar{b}}+f_{\mu^+\mu^-}+f_{\tau^+\tau^-}=1$.}
   \label{tab:bench}
\end{table}

%%%%%%%%%%%%%%%%%%%%%%%%%%%%%%%%%%%%%%%%%%%%%%%%%%%%%%%%%%%%%%%%%%%%

We re-emphasize that our benchmark points do not assume any particular
particle theory model. We thus treat \sigsip, \sigv, and the final
state branching ratios as independent parameters that are only
constrained by experiments. Nor do we impose the relic density
constrain.  Note that although in several models \sigsip\ and \sigv\
can feature some degree of correlation, this is not always warranted. For example, 
in the familiar case of the MSSM \sigsip\ and \sigv\
do not show correlation in global analyses (see, e.g.,\cite{Roszkowski:2014iqa}).

Given four benchmark masses selected across the WIMP mass range
typically considered in the literature ($\mchi=25$, 100, 250,
1000\gev), we choose the benchmark values of \sigsip\ and \sigv\ such
that they can lead to a strong signal in either direct or indirect detection
experiments, or both, given realistic expectations for the projected future
sensitivity. At the same time, we make sure these points are not excluded
by the current experimental bounds.  In particular, we choose \sigv\
to lie on the published 95\%~C.L. exclusion line of the most recent
analysis that combines the 6 year PASS 8 data for 15 dSphs at Fermi-LAT
with observations from the GC at MAGIC\cite{Ahnen:2016qkx}.

Here we briefly comment on our benchmark points.

\begin{itemize}
\item {\bf BP1}: This is the case of a relatively light WIMP, $\mchi=25\gev$,
  annihilating exclusively (100\% branching ratio) into hadronic
  products ($b\bar{b}$ final state). 
%lr *** WIMPs with this characteristics
%lr ***   are somewhat motivated by the GC excess~[Gherghetta] and by some
%lr ***   models (\es{any ideas which?}).  
The strongest
  sensitivity to \sigsip\ in Xenon underground detectors is achieved
  approximately at this mass, and the current direct detection bound
  is $\sigsip\lesssim 10^{-45}\textrm{ cm}^2$.

\item {\bf BP2}: A benchmark characterized by $\mchi=100\gev$, possibly motivated by models with rough expectations of EW naturalness. 
The sensitivity of direct detection experiments currently implies $\sigsip\lesssim2\times 10^{-45}\textrm{ cm}^2$. 
The annihilation final state is set to 100\% $b\bar{b}$.

\item {\bf BP3}: A benchmark characterized by $\mchi=250\gev$, situated at the possible onset of the Fermi-LAT/CTA interplay for mass and final state reconstruction. 
The annihilation final state is set to 100\% $b\bar{b}$.

\item {\bf BP4(a)}: As was mentioned in \refsec{sec:intro}, 
a $\sim 1\tev$ WIMP is characteristic of many GUT-constrained models based on supersymmetry with neutralino DM.

Even if we do not make any assumptions on the relic density, it is worth pointing out that the Fermi-LAT/MAGIC bound\cite{Ahnen:2016qkx} for an $\mchi=1000\gev$ WIMP with 100\% $b\bar{b}$ 
final state reads $\sigv=2\times10^{-25}\textrm{ cm}^3/\textrm{s}$. This is a value that
is larger than the cross section expected for a ``canonical'' thermal candidate. 
However, it is well known that the present-day annihilation cross section does not need to coincide with 
the thermal annihilation cross section in the early Universe. Examples can be easily found in which
the WIMP has the correct relic abundance but the present-day annihilation cross section is slightly enhanced, for example
by an $s$-channel resonance with the exchange of a particle with mass $\approx 2\mchi=2\tev$ (in the case of 
supersymmetry this particle is often the pseudoscalar Higgs).

\item {\bf BP4(b, c, d)}: We investigate the ability of CTA to
  reconstruct annihilation final states different from pure $b\bar{b}$ for
  the same \sigv.

\item {\bf BP5}: We also consider the case of a 1\tev\ WIMP with the
  canonical thermal $\sigv\approx3\times 10^{-26}\textrm{
    cm}^3/\textrm{s}$ and 100\% braching ratio to $W^+W^-$ 
(typical, again, of some popular supersymmetric candidates but present also in other models\cite{Goudelis:2013uca}). 

\end{itemize} 

%%%%%%%%%%%%%%%%%%%%%%%%%%%%%%%%%%%%%%%%%%%%%%%%%%%%%%%%%%%%%%%%%%
\begin{table}[t]
   \centering\footnotesize
   %\topcaption{Table captions are better up top} % requires the topcapt package
   \begin{tabular}{|c|c|c|c|} % Column formatting, @{} suppresses leading/trailing space
      \hline
       & \textbf{Acceptance} & \textbf{Recoil energy range} & \textbf{Exposure} \\
      \hline 
      \hline
       XENON-1T (Xe) & 40\% & $4-50\kev$ & 730~ton days (all BPs)\\
      \hline
       & & & 1150~ton days (BP1)\\
      SCDMS-Snolab & 50\% & $8-115\kev$ & 1100~ton days (BP2)\\
       (Ge) & & & 950~ton days (BP3)\\
       & & & 900~ton days (BP4)\\
      \hline
    & & & $-$ (BP1)\\
      DarkSide-G2 & Ref.\cite{Agnes:2014bvk} & $38-200\kev$ & 7300~ton days (BP2)\\
       (Ar) & & & 4300~ton days (BP3)\\
       & & & 3300~ton days (BP4)\\
      \hline
   \end{tabular}
   \caption{The experiments for which we consider a putative signal in direct detection. 
   Exposure has been adjusted in Ge and Ar experiments to reach a sensitivity 
to the benchmark points equal to the one in XENON-1T with 730~ton~days.
For DarkSide-G2 we consider a recoil-energy dependent acceptance, see Fig. 6 in Ref.\cite{Agnes:2014bvk}.}
   \label{tab:exp}
\end{table}

%%%%%%%%%%%%%%%%%%%%%%%%%%%%%%%%%%%%%%%%%%%%%%%%%%%%%%%%%%%%%%%%%%%%   

For each of the benchmark points we perform a global scan of the
parameter space, where the scans are guided by a global likelihood
function that includes the different pieces described in
\refsec{sec:like}.  To each piece there corresponds a different experiment:
three direct detection experiments and two gamma-ray telescopes.  
To make contact with current and realistic or planned experiments we
indicatively select XENON-1T\cite{Aprile:2015uzo} for a Xe target, SuperCDMS-Snolab\cite{Sander:2012nia,Brink:2012zza} 
for a Ge target, and for Ar the planned experiment DarkSide-G2\cite{Aalseth:2015}.  
The schematic characteristics of the three experiments considered here are
given in \reftable{tab:exp}.  
In this paper we use units of ton~days to quatify exposure in direct detection experiments.
Note that the exposure has been adjusted
in Ge and Ar experiments to reach the sensitivity to the signal
generated by the benchmark points equal to the one in XENON-1T with $\sim2$
years of data.  Note that for DarkSide-G2 we consider a recoil-energy
dependent acceptance, which we borrow from Fig. 6 of Ref.\cite{Agnes:2014bvk}.
For indirect detection, we consider signals at Fermi-LAT and CTA, as
described in \refsec{sec:like}. We consider a default target observation time of
500~hours for CTA and 15 years 46 dSphs in Fermi-LAT. 

%%%%%%%%%%%%%%%%%%%%%%%%%%%%%%%%%%%%%%%%%%%%%%%%%%%%%%%%%%%%%%%%%%
\begin{table}[t]
   \centering\footnotesize
   %\topcaption{Table captions are better up top} % requires the topcapt package
   \begin{tabular}{|c|c|c|c|} % Column formatting, @{} suppresses leading/trailing space
      \hline
      \textbf{Symbol} & \textbf{Parameter} & \textbf{Scan range} & \textbf{Prior distribution} \\
      \hline
      \hline
       \mchi & WIMP mass & $10-10000\gev$ & log \\
       \hline
       \sigv & Annihilation cross section & $10^{-30}-10^{-21}\textrm{ cm}^3/\textrm{s}$ & log \\
       \hline
       \sigsip & Spin-independent cross section & $10^{-48}-10^{-42}\textrm{ cm}^2$ & log \\
       \hline
       \hline
        & \textit{Hadronic benchmark points} & & \\
       \hline 
       $f_{b\bar{b}}$ & Branching ratio $b\bar{b}$ final state & $0-1^{\ast}$ & See text \\
       \hline
       $f_{WW}$ & Branching ratio $WW$ final state & $0-1$ & See text \\
       \hline
       $f_{hh}$ & Branching ratio $hh$ final state & $0-1$ & See text \\
       \hline
       $f_{\tau\tau}$ & Branching ratio $\tau\tau$ final state & $0-1$ & See text \\
       \hline
       \hline 
        & \textit{Leptonic benchmark point} --BP4(d)& & \\
       \hline 
       $f_{\textrm{lep}}$ & Branching ratio leptons & $0-1^{\ast}$ & See text \\
       \hline
       $f_{\textrm{had}}$ & Branching ratio hadrons & $0-1$ & See text \\
       \hline
       $f_{\tau\tau}$ & Branching ratio $\tau\tau$ final state & $0-1$ & See text \\
       \hline
       \hline 
        & \textit{Nuisance parameters} & & \\
       \hline
       $v_0$ & Circular velocity & $220\pm 20$ km/s & Gaussian \\
       \hline
       $v_{\textrm{esc}}$ & Escape velocity & $544\pm 40$ km/s & Gaussian  \\
       \hline
       $\rho_0$ & Local DM density & $0.3\pm 0.1\gev/\textrm{cm}^3$ & Gaussian \\
       \hline
       $\gamma_{\textrm{NFW}}$ & NFW slope parameter & $1.20\pm 0.15$ & Gaussian \\
       \hline   
   \end{tabular}
   
   $^{\ast}$The sum of the branching ratios is 1 and the prior is a modified Dirichlet distribution (see text).

   \caption{Input parameters in our scans.}
   \label{tab:params}
\end{table}
%%%%%%%%%%%%%%%%%%%%%%%%%%%%%%%%%%%%%%%%%%%%%%%%%%%%%%%%%%%%%%%%%%%%

The scanning sampling is performed with \texttt{MultiNest}\cite{Feroz:2008xx}.
Recoil energy rates $dR/dE_R$ are computed with $\tt micrOMEGAs\ v.4.1.5$\cite{Belanger:2013oya}, while annihilation spectra for 
gamma-ray fluxes, $d\Phi/dE$, are obtained using\cite{Cirelli:2010xx,Ciafaloni:2010ti,Buch:2015iya}.
The uncertainties in Fermi-LAT are calculated as described in \refsec{sec:LAT} with the help of the Fermi Science Tools\cite{Fermi2}.
The scanned input parameters, their ranges, and their prior distributions 
are given in \reftable{tab:params}.    
The nuisance parameters are fixed at their central values in the benchmark points.

In each scan the global likelihood function is given by the product of Eqs.~(\ref{likeDD}), (\ref{likeFL}), (\ref{binlike}), and the Gaussian 
likelihood functions associated with the nuisance parameters (see \reftable{tab:params}).
A global test statistics is then constructed according to the profile likelihood technique.
We remind the reader that inference on a subset of $r\leq n$ specific model
parameters or observables, or a combination of both (collectively
denoted by $\psi_i$) is drawn by ``profiling'' the global likelihood function along the other
directions in the parameter space,
\begin{equation}
\mathcal{L}(\psi_{i=1,..,r})=\max_{\psi_i\in\mathbb{R}^{n-r}}\mathcal{L}(\psi_{i=1,..,n})\,.\label{profilike}
\end{equation}

Confidence intervals are calculated from tabulated values
of $\Delta\chi^2\equiv -2\ln(\mathcal{L}/\mathcal{L}_{\textrm{max}})$.
For example, in $r=2$ dimensions, 95.0\% confidence regions are given by $\Delta\chi^2=5.99$.  

In the case of BP4, to which CTA shows the highest sensitivity,
we run two different types of scans. ``Hadronic'' scan inputs (for fitting to BP4(a)-(c)) 
use a combination of 4 types of 
hadronic spectra, which present similar shapes and normalizations:
$b\bar{b}$, $W^+W^-$, $hh$, and $\tau^+\tau^-$. 
In the ``leptonic'' scan (BP4(d)) we investigate whether distinguishing between hadronic and leptonic spectra is easier 
than distinguishing among 4 different hadronic spectra. We thus consider 3 final-state input channels: $b\bar{b}$, $\mu^+\mu^-$, and $\tau^+\tau^-$. 

In order to avoid introducing statistical bias towards points characterized by pure final states or, alternatively, 
by maximal admixtures of different final states, which would create unphysical inhomogeneities in the density of points appearing 
in the plots, we modify a prior Dirichlet distribution for the branching ratios so that eventually all marginal one-dimensional distributions 
for the individual branching ratios are close to being uniform. We remind the reader that 
from a frequentist point of view the priors should not matter in the results, so that this choice is merely driven by aesthetic reasons.

It is also possible to study model-dependent priors that introduce specific fixed relations between different branching ratios, e.g., between $\mu^+\mu^-$ and $e^+e^-$ or between $W^+W^-$ and $hh$, for priors motivated by Minimal Flavor Violation or by gauge symmetry, respectively. By effectivelty being equivalent to a redefinition of the number of independent parameters, 
this could eventually affect the DM reconstruction. We will not investigate this issue in more detail in the present paper.

%%%%%%%%%%%%%%%%%%%%%%%%%%%%%%%%%%%%%%%%%%%%%%%%%%%%%%%%%%%%%%%%%%%%%%%%%%%%%%%%%%%%%%%%%%%%
\section{Discussion of results\label{sec:results}}
%%%%%%%%%%%%%%%%%%%%%%%%%%%%%%%%%%%%%%%%%%%%%%%%%%%%%%%%%%%%%%%%%%%%%%%%%%%%%%%%%%%%%%%%%%%%%%%%%%

We present here the results of fitting the mock data with the likelihood functions 
defined in Secs.~\ref{sec:like} and \ref{sec:bench}. 
In all the plots we show points that are consistent at the 95\%~C.L. with the mock data in different experiments and with the current 
measurements of the nuisance parameters. 

To make contact with the existing literature, we first re-derive the case of direct detection, 
which has been extensively discussed\cite{Green:2008rd,Strigari:2009zb,Pato:2010zk,Pato:2012fw,Chou:2010qt,Shan:2011jz,Shan:2011ct,
Kavanagh:2012nr,Kavanagh:2013wba,Peter:2013aha,Newstead:2013pea}. 
As was mentioned in \refsec{sec:like}, for a given exposure direct detection experiments based on different
target materials and/or characterized by different detector acceptances 
will not reach the same sensitivity.
We showed in \reftable{tab:exp} that, if one assumes that a WIMP discovery will be 
made in a Xenon detector of $\sim1\textrm{ ton}$ fiducial mass with 2 years of data,
it will in general take a longer time or a greater mass to reach the same sensitivity in detectors based on Ge or Ar.
Experiments based on Argon, in particular, need large exposures.

On the other hand, the relative importance of the current astrophysical uncertainties for the inability to fully reconstruct WIMP mass or cross section 
in one or another direct detection experiment can only be gauged properly when all experiments reach comparable sensitivity. 
We thus make this assumption in what follows. 
Note that for BP1 in \reftable{tab:bench}, characterized by $\mchi=25\gev$, the required exposure in an Argon experiment to be 
competitive with detectors based on Xe or Ge becomes much greater than 20~ton~years, 
and for this reason we do not try to reconstruct BP1 in an Ar experiment.

%%%%%%%%%%%%%%%%%%%%%%%%%%%%%%%%
\begin{figure}[t]
\centering
\subfloat[]{%
\label{fig:a}%
\includegraphics[width=0.47\textwidth]{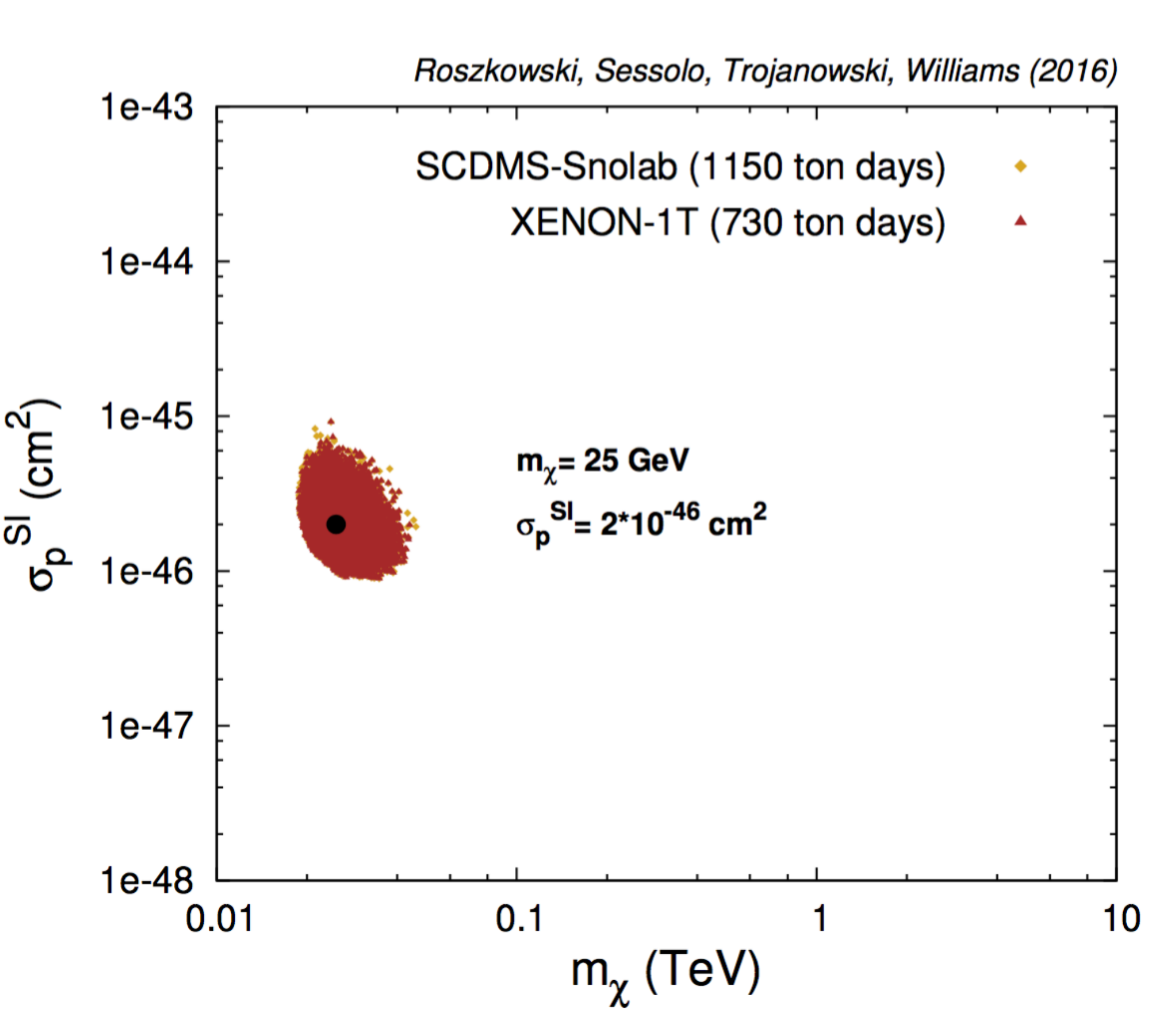}
}%
\hspace{0.02\textwidth}
\subfloat[]{%
\label{fig:b}%
\includegraphics[width=0.47\textwidth]{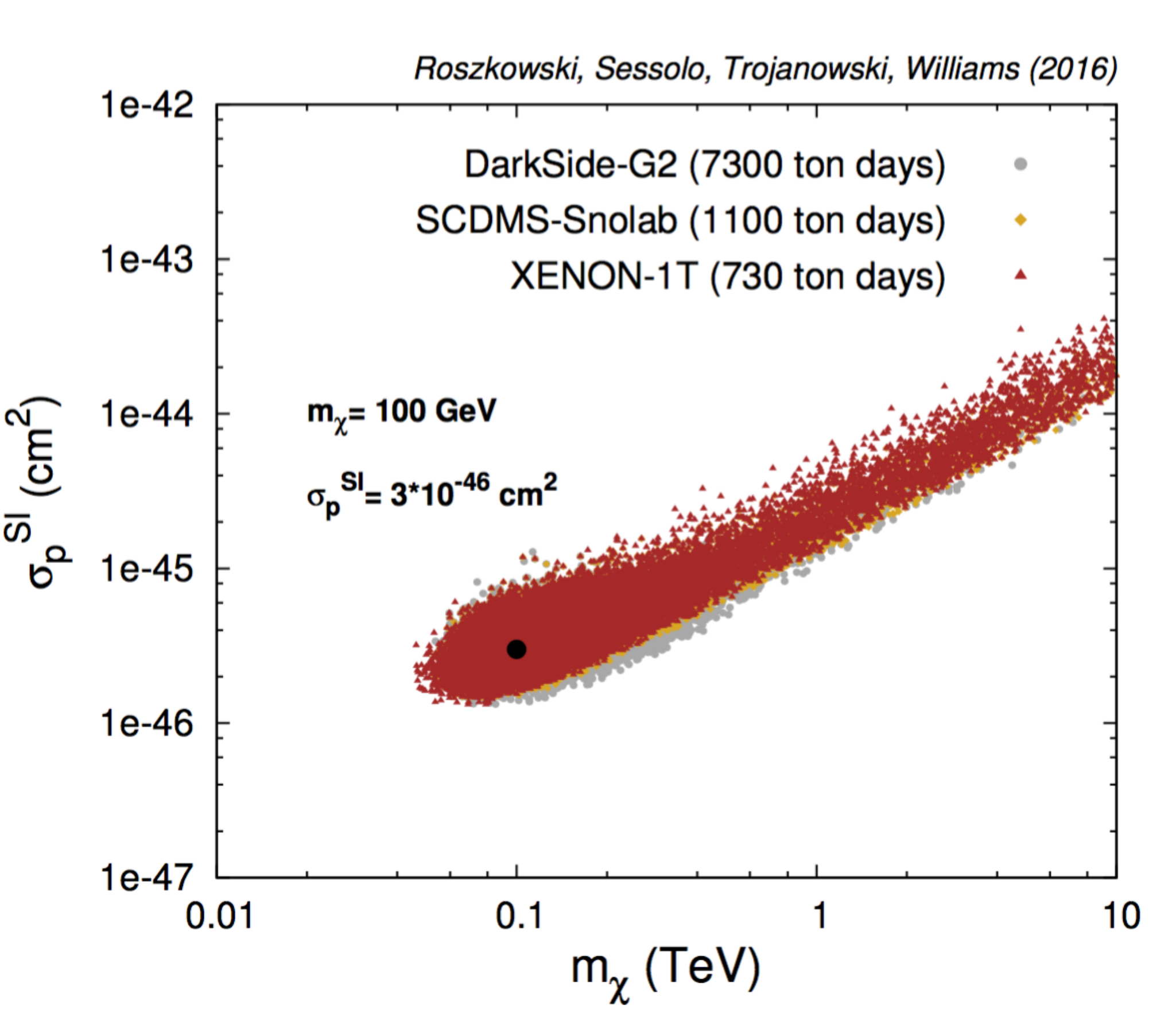}
}%
\hspace{0.07\textwidth}
\subfloat[]{%
\label{fig:c}%
\includegraphics[width=0.47\textwidth]{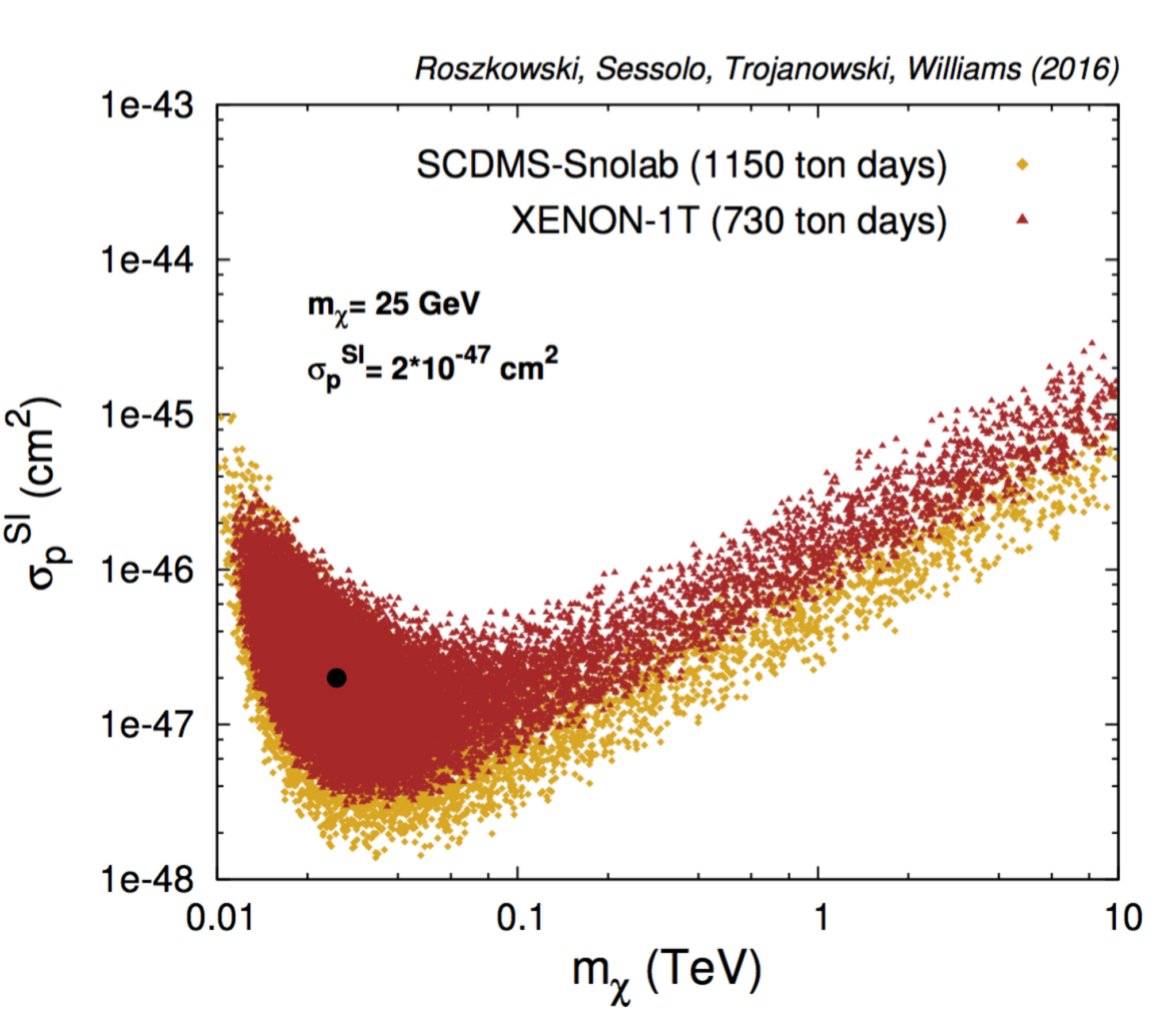}
}%
\hspace{0.02\textwidth}
\subfloat[]{%
\label{fig:d}%
\includegraphics[width=0.47\textwidth]{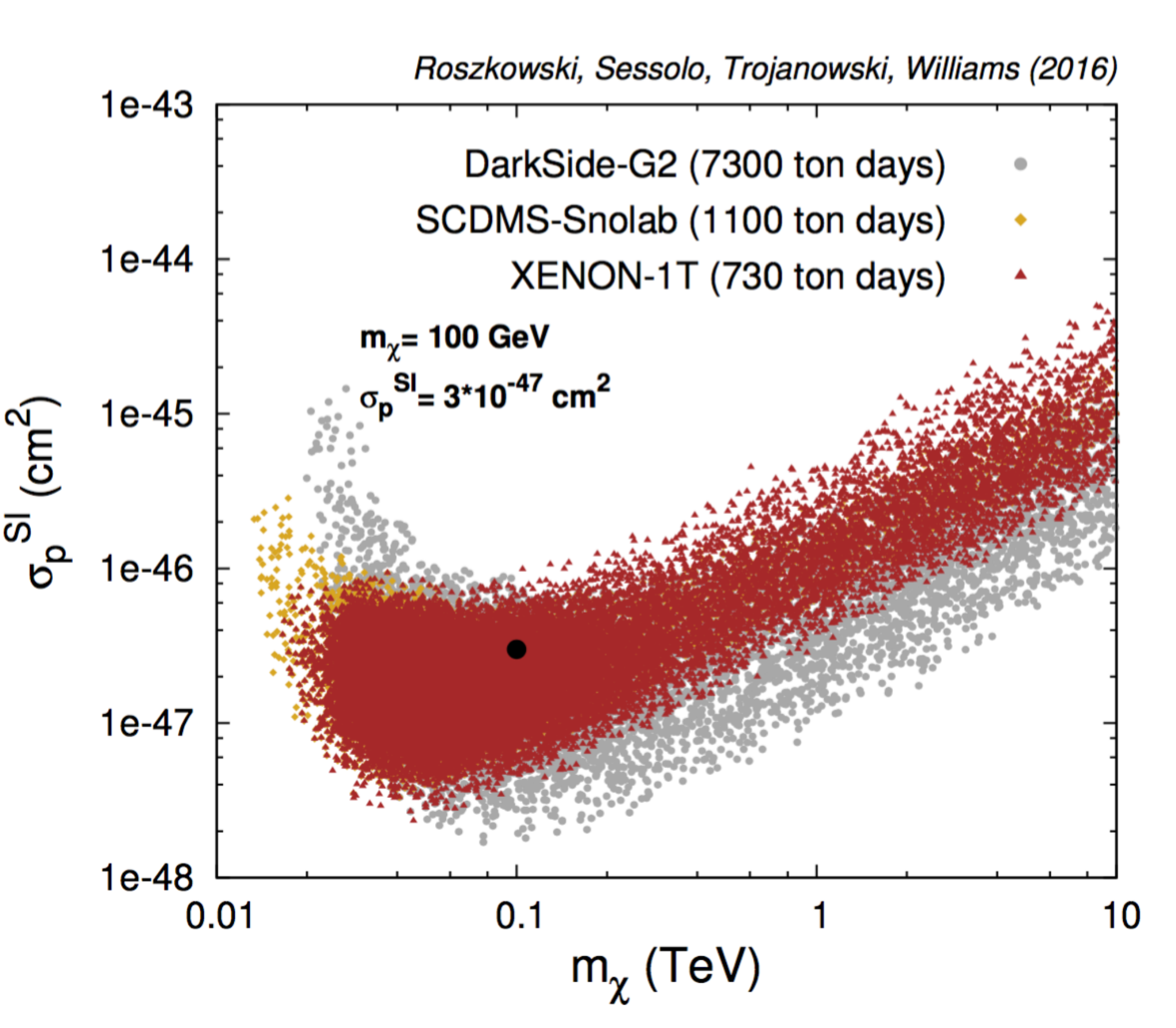}
}%
\caption{(a) Brown triangles show the 95\%~C.L. fit in the (\mchi, \sigsip) plane to BP1 data in a Xenon experiment (XENON-1T) with 
730~ton~days exposure (the benchmark is shown in black). The same sensitivity can be reached in a Germanium experiment (SuperCDMS-Snolab) with 
approximately 1150~ton~days. The corresponding 95\%~C.L. region is shown with golden-rod diamonds. 
(b) The 95\%~C.L. region fit (brown triangles) to BP2 data in XENON-1T. 
Fit to data in SuperCDMS-Snolab is shown with golden-rod diamonds, and the 95\%~C.L. 
region in DarkSide-G2, as a proxy for a generic Argon experiment, is shown with dark gray circles. 
(c) The 95\%~C.L. region fit to a point characterized 
by \sigsip\ lower than in BP1 by one order of magnitude. The color code is the same as in (a). 
(d) The 95\%~C.L. region fit to a point characterized by \sigsip\ 
lower than in BP2 by one order of magnitude. The color code is the same as in (b).}
\label{fig:DD_rec}
\end{figure}
%%%%%%%%%%%%%%%%%%%%%%%%%%%%%%%%

In \reffig{fig:DD_rec}(a) we show the 95\%~C.L. reconstruction of BP1 in the (\mchi, \sigsip) plane. 
We take XENON-1T as a representative of a Xe experiment and SuperCDMS-Snolab as a proxy for a Ge experiment.
The 95\%~C.L. regions, shown with brown triangles for Xenon and golden-rod diamonds for Germanium, overlap when the experiments 
reach the same sensitivity. The WIMP mass and scattering cross section can be 
reconstructed to good precision, in agreement with what has been shown in the literature (see, e.g.,\cite{Newstead:2013pea}).  
The residual uncertainties of the mass amount to approximately a factor 2 and of the cross section to approximately one order 
of magnitude. They are due to the uncertainties in the nuisance parameters $\rho_0$, $v_0$, and $v_{\textrm{esc}}$.

Good reconstruction for WIMP mass and cross section in direct detection is expected as long as the WIMP mass 
is not much larger than the mass of the nuclei of the target material. The experiments lose sensitivity,
irrespective of the target material, when \mchi\ becomes greater or equal to approximately 100\gev. 
This is shown in \reffig{fig:DD_rec}(b), where we present the 95\%~C.L. reconstruction of point BP2, characterized by 
$\mchi=100\gev$. Note that we add an Argon experiment to the lot (dark gray circles), which we model after DarkSide-G2, and which requires 
an exposure 10 times as large as XENON-1T to reach a comparable sensitivity. 
However, again one can see that equivalent sensitivities produce equivalent regions in the (\mchi, \sigsip) plane.
We neglect to show here the reconstructions of the heavier benchmark points, BP3 (250\gev) and BP4 (1000\gev) in \reftable{tab:bench}, as the 
confidence regions look like in \reffig{fig:DD_rec}(b) despite the points being further up the 
reconstructed band. 

We remind the reader here that, because of the near absence of background, the reconstruction quality 
is strongly affected by the Poisson uncertainties so that increasing exposure plays a pivotal 
role in improving the quality of reconstruction.  
In this regard, one can see in Figs~\ref{fig:DD_rec}(c) and \ref{fig:DD_rec}(d), where we show the 95\%~C.L. regions 
for points featuring \sigsip\ exactly one order of magnitude lower than in BP1 and BP2, 
that, given the designed exposures, the experiments lose much if not all 
of their reconstruction power, even when $\mchi=25\gev$.

As reconstructing the properties of the DM particle in direct detection becomes
difficult for WIMPs characterized by mass around and above the electroweak scale,
we here proceed to investigate how an indirect detection in one or more of the gamma-ray observatories 
introduced in \refsec{sec:like} can provide enough 
information to derive the WIMP properties, or at least improve over 
a detection in underground laboratories. 
While there is no doubt that a real signal should be 
detected in all the experiments that are sensitive to the same observables, 
one can find several models of DM that are likely to produce a strong signal only in direct or indirect detection. 
For this reason, although the main purpose of this paper is to investigate the interplay of different experimental strategies in case of concurrent 
detection, it is also important to 
quantify how well information can be reconstructed in each experiment separately.

%%%%%%%%%%%%%%%%%%%%%%%%%%%%%%%%
\begin{figure}[t]
\centering
\subfloat[]{%
\label{fig:a}%
\includegraphics[width=0.47\textwidth]{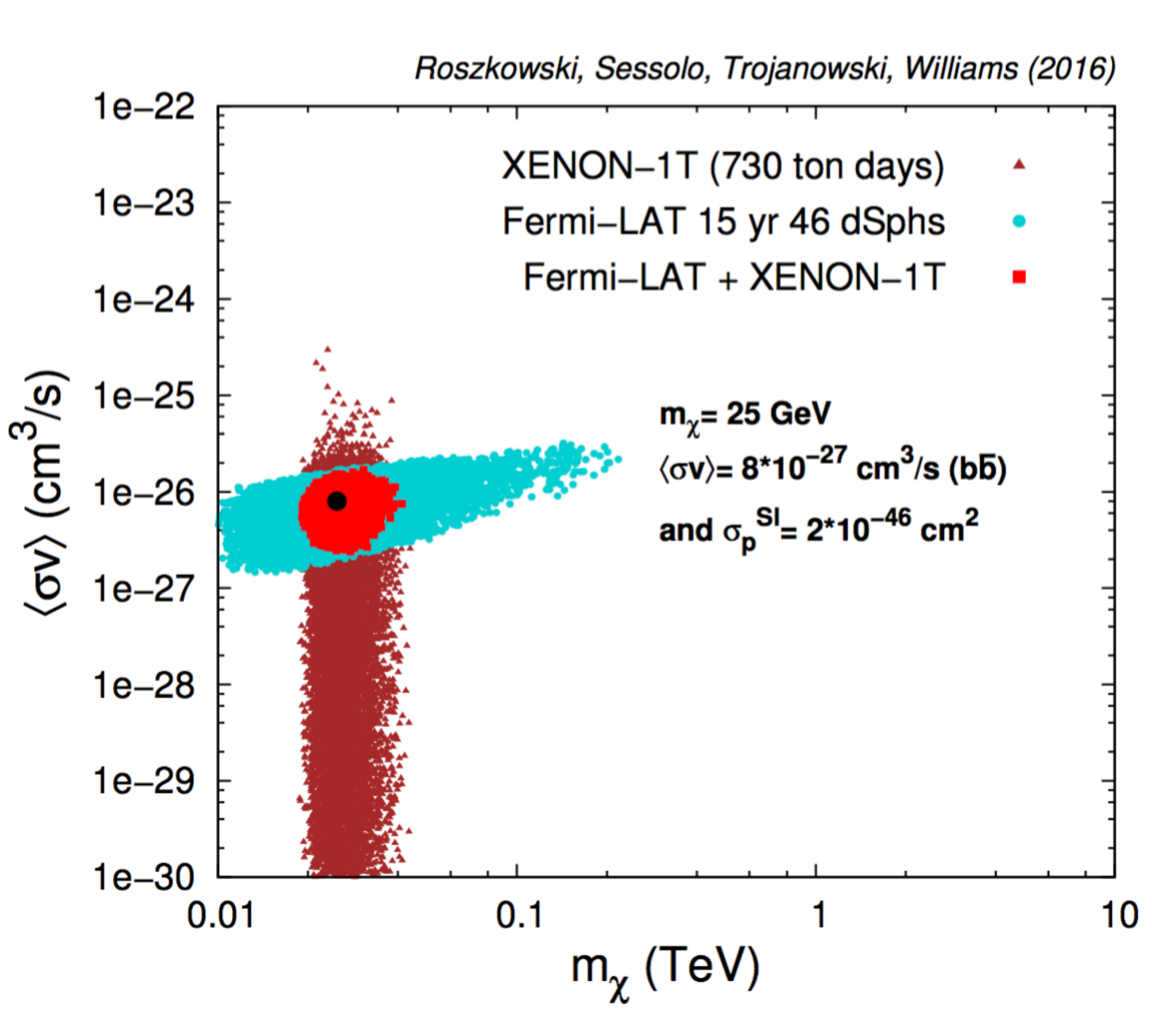}
}%
\hspace{0.02\textwidth}
\subfloat[]{%
\label{fig:d}%
\includegraphics[width=0.47\textwidth]{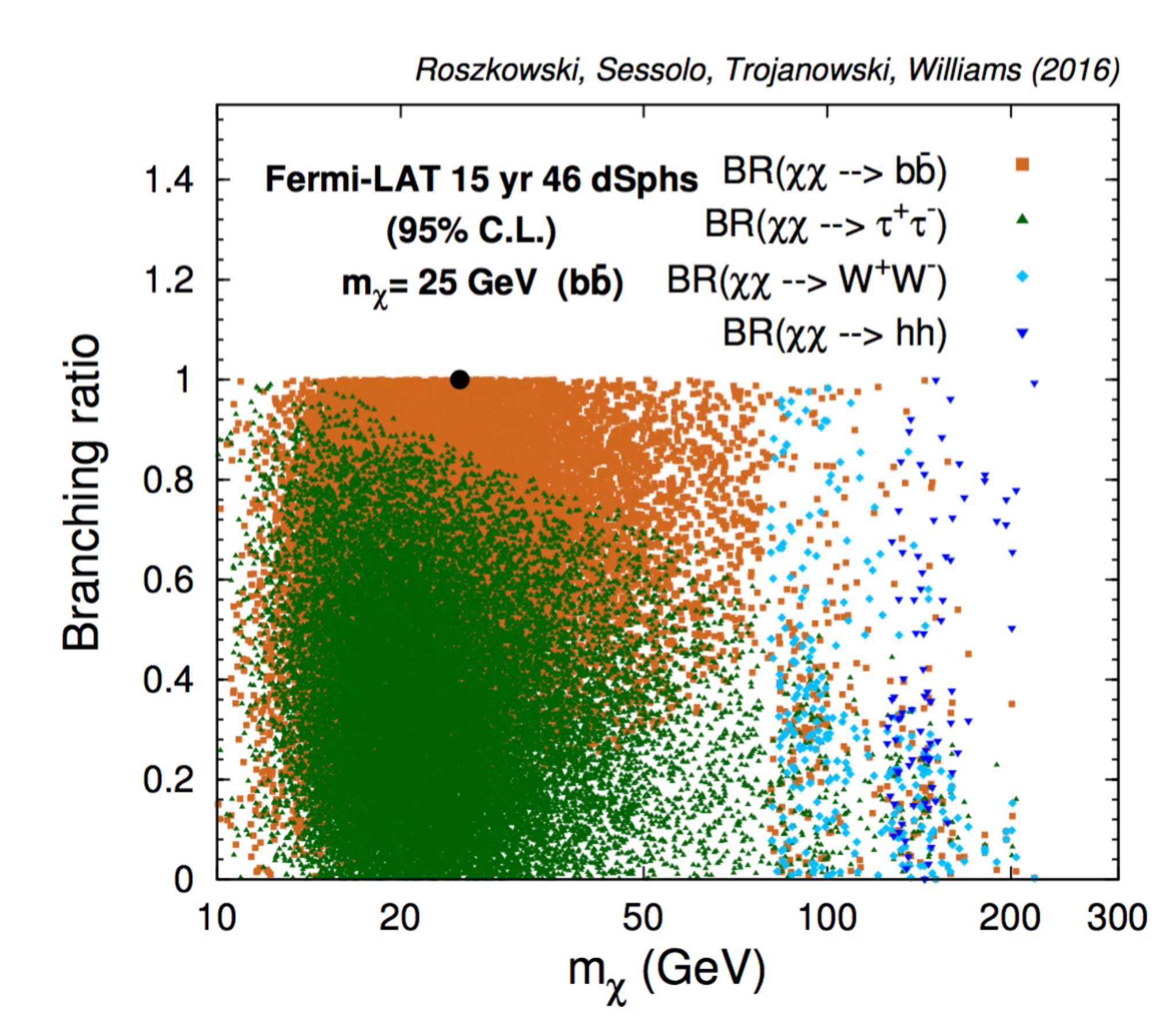}
}%
\caption{(a) Dark turquoise circles show the 95\%~C.L. fit to Fermi-LAT 15 yr 46 dSphs mock BP1 data in the (\mchi, \sigv) plane (the benchmark point is shown in black). 
Brown triangles show the fit to the corresponding XENON-1T data with 730~ton~days exposure.
The 95\%~C.L. combination of the two experiments in the (\mchi, \sigv) plane is shown with red squares. (b) The breakdown 
of the branching ratios to a particular annihilation final state versus the WIMP mass for the points of the 95\%~C.L. fit to Fermi-LAT data
considered in (a). Light brown squares show 
the $b\bar{b}$ branching ratio, dark green triangles the one to $\tau^+\tau^-$,  
deep-sky blue diamonds the one to $W^+W^-$, and blue upside-down triangles the one to $hh$.}
\label{fig:ID25_rec}
\end{figure}
%%%%%%%%%%%%%%%%%%%%%%%%%%%%%%%%

In \reffig{fig:ID25_rec}(a) we show with dark turquoise circles the 
95\%~C.L. region for Fermi-LAT 15 yr 46 dSphs in the (\mchi, \sigv) plane for BP1.
The cross section spread due to the uncertainties described in \refsec{sec:LAT} is about one order of magnitude, 
whereas the derived uncertainty of the mass reconstruction is much larger. 
Note that CTA is not sensitive to a light WIMP signal from the GC, so that 
it cannot help improve on the poor mass reconstruction. Given realistic expectations for the experiments of the near future
little more can be said in gamma-ray experiments for this benchmark point. Possibly 
the next generation of anti-proton observatories will be able to provide some complementarity but this is beyond the scope of this paper.

The mass degeneracy observed in the fit to gamma-ray data is associated with a 
degeneracy in the possible final state products (we remind the reader that for BP1 we fit to 4 possible final states, all yielding in some 
large percentage to hadrons, see \refsec{sec:bench}). 
The benchmark BP1 is characterized by 100\% $b\bar{b}$ branching ratio. 
However, \reffig{fig:ID25_rec}(b) shows that the Fermi-LAT mock data can be fitted 
to the same precision by tau-dominated cases with $\mchi\approx 10-15\gev$ (dark green triangles), or 
by cases with $\mchi\gsim 100\gev$ and admixtures of $b\bar{b}$ (light brown squares) and $W^+W^-$ (deep-sky blue diamonds) or $hh$ (dark blue down-pointing triangles).
We will come back to discussing the degeneracy due to different final states below.

On the other hand, for large enough \sigsip\ as to allow for concurrent signatures in direct and indirect 
detection the complementarity of these two venues can be used to narrow down the mass
and annihilation final state. In \reffig{fig:ID25_rec}(a) we project 
the 95\%~C.L. region for XENON-1T (brown triangles) into the (\mchi, \sigv) plane. 
The 95\%~C.L. combined region for Fermi-LAT and XENON-1T is then shown with red squares.
Intersecting the two regions allows one to strongly narrow down the uncertainties of the mass determination.

Note that if one restricts \reffig{fig:ID25_rec}(b) to the range allowed by the combined Fermi-LAT + XENON-1T region, 
$20\gev\lesssim\mchi\lesssim40\gev$, the final state options for the surviving parameter space are drastically limited 
to only include either the real, pure $b\bar{b}$, or a $b\bar{b}/\tau^+\tau^-$ admixture.  

%%%%%%%%%%%%%%%%%%%%%%%%%%%%%%%%
\begin{figure}[t]
\centering
\subfloat[]{%
\label{fig:a}%
\includegraphics[width=0.47\textwidth]{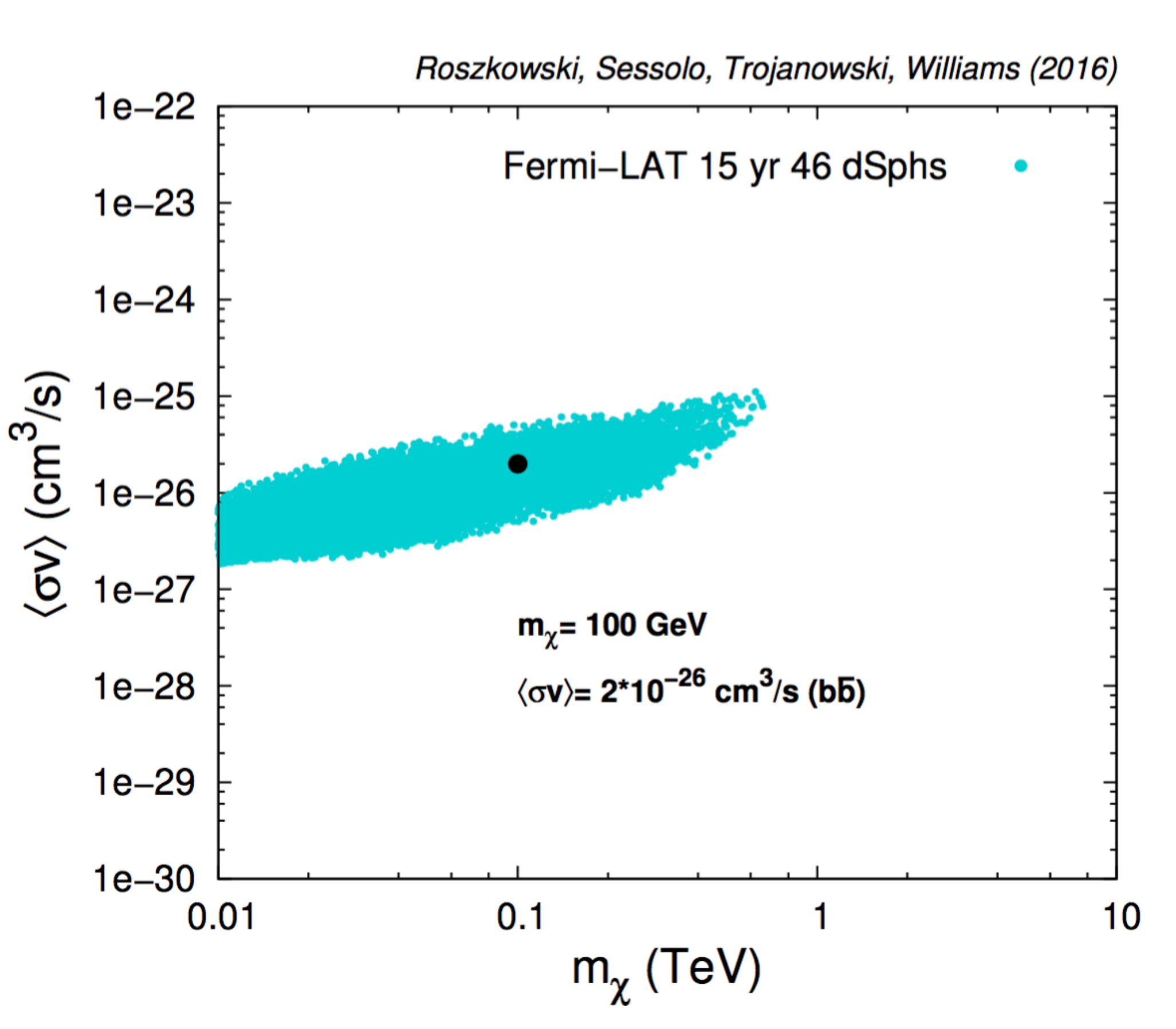}
}%
\hspace{0.02\textwidth}
\subfloat[]{%
\label{fig:d}%
\includegraphics[width=0.47\textwidth]{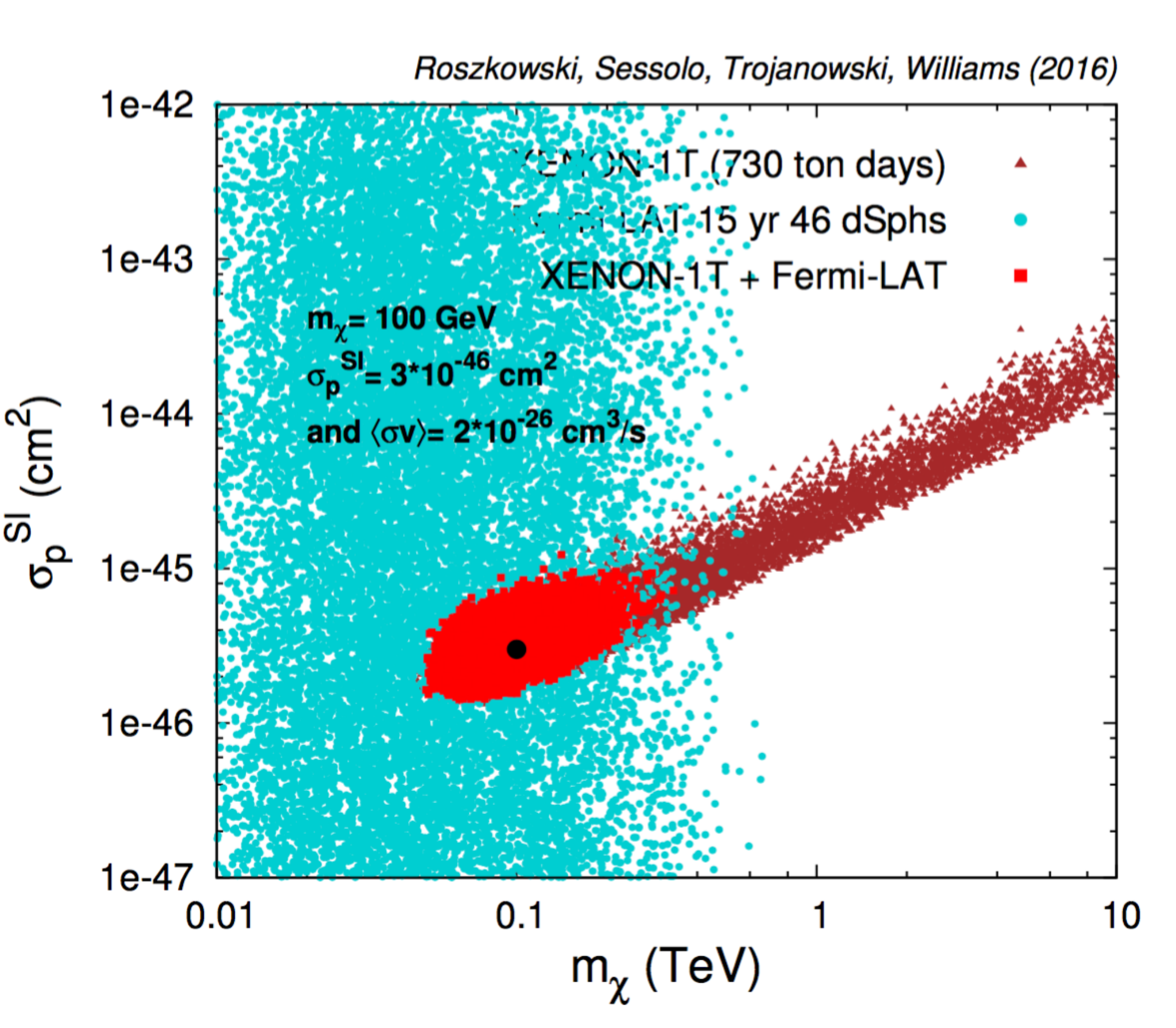}
}%
\caption{(a) The 95\%~C.L. fit to Fermi-LAT 15 yr 46 dSphs mock BP2 data in the (\mchi, \sigv) plane (the benchmark point is shown in black).  
(b) The 95\%~C.L. region fit to BP2 data in XENON-1T in the (\mchi, \sigsip) plane is shown with brown triangles. 
Dark turquoise circles show the 95\%~C.L. fit to Fermi-LAT 15 yr 46 dSphs data. 
The 95\%~C.L. combination of the two experiments in the (\mchi, \sigsip) plane is shown with red squares.}
\label{fig:100}
\end{figure}
%%%%%%%%%%%%%%%%%%%%%%%%%%%%%%%%

The equivalent fit to Fermi-LAT mock data for BP2 is shown in \reffig{fig:100}(a). 
The signal in gamma rays constrains the mass to $\mchi\lesssim 800\gev$. On the other hand, 
we have shown in \reffig{fig:DD_rec}(b) that a hypothetical concurrent signal in one of the direct detection experiments 
would constrain the WIMP mass to $\mchi\gsim 40\gev$. 
By projecting, for instance, the Fermi-LAT constrained region to the (\mchi, \sigsip) plane,
one can visualize the combined 95\%~C.L. mass reconstruction region, 
whose lower bound is determined by XENON-1T and upper bound by Fermi-LAT dSphs. We show
it with red squares in \reffig{fig:100}(b).   

%%%%%%%%%%%%%%%%%%%%%%%%%%%%%%%%
\begin{figure}[t]
\centering
\subfloat[]{%
\label{fig:b}%
\includegraphics[width=0.47\textwidth]{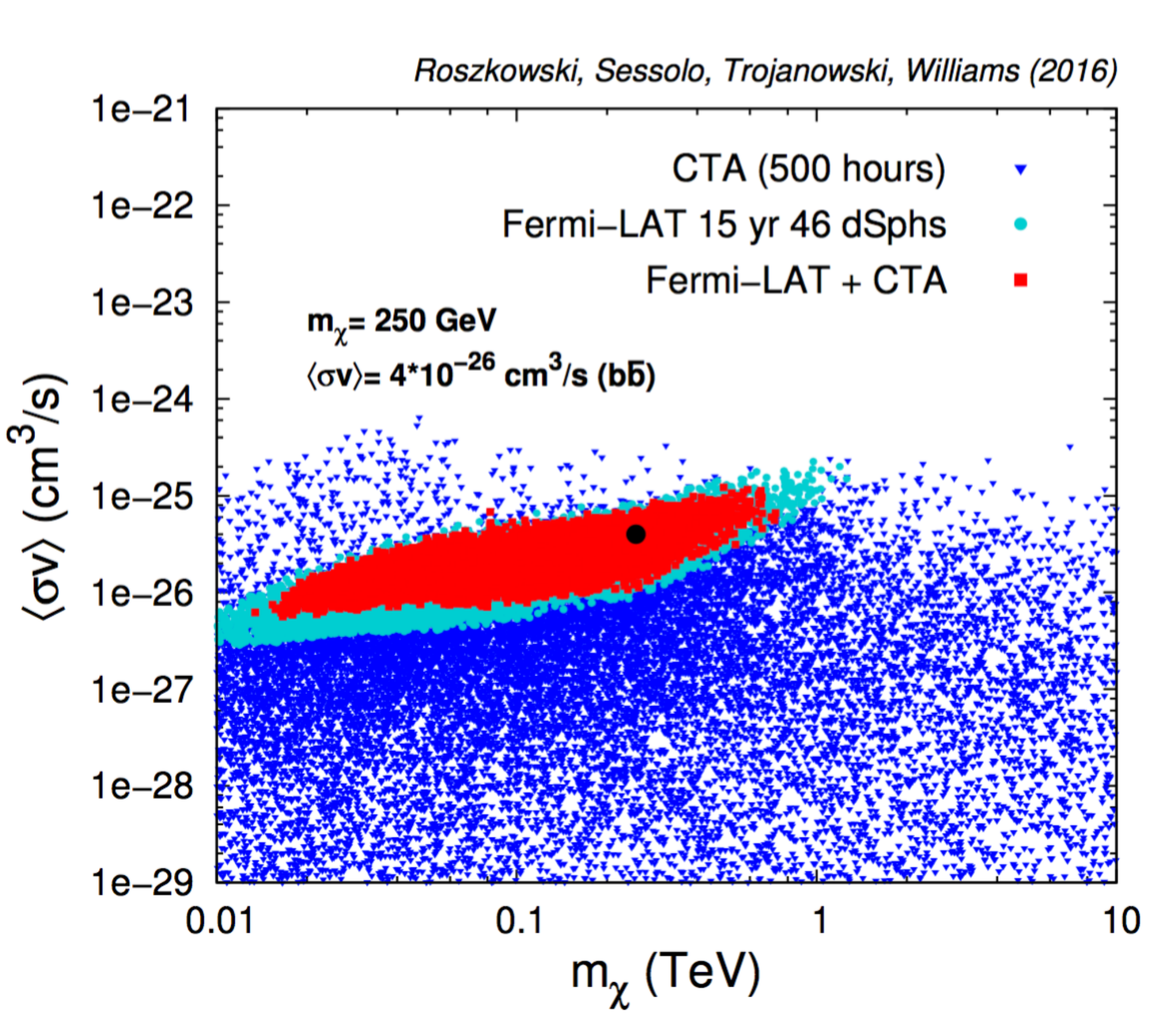}
}%
\hspace{0.02\textwidth}
\subfloat[]{%
\label{fig:c}%
\includegraphics[width=0.47\textwidth]{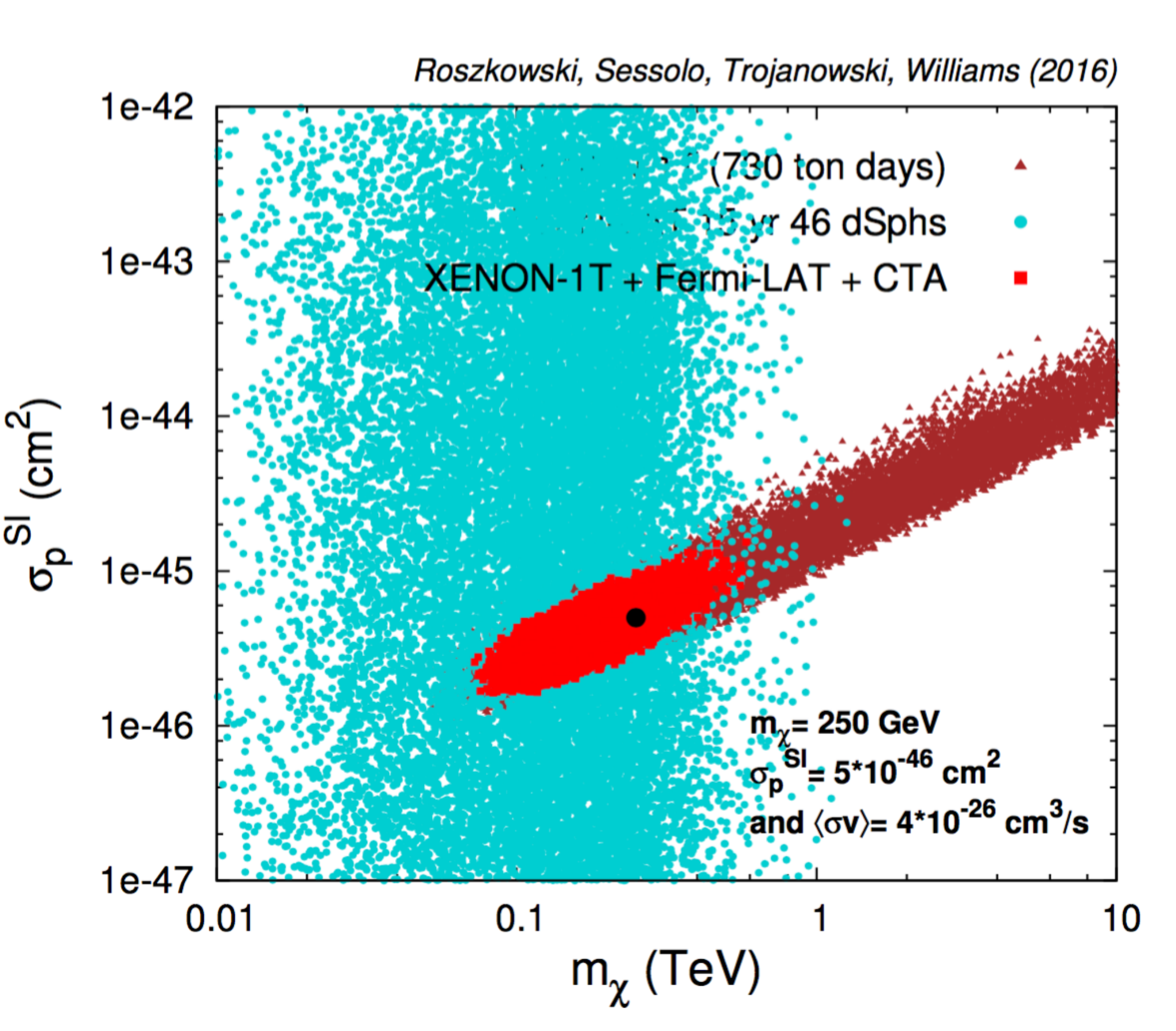}
}%
\caption{(a) Dark turquoise circles show the 95\%~C.L. fit to Fermi-LAT 15 yr 46 dSphs mock BP3 data in the (\mchi, \sigv) plane,
blue upside-down triangles show the 95\%~C.L. reconstruction for the same point in CTA with 500 hours of observation, and the 95\%~C.L. combination of the two experiments is shown with red squares. 
(b) Same as \reffig{fig:100}(b) for BP3. Red squares show here the combination of Fermi-LAT, XENON-1T, and CTA.}
\label{fig:250}
\end{figure}
%%%%%%%%%%%%%%%%%%%%%%%%%%%%%%%%

We can use the same technique to pinpoint the mass of BP3, characterized by $\mchi=250\gev$. 
The BP3 reconstructed region at Fermi-LAT 15 yr 46 dSphs, shown with dark turquoise circles in \reffig{fig:250}(a), 
presents the same qualitative features as for BP1 and BP2, as expected by construction. 
As was mentioned in \refsec{sec:bench}, CTA's sensitivity with approximately 500 hours of observation of the GC
is expected for this mass to start closing in on the sensitivity expected at Fermi-LAT.
In \reffig{fig:250}(a) we plot with blue down-pointing triangles the 95\%~C.L.
reconstruction of BP3 in CTA. One can see that the resulting signal is too weak to provide 
any meaningful information. However, a combination of the Fermi-LAT and CTA likelihood functions
shows for this point a slight improvement (red squares) on the mass reconstruction with respect to Fermi-LAT alone. 
This is a feature that is bound to become increasingly more pronounced as one considers larger DM mass.

The combined effect of Fermi-LAT, CTA, and XENON-1T data on the reconstruction of BP3 properties 
when one considers a concurrent observation in direct and indirect detection experiments is shown with red squares 
in \reffig{fig:250}(b).

%%%%%%%%%%%%%%%%%%%%%%%%%%%%%%%%
\begin{figure}[t]
\centering
\subfloat[]{%
\label{fig:a}%
\includegraphics[width=0.47\textwidth]{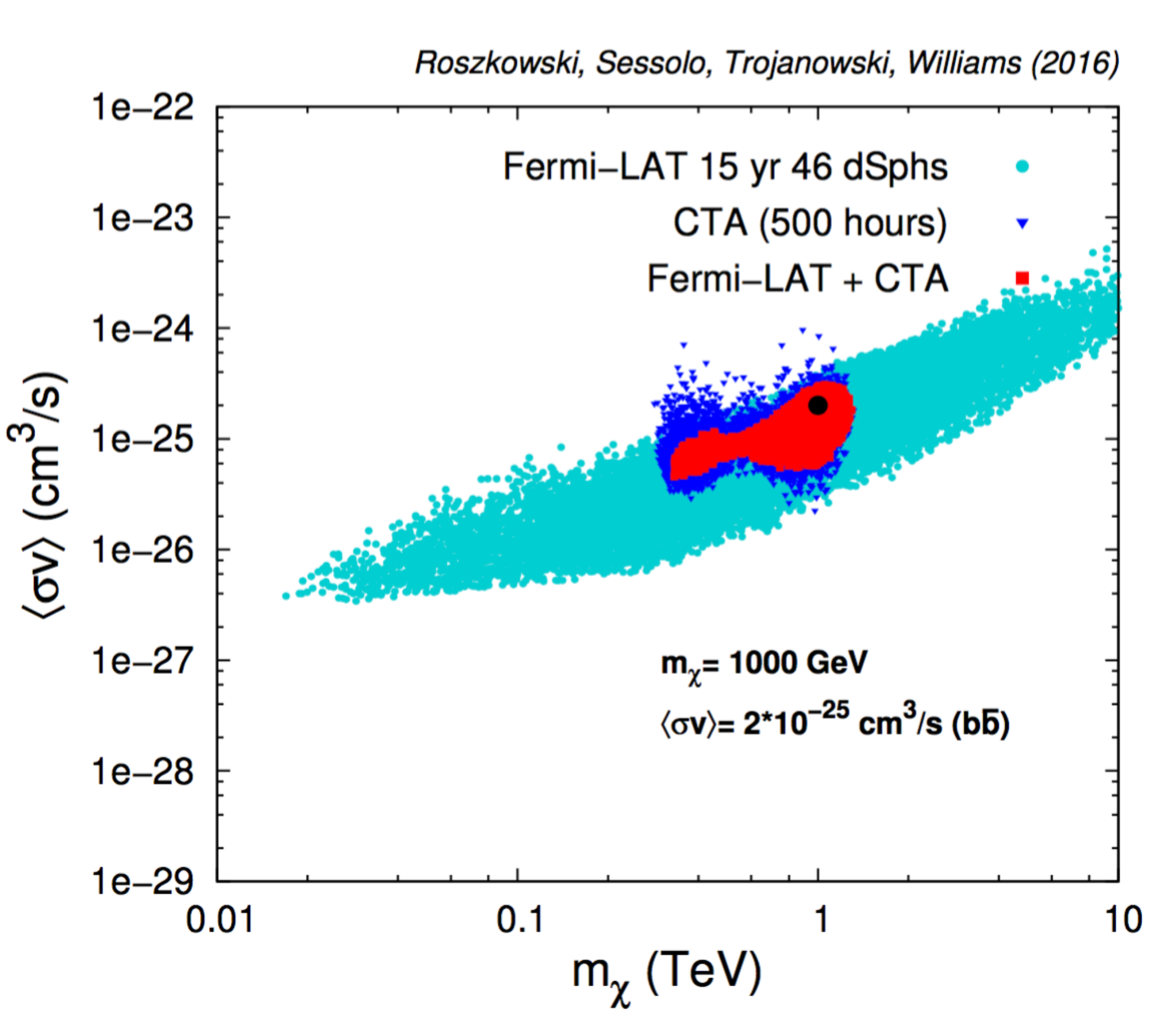}
}%
\hspace{0.02\textwidth}
\subfloat[]{%
\label{fig:d}%
\includegraphics[width=0.47\textwidth]{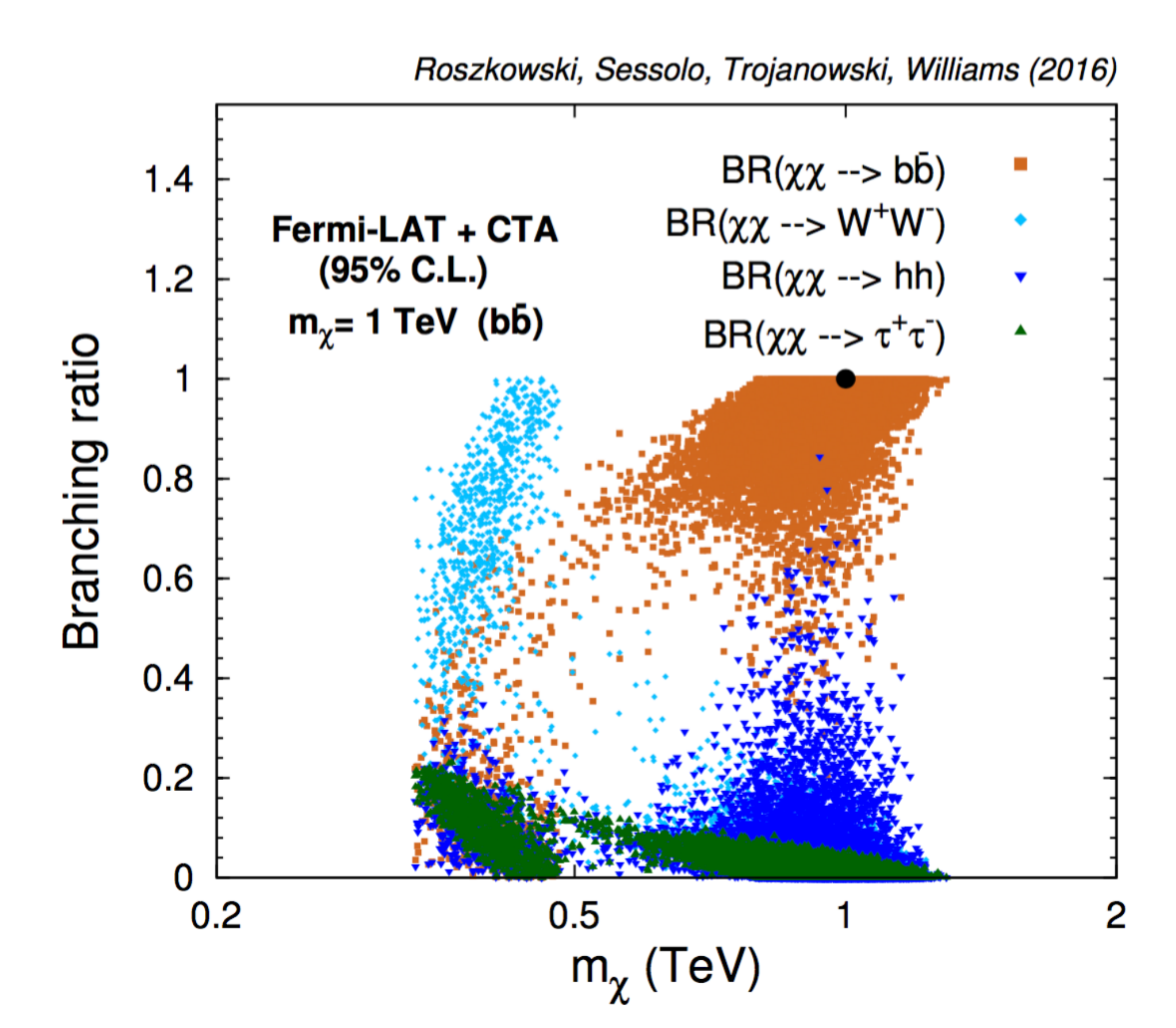}
}%
\caption{(a) Dark turquoise circles show the 95\%~C.L. fit in the (\mchi, \sigv) plane to Fermi-LAT 15 yr 46 dSphs mock BP4(a) data 
(the benchmark is shown in black). 
Blue upside-down triangles show the 95\%~C.L. fit for CTA 500 hours.
The 95\%~C.L. combination of the two results in the (\mchi, \sigv) plane is shown with red squares. (b) The breakdown 
of the branching ratios to a particular annihilation final state versus the WIMP mass for the points of the 95\%~C.L. fit to FermiLAT + CTA data
considered in (a). Light brown squares show 
the $b\bar{b}$ branching ratio, dark green triangles the one to $\tau^+\tau^-$,  
deep-sky blue diamonds the one to $W^+W^-$, and blue upside-down triangles the one to $hh$.}
\label{fig:ID_1000bb}
\end{figure}
%%%%%%%%%%%%%%%%%%%%%%%%%%%%%%%%

We now move on to the 1000\gev\ WIMP case that will allow us to investigate in detail how the considered uncertainties affect 
the mass and cross section reconstruction abilities of CTA. In \reffig{fig:ID_1000bb} 
we present the case of BP4(a), featuring a 1\tev\ WIMP with 100\% branching ratio to $b\bar{b}$ and 
$\sigv=2\times 10^{-25}\textrm{ cm}^3/\textrm{s}$. We show the reconstructed 95\%~C.L. regions in the (\mchi, \sigv) plane 
in \reffig{fig:ID_1000bb}(a). Again, dark turquoise circles show the reconstruction in Fermi-LAT, blue triangles the 
reconstruction in CTA, whose precision is now much higher than for the previous benchmark points, and red squares the combined 95\%~C.L. region.
Note how CTA can narrow down the mass range by almost two orders of magnitude with respect to Fermi-LAT alone, although 
substaintial degeneracy among different reconstructed values of \mchi\ and \sigv\ remains. 

Much of this degeneracy is due to the fact that the scan has the freedom to adjust 
the final state branching fraction of the DM candidate, giving a very similar energy spectrum, so that different options can equally fit in 
the wiggle room left by the large astrophysical and background uncertainties described in \refsec{sec:CTA}.   
This is shown in \reffig{fig:ID_1000bb}(b) where we plot the branching ratios to the 4 different 
final states considered here for the points that belong to the combined CTA + Fermi-LAT 
confidence region shown in \reffig{fig:ID_1000bb}(a). Interestingly, one can see 
that the data originating from a 1\tev\ $b\bar{b}$ benchmark point can be 
equally well fitted by that of a $400-500\gev$ WIMP annihilating almost entirely to $W^+W^-$ with 
a cross section \sigv\ approximately 2 to 5 times smaller than the benchmark $2\times 10^{-25}\textrm{ cm}^3/\textrm{s}$.
  
%%%%%%%%%%%%%%%%%%%%%%%%%%%%%%%%
\begin{figure}[t]
\centering
\subfloat[]{%
\label{fig:a}%
\includegraphics[width=0.47\textwidth]{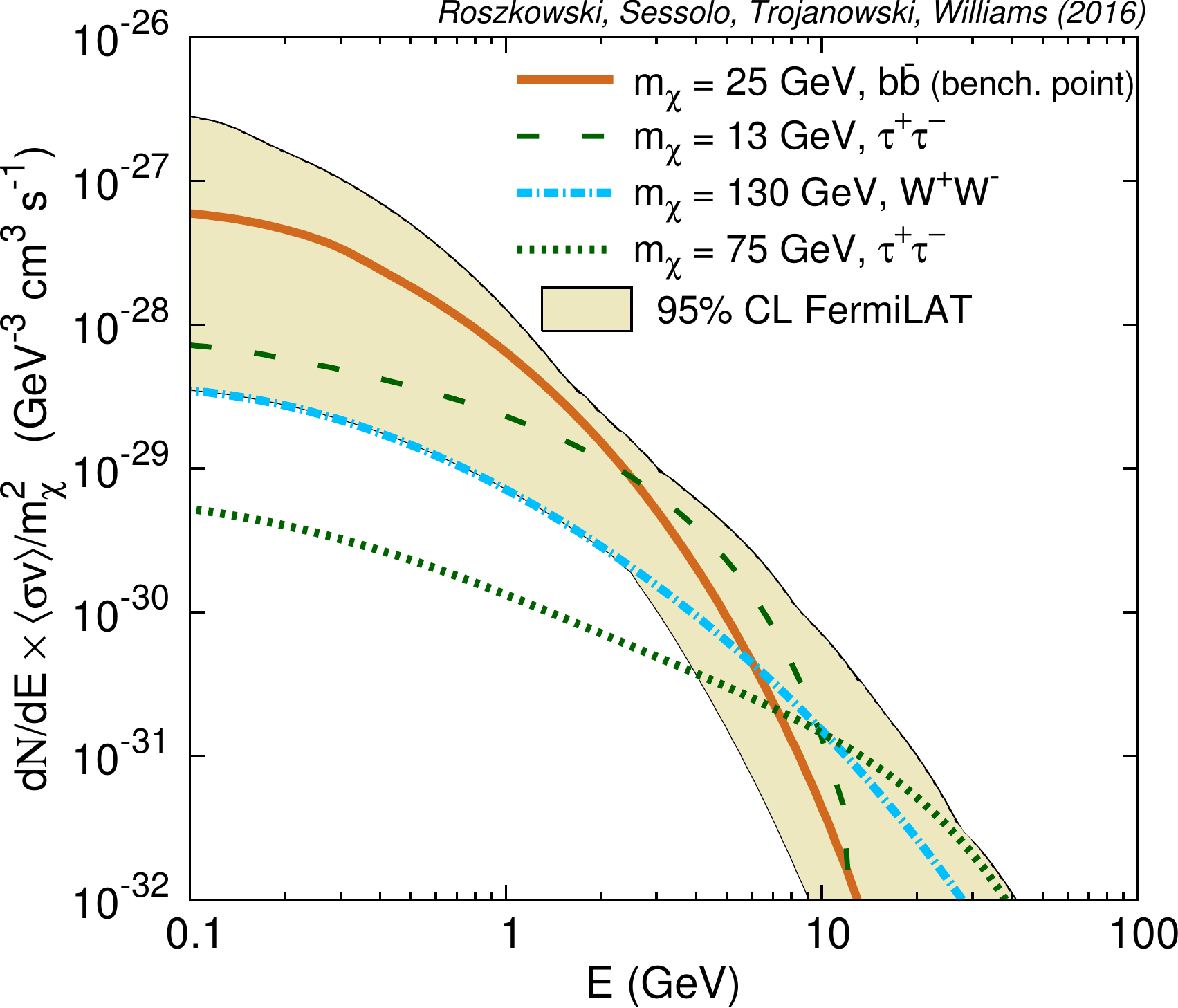}
}%
\hspace{0.02\textwidth}
\subfloat[]{%
\label{fig:d}%
\includegraphics[width=0.47\textwidth]{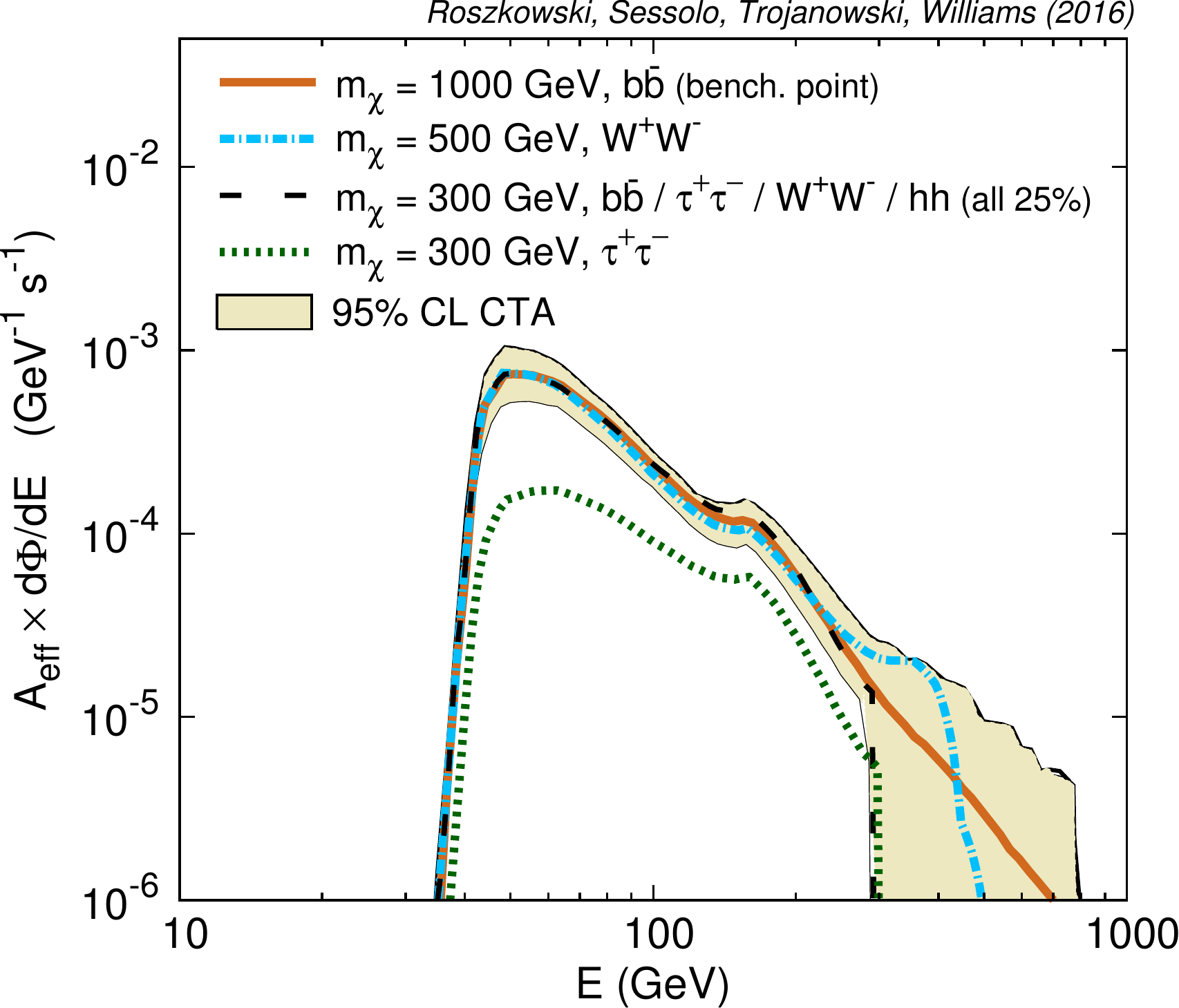}
}%
\caption{(a) A plot of several differential gamma-ray fluxes, $dN_{\gamma}/dE\times\sigv/\mchi^2$, that can fit the spectrum produced by BP1. 
The uncertainties around the reconstruction of the benchmark point spectrum in Fermi-LAT are shown with a light beige band. 
(b) Several effective gamma-ray fluxes, $d\Phi/dE\times A_{\textrm{eff}}$, that can fit the spectrum produced by BP4(a) in CTA. 
The uncertainties around the reconstruction of the benchmark point spectrum are shown with a light beige band.}
\label{fig:spectra}
\end{figure}
%%%%%%%%%%%%%%%%%%%%%%%%%%%%%%%%

We pictorially translate the 95\%~C.L. uncertainty found in the (\mchi, \sigv) plane to an uncertainty in the 
gamma-ray fluxes in \reffig{fig:spectra}. 
Figure~\ref{fig:spectra}(a) shows the uncertainties in the reconstruction of BP1 spectra in Fermi-LAT, for 
the case that was presented in \reffig{fig:ID25_rec}.
The uncertainties, which we have assumed to be given 
by a combination of the uncertainty of the measurement of dSphs $J$-factors and the statistical 
Poisson uncertainty of the number of gamma-ray events from fluxes, 
can be shown as a light beige band allowing different shapes for $dN_{\gamma}/dE\times\sigv/\mchi^2$. 
The band can for instance accommodate at the same time the spectra for 13\gev\ WIMPs annihilating 
into $\tau^+\tau^-$, 25\gev\ WIMPs going to $b\bar{b}$, or 130\gev\ WIMPs yielding predominantly $W^+W^-$. 
Note that \reffig{fig:spectra}(a) also pictorially shows how, for instance, the spectrum of a 75\gev\ WIMP with 100\% branching ratio 
into $\tau^+\tau^-$ falls outside of the considered uncertainties, but that one could manage to bring it back into the allowed band
by reducing the branching ratio to $\tau^+\tau^-$ and at the same time increasing $b\bar{b}$, in agreement with what \reffig{fig:ID25_rec}(b) shows.
    
In \reffig{fig:spectra}(b) we show the case of BP4(a). The uncertainties of CTA can be translated into a light beige band
about the gamma-ray flux times effective area, where the latter is responsible for effectively 
cutting out all signal in the bins with $E<30\gev$. 
As one can see, the benchmark point spectrum corresponding to pure $b\bar{b}$ final state and
$\mchi=1\tev$ can be mimicked over a wide range of energies. A $\sim500\gev$
DM particle annihilating into $W^+W^-$ or, for even lower \mchi, a DM particle
annihilating into all four considered final states with branching ratios close to 25\% 
can reproduce the same spectrum as BP4(a).

%Note that effective fluxes $d\Phi/dE\times A_{\textrm{eff}}<10^{-5}\gev^{-1}\textrm{ s}^{-1}$ 
%give very little contribution to the integral in \refeq{DMsign}, thus enlarging the uncertainty shown in the lower part of \reffig{fig:spectra}(b).

%%%%%%%%%%%%%%%%%%%%%%%%%%%%%%%%
\begin{figure}[t]
\centering
\subfloat[]{%
\label{fig:a}%
\includegraphics[width=0.47\textwidth]{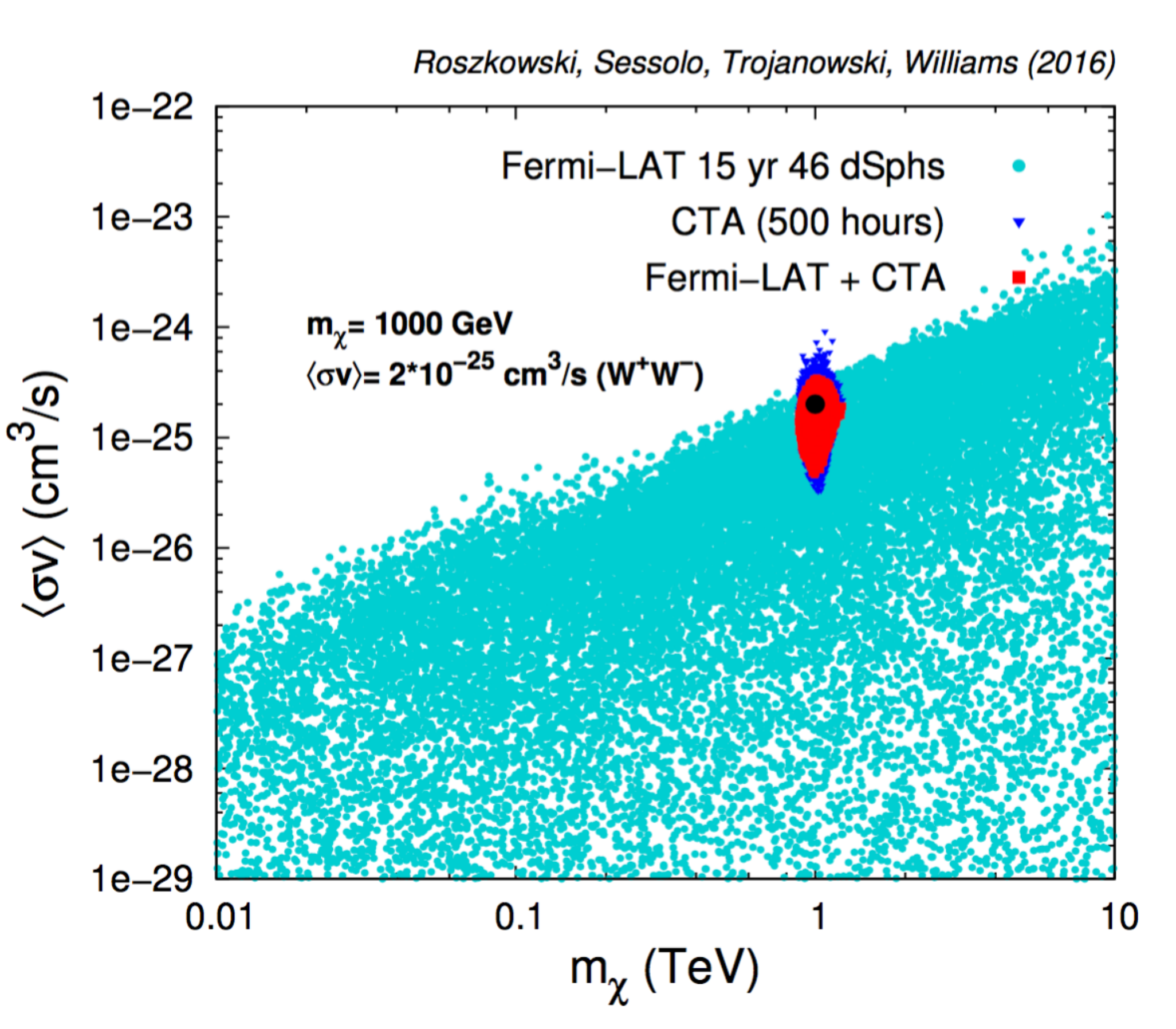}
}
\hspace{0.02\textwidth}
\subfloat[]{%
\label{fig:c}%
\includegraphics[width=0.47\textwidth]{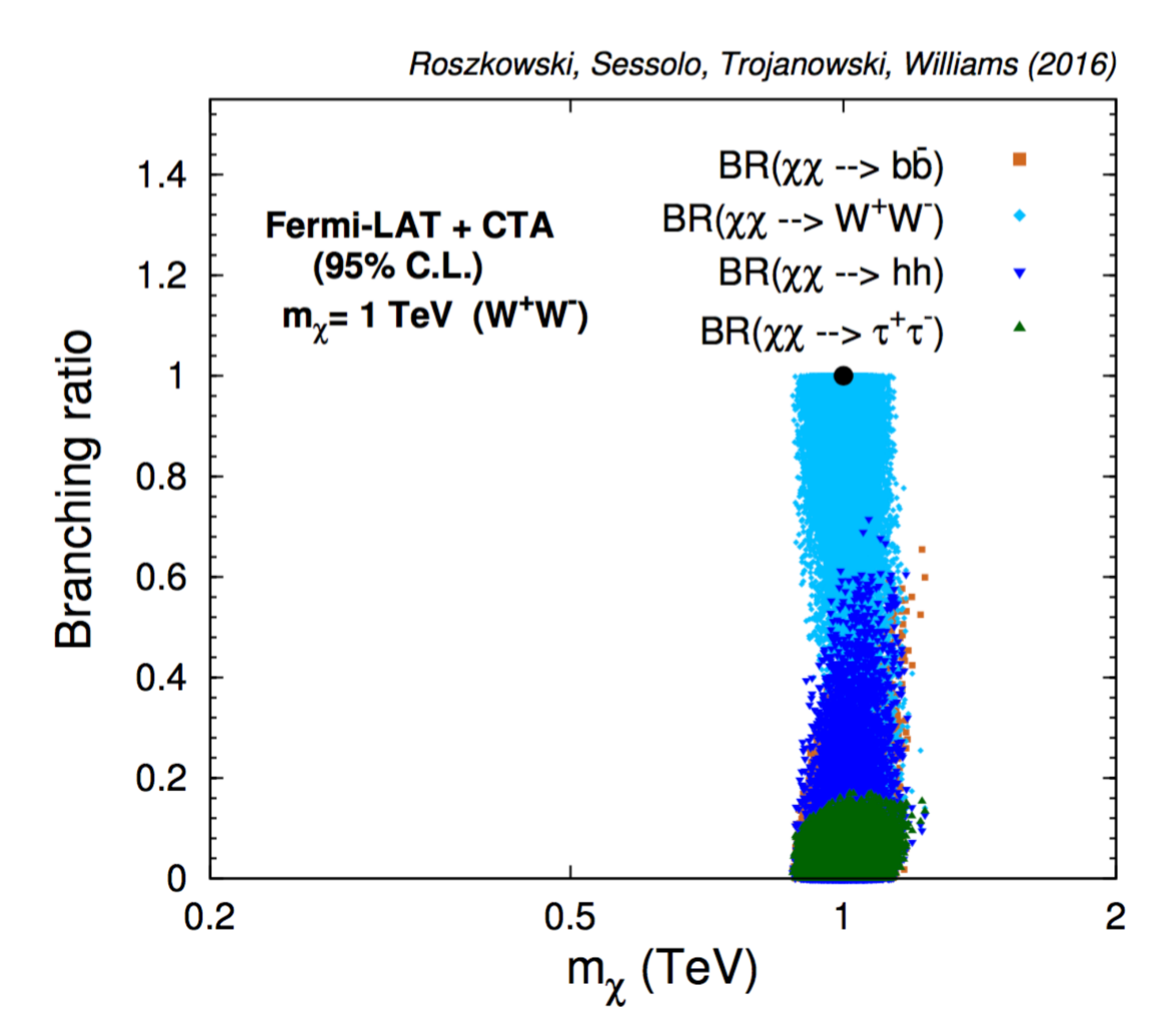}
}%
\caption{Same as \reffig{fig:ID_1000bb}, but for the benchmark BP4(b).}
\label{fig:ID_1000WW}
\end{figure}
%%%%%%%%%%%%%%%%%%%%%%%%%%%%%%%%%

While the reconstruction in CTA of a point with the pure $b\bar{b}$ final state 
is still lacking in precision and allows for degeneracies in mass and \sigv, 
the same is not necessarily true if the benchmark 
point is characterized by a different final state.
We present in \reffig{fig:ID_1000WW}(a) the reconstructed confidence regions for BP4(b), a point characterized by the same mass and \sigv\ as BP4(a)
but with final state 100\% $W^+W^-$.
Note that our projection of 15 yr 46 dSphs Fermi-LAT data does not show enough 
sensitivity to reconstruct a $W^+W^-$ point with $\sigv=2\times 10^{-25}\textrm{ cm}^3/\textrm{s}$, so that
the 95\%~C.L. region (dark turquoise circles) does not provide useful information.
However, CTA has a strong sensitivity to the features of a gamma-ray spectrum originating in 
a 1\tev\ WIMP with 100\% branching fraction to $W^+W^-$, especially because of 
a characteristic spectral ``spike'' appearing at about $E\approx \mchi$, which has its origin in the splitting 
$W^{\pm}\rightarrow W^{\pm}\gamma$ when $E\gg M_W$ (see, e.g.,\cite{Lefranc:2015pza}).
The mass reconstruction becomes eventually very precise.

The final state can also be reconstructed very precisely for BP4(b), as is shown 
in \reffig{fig:ID_1000WW}(b), where one can see that the only allowed pure state is $W^+W^-$ (deep-sky blue diamonds), 
although some 50\% admixtures, especially with $hh$ (dark blue down-pointing triangles) or $b\bar{b}$ (light brown squares), are also possible.  
 
%%%%%%%%%%%%%%%%%%%%%%%%%%%%%%%%
\begin{figure}[t]
\centering
\subfloat[]{%
\label{fig:a}%
\includegraphics[width=0.47\textwidth]{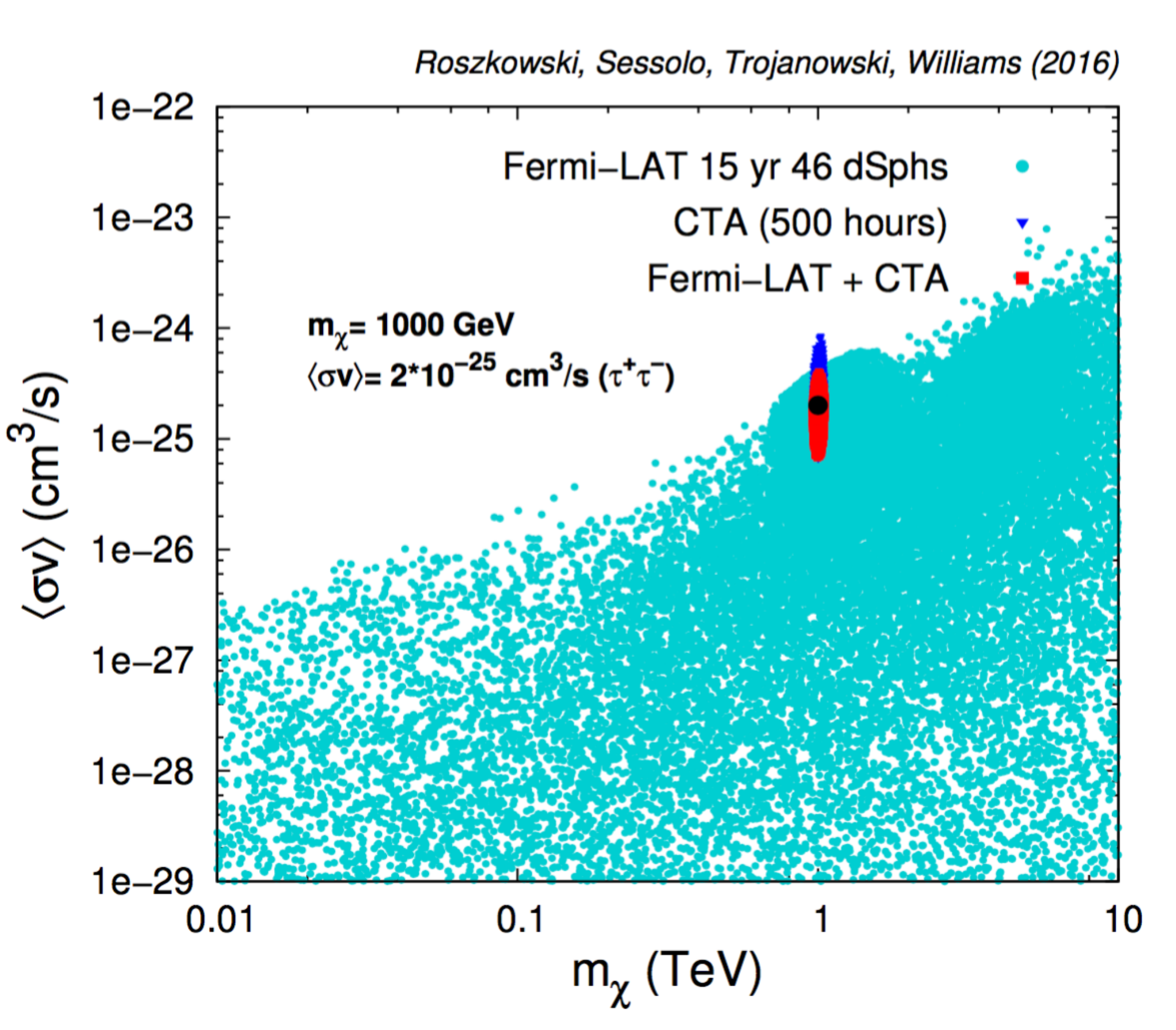}
}
\hspace{0.02\textwidth}
\subfloat[]{%
\label{fig:c}%
\includegraphics[width=0.47\textwidth]{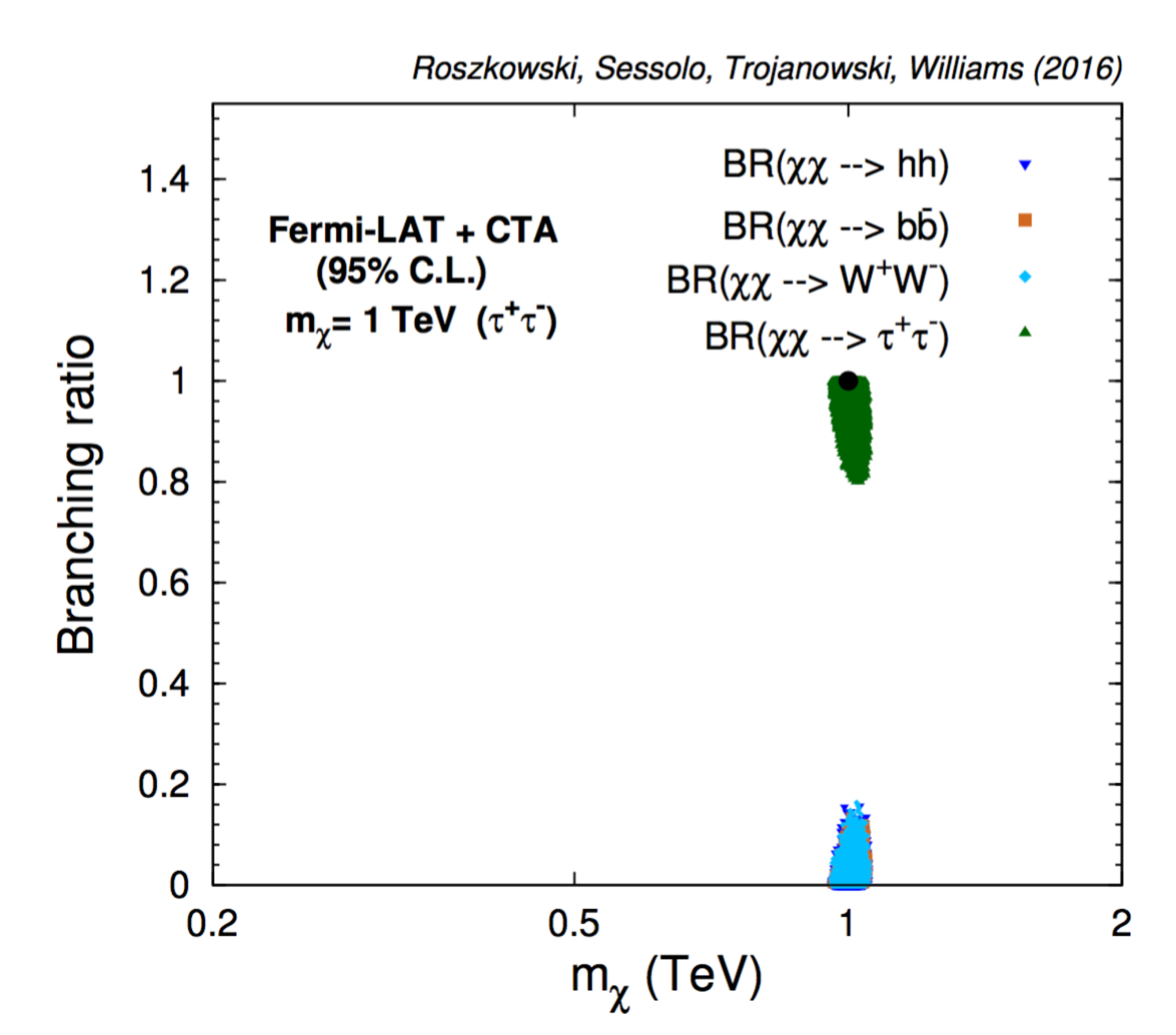}
}%
\caption{Same as Figs.~\ref{fig:ID_1000bb} and \ref{fig:ID_1000WW}, but for the benchmark BP4(c).}
\label{fig:ID_1000tata}
\end{figure}
%%%%%%%%%%%%%%%%%%%%%%%%%%%%%%%%%

An analogous situation is encountered for BP4(c), characterized by a 1\tev\ WIMP with 100\% $\tau^+\tau^-$ final state. 
The reconstruction is presented in \reffig{fig:ID_1000tata}(a). 
The CTA likelihood is responsible for the highly precise mass reconstruction, to which correspond an equally precise final state reconstruction, 
shown in \reffig{fig:ID_1000tata}(b).

%%%%%%%%%%%%%%%%%%%%%%%%%%%%%%%%
\begin{figure}[t]
\centering
\subfloat[]{%
\label{fig:a}%
\includegraphics[width=0.47\textwidth]{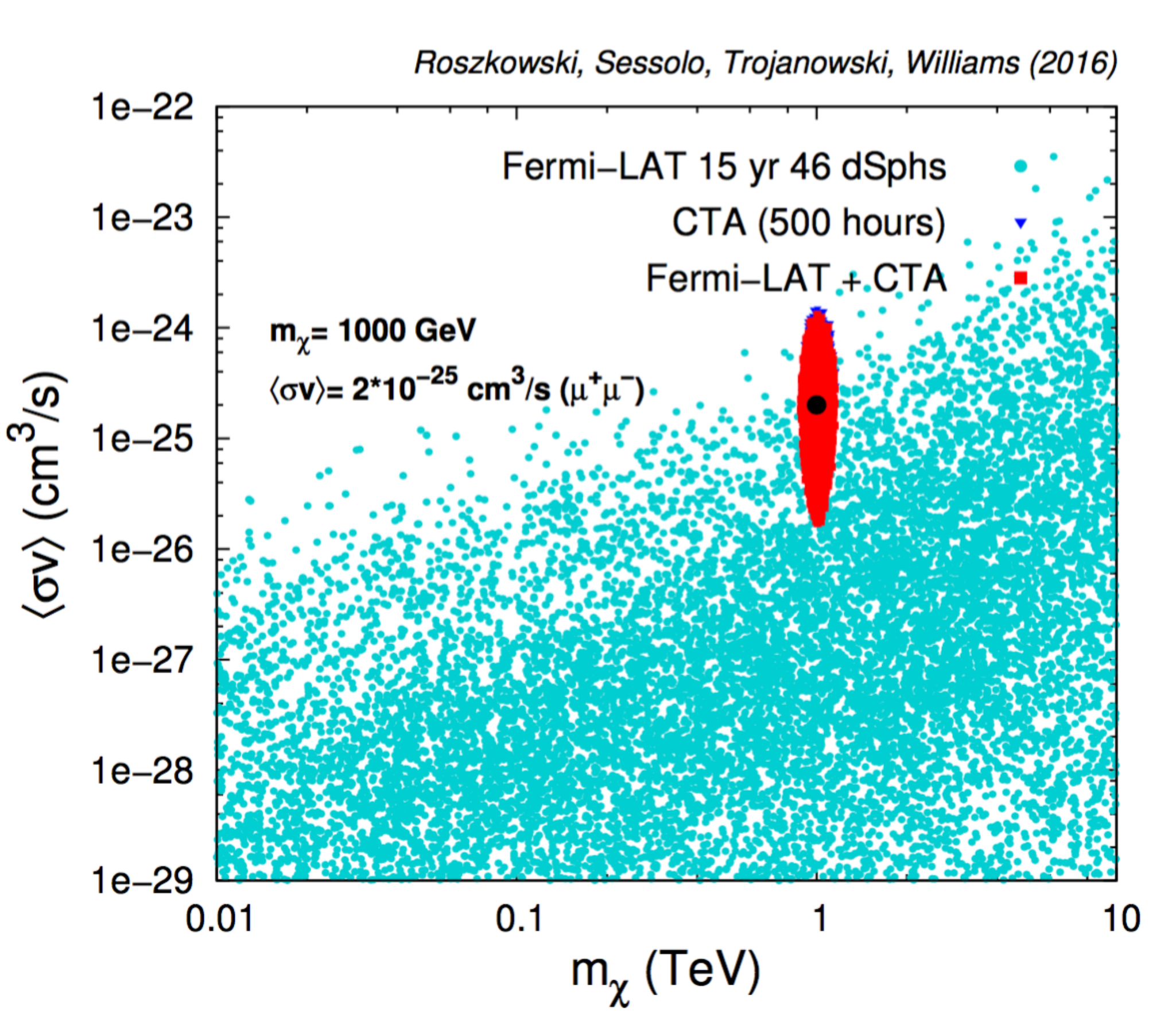}
}
\hspace{0.02\textwidth}
\subfloat[]{%
\label{fig:c}%
\includegraphics[width=0.47\textwidth]{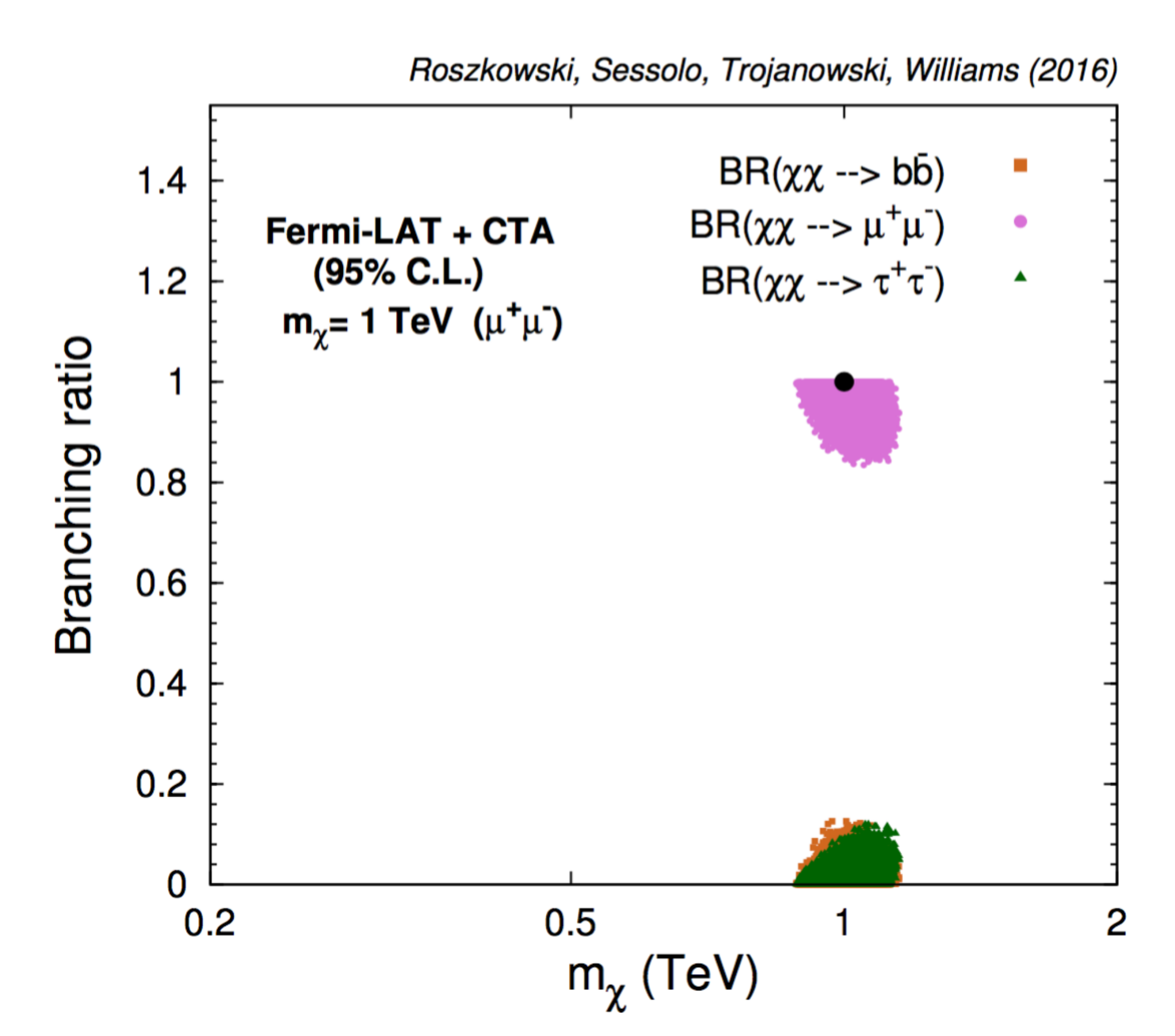}
}%
\caption{(a) Same as \reffig{fig:ID_1000bb}(a), but for the benchmark BP4(d). 
(b) The breakdown of the branching ratios to a particular annihilation final state versus the WIMP mass for the points of the 95\%~C.L. combination of FermiLAT and CTA data
considered in (a). Light brown squares show 
the $b\bar{b}$ branching ratio, dark green triangles the one to $\tau^+\tau^-$,
and orchid circles the one to $\mu^+\mu^-$.}
\label{fig:ID_had_lep}
\end{figure}
%%%%%%%%%%%%%%%%%%%%%%%%%%%%%%%%%

We next move to the reconstruction of the gamma-ray spectra of purely leptonic origin. 
Consider the case of BP4(d), featuring a 1\tev\ WIMP with the same \sigv\ as the previous benchmark points and 100\% $\mu^+\mu^+$ final state. 
As was described in \refsec{sec:bench}, we limit the final-state channels for this scan to 3:
points are assumed to correspond to 
leptonic ($\mu^+\mu^+$ final state), hadronic ($b\bar{b}$), or mixed ($\tau^+\tau^+$ and combinations) spectra.
The reconstruction in the (\mchi, \sigv) plane is shown in \reffig{fig:ID_had_lep}(a) and the final state reconstruction is shown in \reffig{fig:ID_had_lep}(b). 
Again the mass reconstruction is very good, although the weakness of the signal for the $\mu^+\mu^+$ mode slightly spoils the reconstruction in 
\sigv.

Finally, for BP5, characterized by $\mchi=1000\gev$, a canonical thermal $\sigv=3\times 10^{-26}\textrm{ cm}^3/\textrm{s}$, 
and 100\% branching ratio to $W^+W^-$, we found in our analysis that, with 500 hours of observation, CTA fairs relatively poorly (albeit better than Fermi-LAT 15 yr 46 dSphs)
on the reconstruction of the WIMP mass or cross section, so that we do not present a plot for this case.  

%%%%%%%%%%%%%%%%%%%%%%%%%%%%%%%%
\begin{figure}[t]
\centering
\subfloat[]{%
\label{fig:a}%
\includegraphics[width=0.47\textwidth]{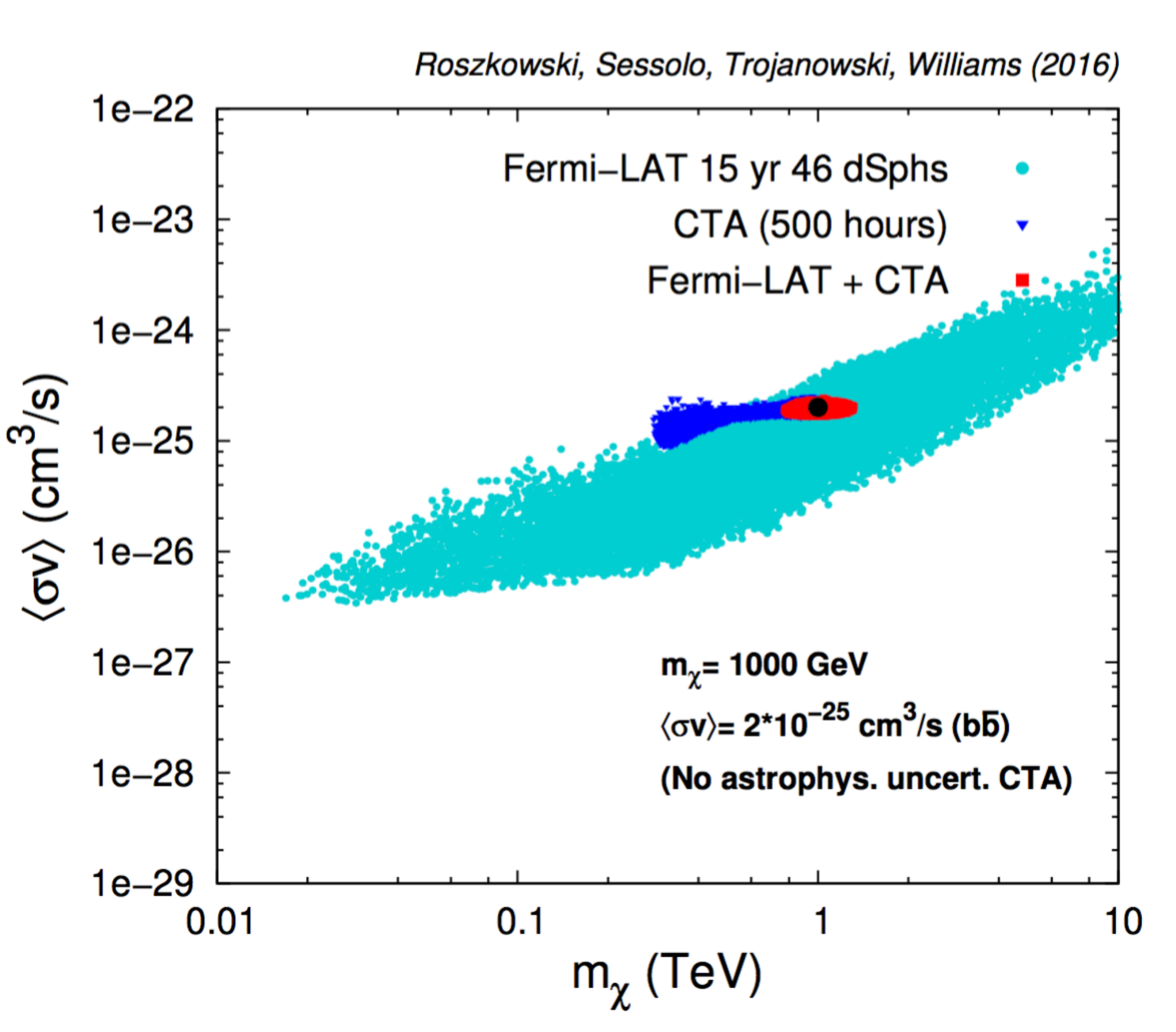}
}
\hspace{0.02\textwidth}
\subfloat[]{%
\label{fig:c}%
\includegraphics[width=0.47\textwidth]{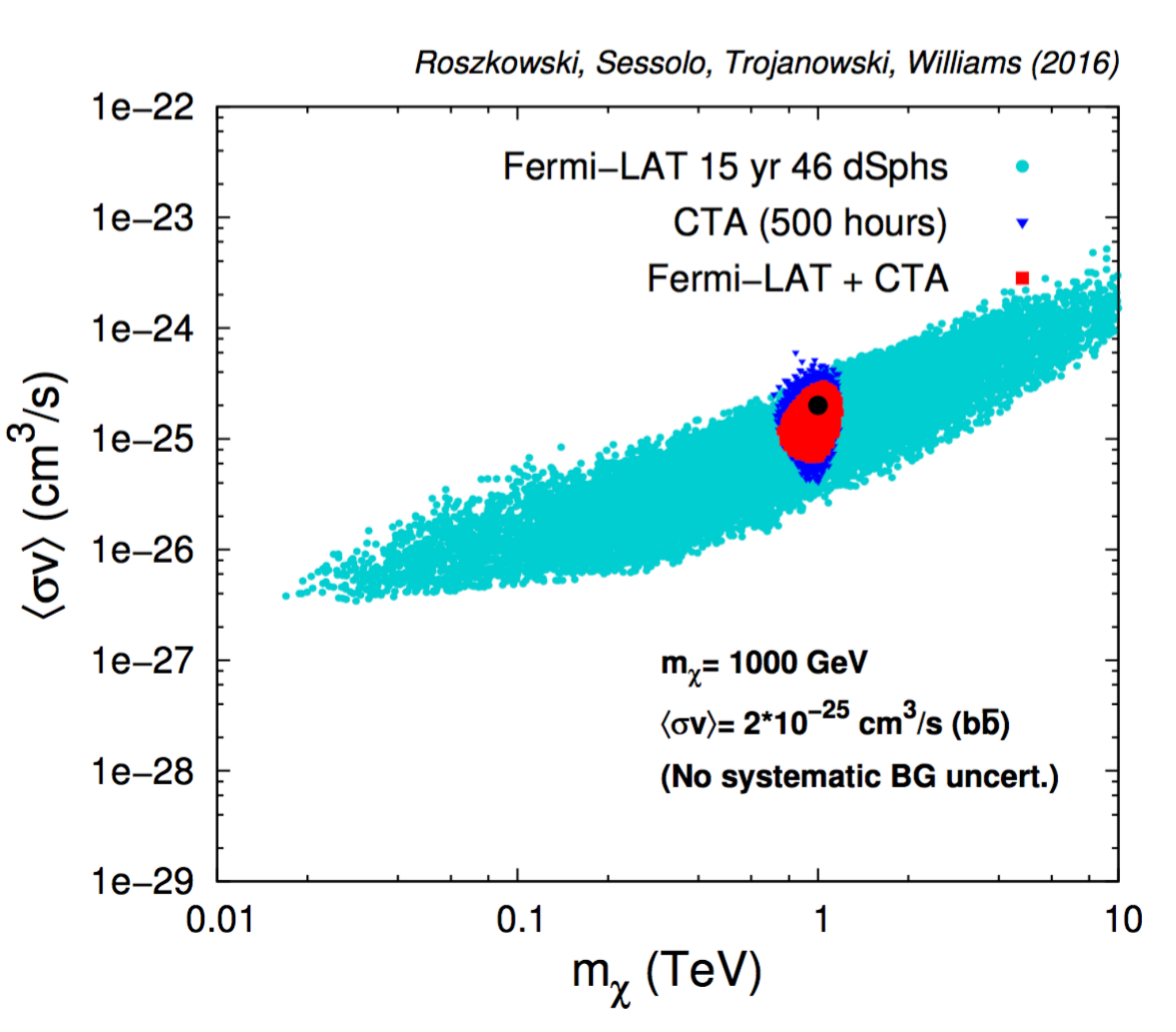}
}%
\caption{(a) The effect of neglecting the nuisance parameters when fitting for point BP4(a) in the (\mchi, \sigv) plane. The color code is the same as in \reffig{fig:ID_1000bb}(a). (b) The effect of neglecting the uncertainty in the $R_i^{\textrm{CR}}$ and $R_i^{\textrm{GDE}}$ parameters in \refeq{BGunc} when fitting for BP4(a), compare \reffig{fig:ID_1000bb}(a).}
\label{fig:ID_red_unc}
\end{figure}
%%%%%%%%%%%%%%%%%%%%%%%%%%%%%%%%%

Given the excellent mass reconstruction reached in CTA for most benchmark points, 
we break down in \reffig{fig:ID_red_unc} the effect of the different uncertainties that can 
spoil it. We are confronted with uncertainties of several kinds. 
There are astrophysical uncertainties, due to the unknown shape of the Galactic DM halo profile,
which are included in our scans through the nuisance parameters $\rho_0$ and $\gamma_{\textrm{NFW}}$. 
Note that the ranges considered in \reftable{tab:params} for these parameters do not extend as much as to include 
less steep solutions\cite{Salucci:2007tm,Nesti:2013uwa} for the halo profile, which are therefore neglected in this study.
We then consider systematic uncertainties due the normalization of the isotropic 
cosmic-ray background and to the diffuse astrophysical gamma-ray background around the GC, 
parametrized, respectively, by the $R_i^{\textrm{CR}}$ and $R_i^{\textrm{GDE}}$ parameters in \refeq{BGunc}.
We remind the reader that we have not included systematic uncertainties in the detector response\cite{Silverwood:2014yza}. 

In \reffig{fig:ID_red_unc}(a) we show the effect of neglecting the astrophysical uncertainties by setting the nuisance parameters to their central values
when reconstructing BP4(a) in the (\mchi, \sigv) plane. When comparing to \reffig{fig:ID_1000bb}(a) one notices a significant improvement in the cross section reconstruction,
not accompanied by an equally significant mass reconstruction improvement (the 95\%~C.L. region for CTA is indicated with blue upside-down triangles).
This is due to the fact that the $b\bar{b}$ spectrum for a $\sim1\tev$ WIMP and the $W^+W^-$ spectrum for a $\sim500\gev$ WIMP are very similar, see
\reffig{fig:spectra}(b). Our extension of the Ring Method effectively allows one to somewhat constrain the degrees of freedom associated with the 
$R_i^{\textrm{CR}}$ and $R_i^{\textrm{GDE}}$ parameters in \refeq{BGunc} and, at the same time,  
the $J$-factors are in this scan fixed by construction. 
The strength of the signal, driven by \sigv, is then also strongly constrained, but the remaining final state freedom 
introduces the residual uncertainty in the mass reconstruction.   

In \reffig{fig:ID_red_unc}(b) we show instead the effect of setting $R_i^{\textrm{CR}}=R_i^{\textrm{GDE}}=1$ in \refeq{BGunc}, while
maintaining the freedom to scan over the nuisance parameters. One can see the persistence of a small degeneracy 
between $\mchi=1000\gev$ and some points with a smaller mass, characterized by a mixed final-state spectrum mimicking the features of a 1\tev\ $b\bar{b}$ case.
The main uncertainty however is now in \sigv, centered about the benchmark point. It is due to the fact that
the signal depends on the product $\sigv/\mchi^2\times J$, whose factors can be adjusted independently to produce a better fit. 
   
%%%%%%%%%%%%%%%%%%%%%%%%%%%%%%%%
\begin{figure}[t]
\centering
\subfloat[]{%
\label{fig:a}%
\includegraphics[width=0.47\textwidth]{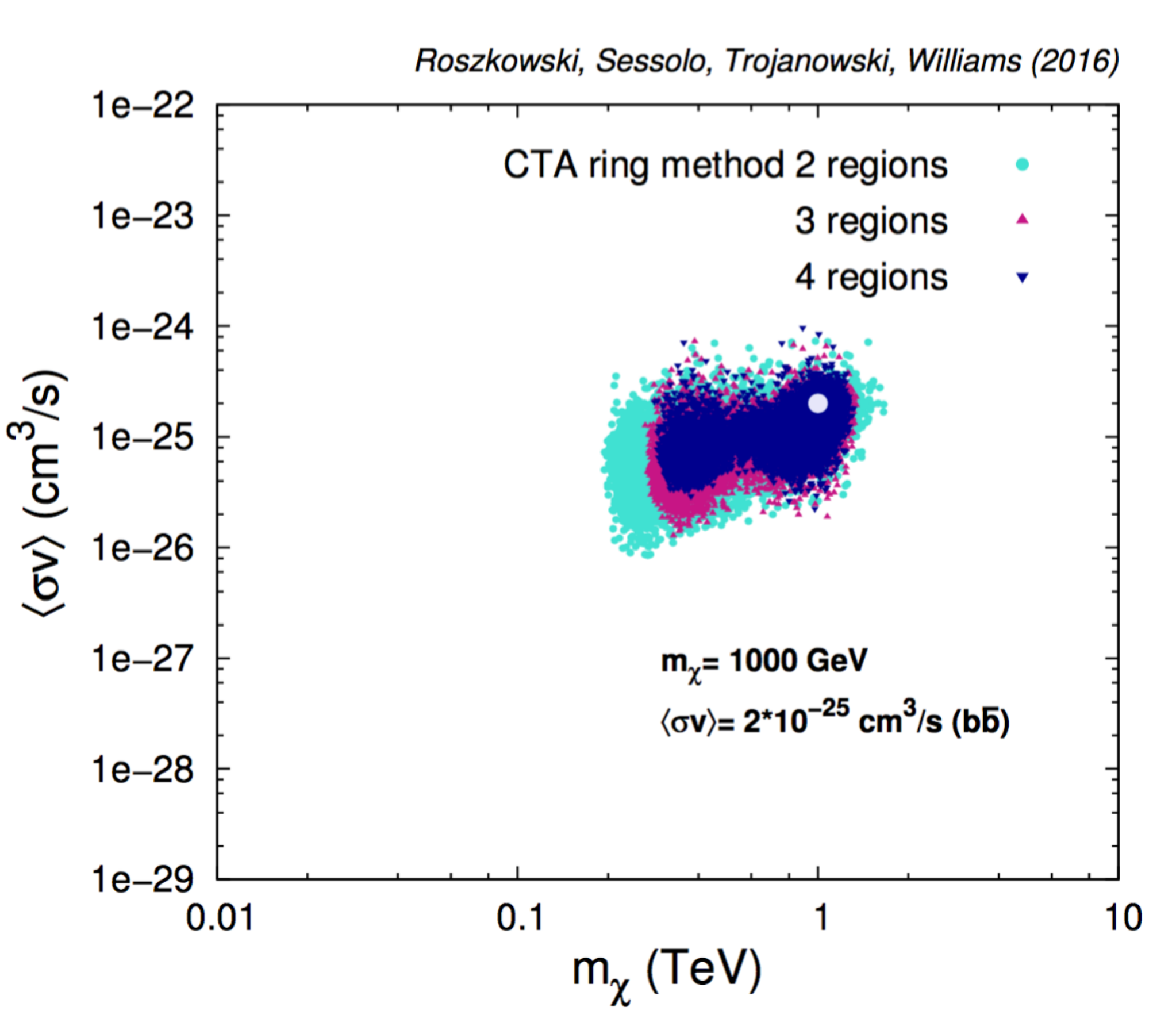}
}
\hspace{0.02\textwidth}
\subfloat[]{%
\label{fig:c}%
\includegraphics[width=0.47\textwidth]{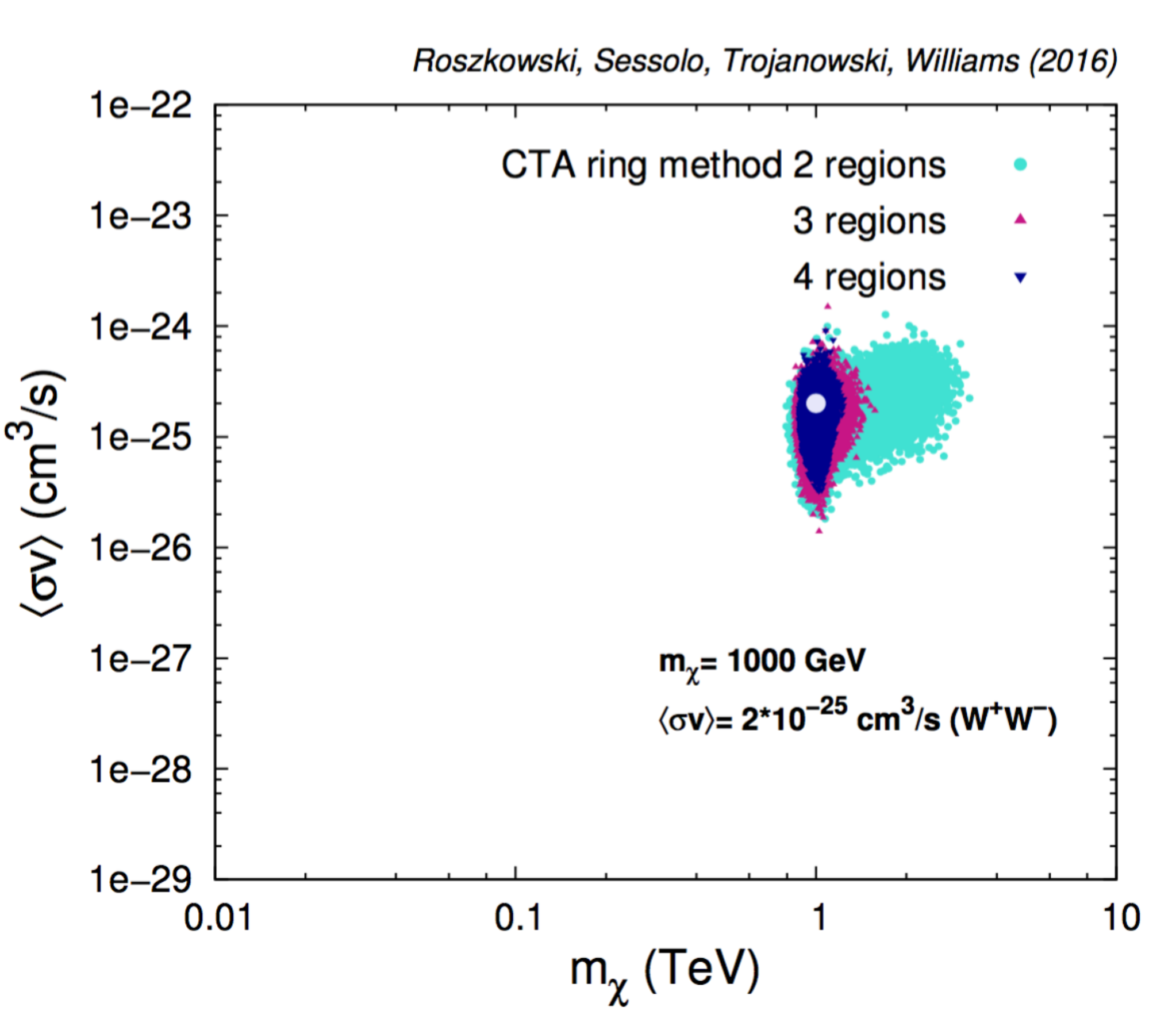}
}%
\caption{(a) Turquoise circles show the fit to CTA 500 hour mock BP4(a) data when considering the Ring Method with 2 spatial regions as, e.g., in 
Ref.\cite{Roszkowski:2014iqa}. 
Violet triangles show the fit when considering 3 regions in the sky, and dark blue upside-down triangles the fit with 4 regions, designed as in \reffig{fig:regions}. 
(b) Same as (a) but for BP4(b).}
\label{fig:ID_regions}
\end{figure}
%%%%%%%%%%%%%%%%%%%%%%%%%%%%%%%%%

As was mentioned in \refsec{sec:CTA}, our extension of the Ring Method plays an important role
in reducing the background systematic uncertainties in CTA. In this paper 
we have presented results based on the 4 regions shown in \reffig{fig:regions}. 
However, we also mentioned in \refsec{sec:CTA} that what is really important is to 
perform the analysis in at least 3 spatial regions, as 
there are 3 separate components producing a signal in \refeq{BGunc}.

In \reffig{fig:ID_regions} we present the 95\%~C.L. reconstruction regions at CTA for a 
progressively increasing number of regions. We show it for BP4(a) in \reffig{fig:ID_regions}(a) and for BP4(b) in \reffig{fig:ID_regions}(b). 
Marked with turquoise circles, one can see the reconstruction obtained using the traditional 
2-region method of, e.g.,\cite{Doro:2012xx,Wood:2013taa,Pierre:2014tra,Roszkowski:2014iqa}. 
The region in violet triangles shows that most of the improvement can be obtained by adding 
one additional region to the Ring Method (3 regions result when Regions 1 and 2 in \reffig{fig:regions} are unified into a single patch).
One can also see that adding the fourth region brings about a moderate additional improvement (dark blue upside-down triangles as in the other plots). 
It is important to point out that, as the uncertainties in the background normalization become increasingly constrained by adding 
additional regions in the sky, systematic uncertainties of the detector response, which we do not consider here, can become the dominant source.

%%%%%%%%%%%%%%%%%%%%%%%%%%%%%%%%
\begin{figure}[t]
\centering
\subfloat[]{%
\label{fig:a}%
\includegraphics[width=0.47\textwidth]{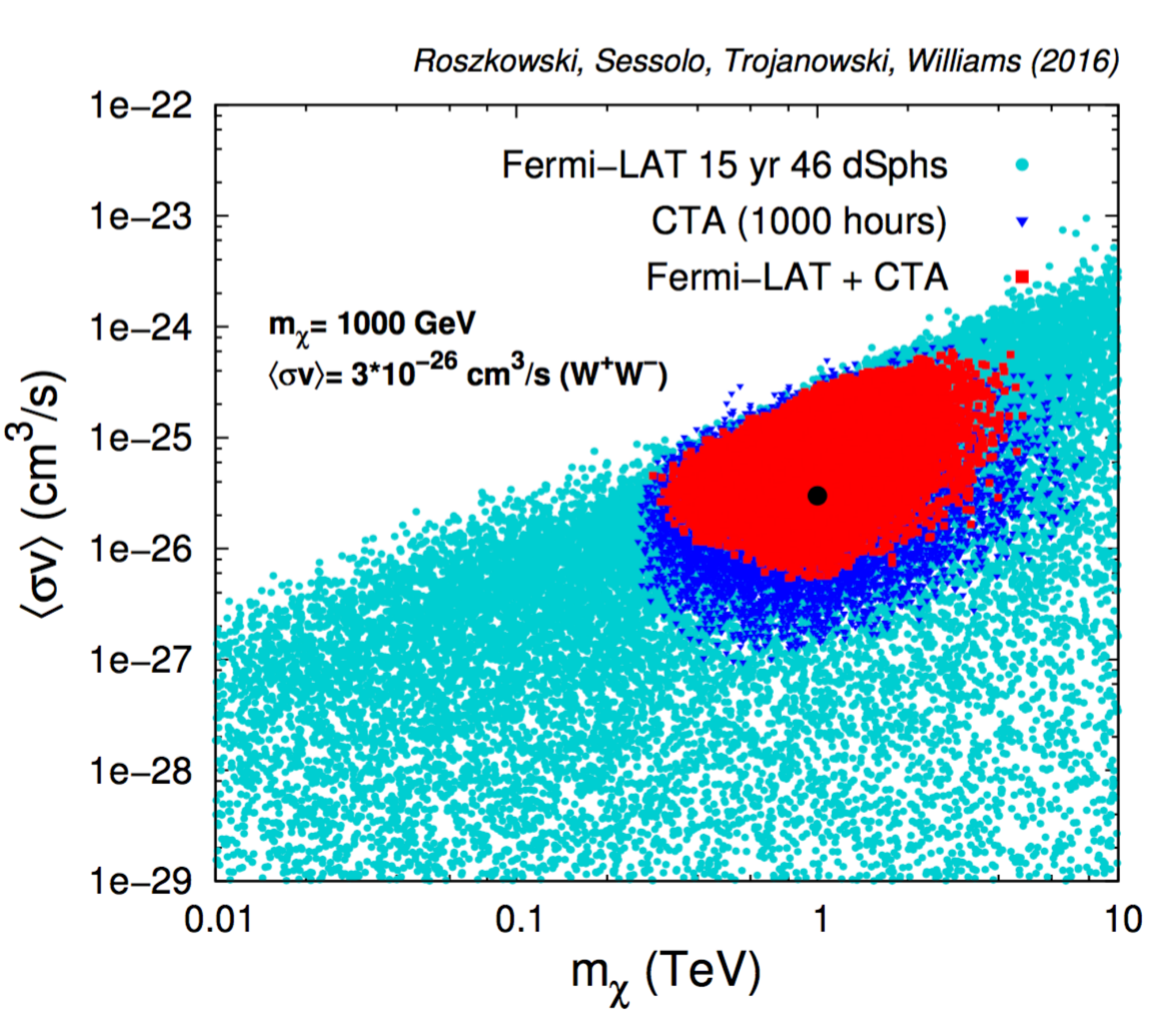}
}
\hspace{0.02\textwidth}
\subfloat[]{%
\label{fig:c}%
\includegraphics[width=0.47\textwidth]{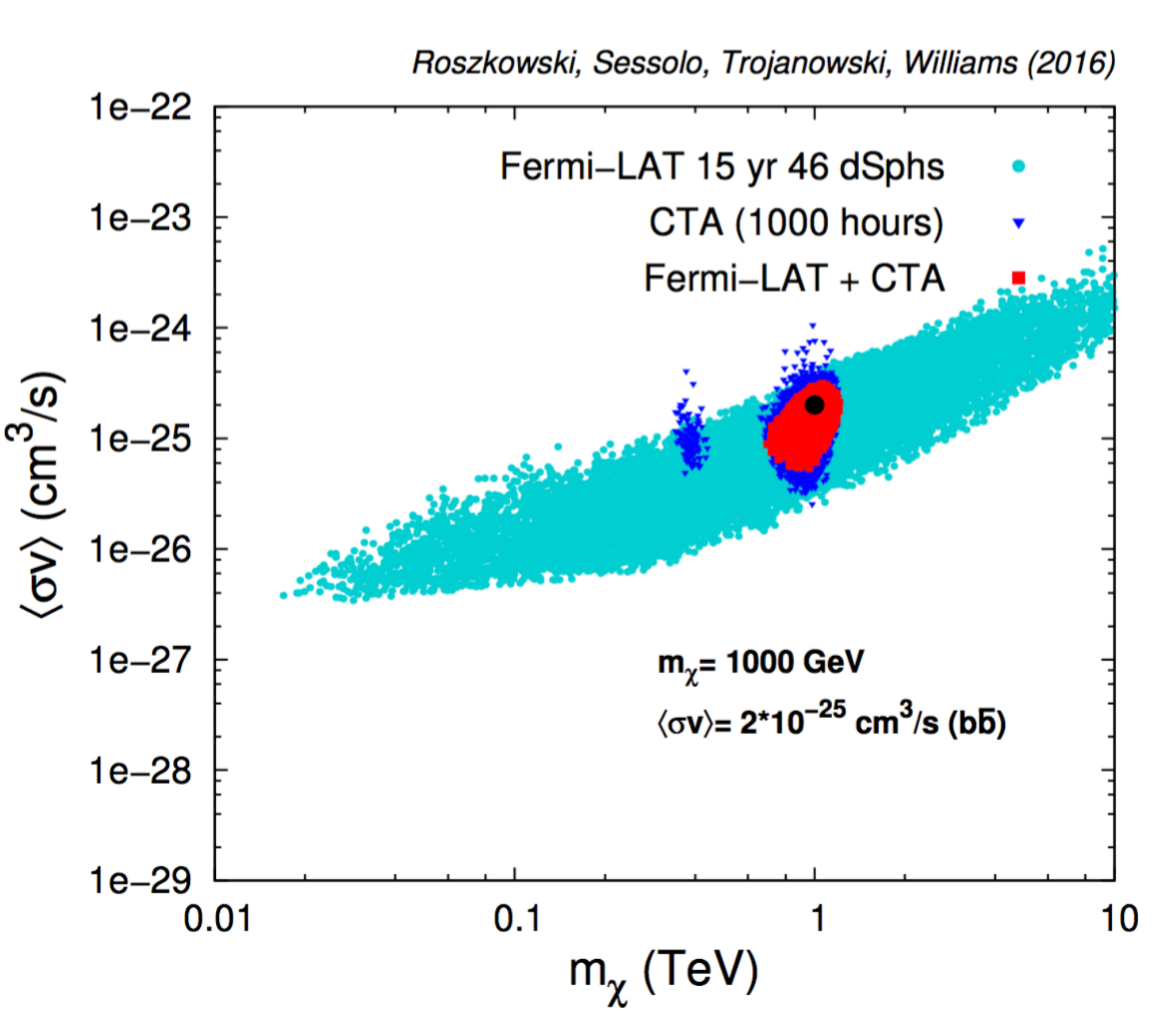}
}%
\caption{(a) Same as \reffig{fig:ID_1000bb}(a), but for the benchmark BP5, 
where the exposure of CTA has been extended to 1000 hours. (b) Same as (a) but for BP4(a), compare \reffig{fig:ID_1000bb}(a).}
\label{fig:ID_1000}
\end{figure}
%%%%%%%%%%%%%%%%%%%%%%%%%%%%%%%%%

In \reffig{fig:ID_1000} we present the results of extending the exposure of CTA to 1000 hours of observation, which could be possibly allocated 
in the optimistic case of a WIMP discovery. Specifically, the reconstruction would improve in the case of BP5,
a point characterized by the canonical 
$\sigv=3\times10^{-26}\textrm{ cm}^3/\textrm{s}$ for which the reconstruction with 500 hours was quite poor.  
One can see in \reffig{fig:ID_1000}(a) that even with increased exposure, this case remains tricky and lies at the borderline of the
instrument reconstruction abilities, given the current state of uncertainties. 
We show for completeness in \reffig{fig:ID_1000}(b) the reconstruction with 1000 hours in the case of BP4(a), with a larger \sigv.
The 95\%~C.L. combined region shrinks here significantly relative to \reffig{fig:ID_1000bb}(a).

We conclude this section by discussing in brief a few cases characterized by annihilation final states different 
from the ones described above. It was recently pointed out in\cite{Queiroz:2016zwd} that gamma-ray telescopes could be sensitive to
the $\nu_{\alpha}\nu_{\alpha}$ annihilation final states for a cross section not much larger than 
the typical values considered here. We have explicitly checked numerically that, 
for a benchmark point characterized $\mchi=1000\gev$, $\sigv=2\times10^{-25}\textrm{ cm}^3/\textrm{s}$, and a 100\% branching 
ratio to $\nu_{\tau}\nu_{\tau}$, our likelihood function for CTA cannot reconstruct the WIMP properties.
By increasing \sigv, some loose 95\%~C.L. contours in the (\mchi, \sigv) plane would start to appear for $\sigv>5\times10^{-25}\textrm{ cm}^3/\textrm{s}$.   
  
\begin{figure}[t]
\centering
\subfloat[]{%
\label{fig:a}%
\includegraphics[width=0.47\textwidth]{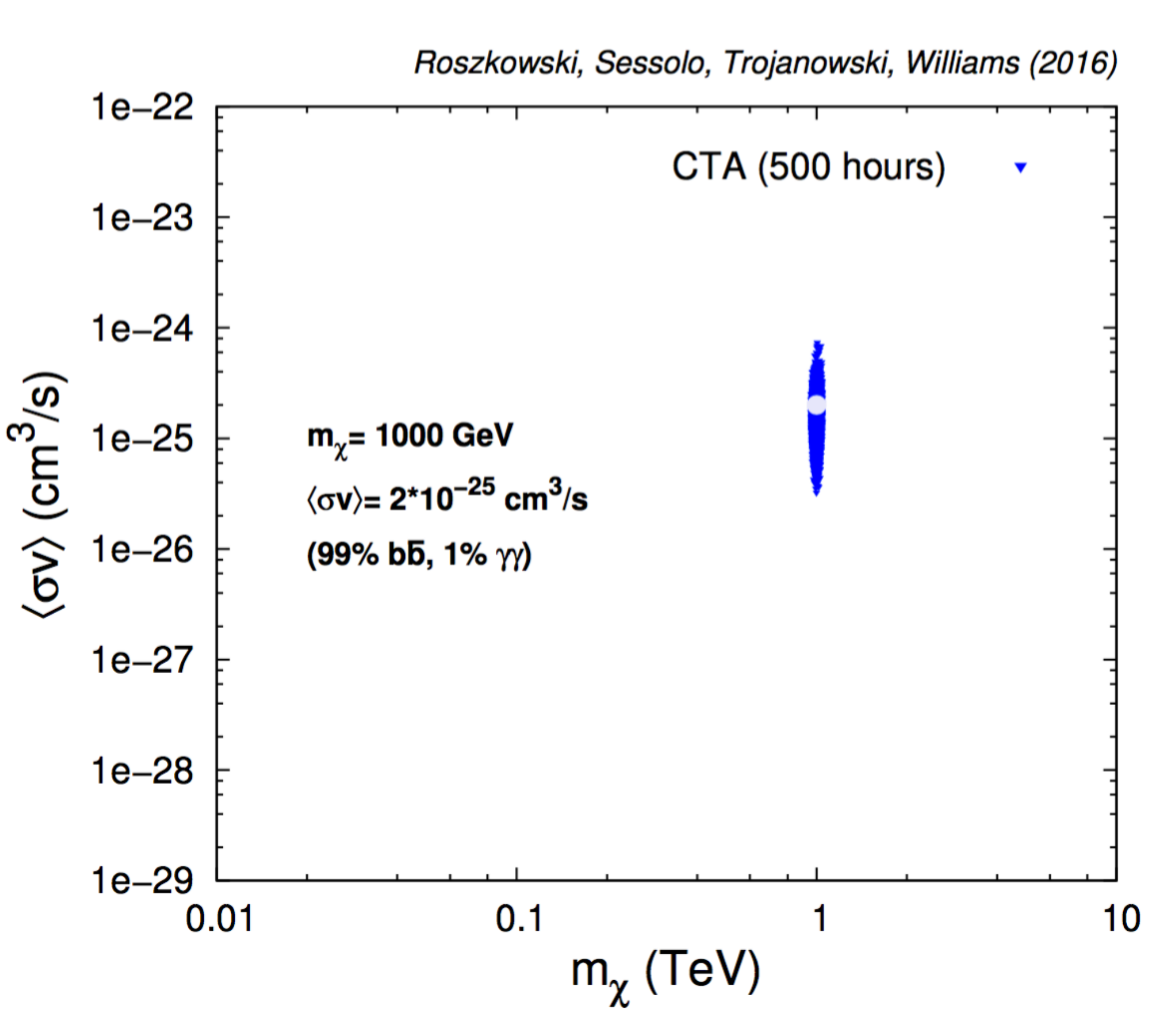}
}
\hspace{0.02\textwidth}
\subfloat[]{%
\label{fig:c}%
\includegraphics[width=0.47\textwidth]{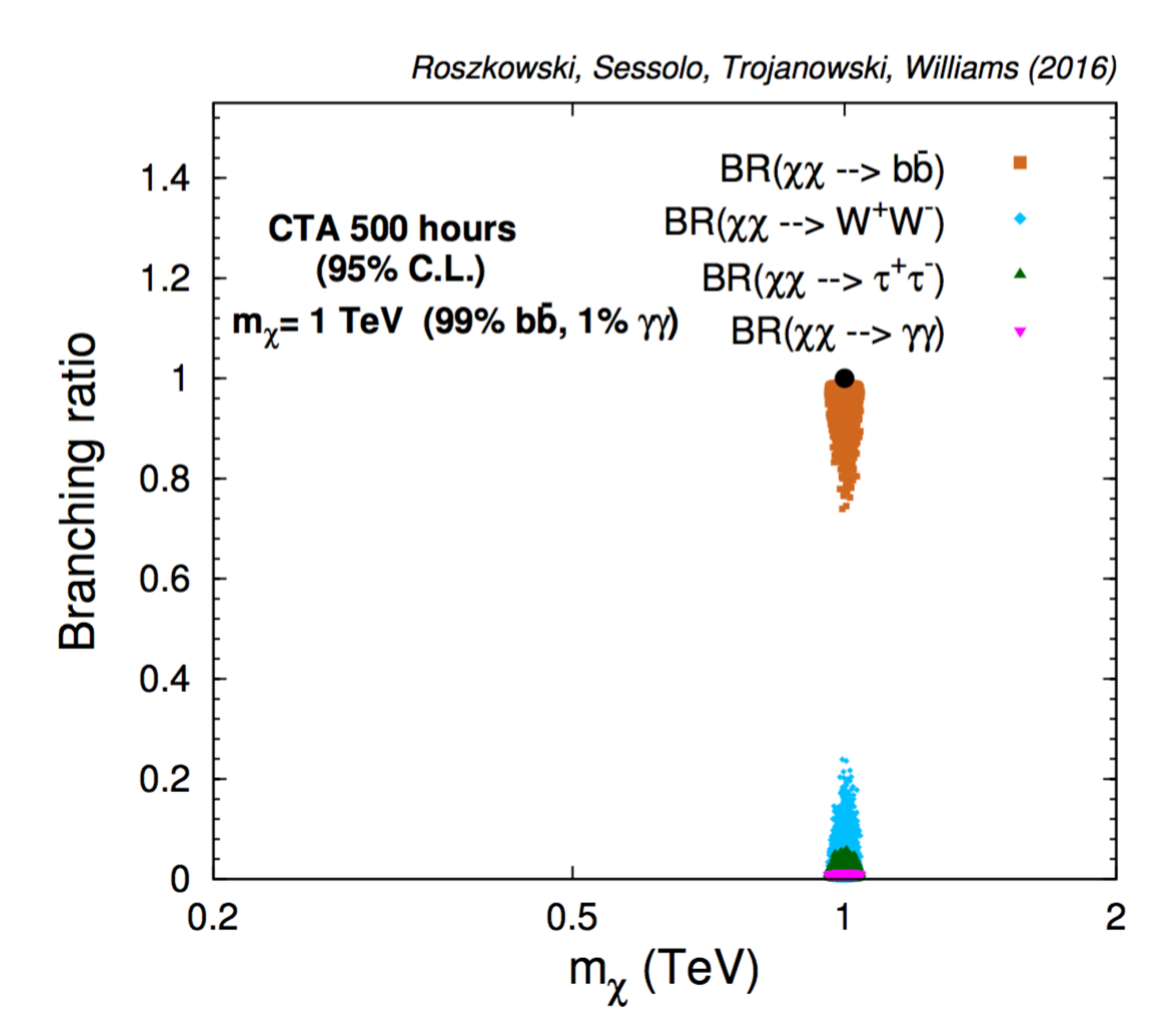}
}%
\caption{(a) The 95\%~C.L. CTA reconstructed region in the (\mchi, \sigv) plane for a point similar to BP4(a) but with 
$\textrm{Br}(\chi\chi\rightarrow b\bar{b})=99\%$ and $\textrm{Br}(\chi\chi\rightarrow \gamma\gamma)=1\%$, compare \reffig{fig:ID_1000bb}(a). 
(b) The breakdown of the reconstructed branching ratios for the points shown in (a).}
\label{fig:ID_gammaline}
\end{figure}

On the other hand, CTA is expected to be very sensitive to even small branching ratios to the monochromatic $\gamma\gamma$ final state.  
To give a feeling of how this effectively affects the reconstruction of WIMP properties at CTA, 
we present in \reffig{fig:ID_gammaline}(a) the 95\%~C.L. region in the (\mchi, \sigv) plane for the fit to a point 
similar to BP4(a) (see \reffig{fig:ID_1000bb}), but characterized by $\textrm{Br}(\chi\chi\rightarrow b\bar{b})=99\%$ and 
$\textrm{Br}(\chi\chi\rightarrow \gamma\gamma)=1\%$. One can see that even a small branching fraction to gamma-ray lines 
has the power to erase the mass degeneracy that appears in the case of 100\% $b\bar{b}$, thus leaving behind  
only the significant uncertainty in the reconstruction of \sigv, which is due to the unknown real value of the nuisance parameters, in particular $\rho_0$.

%%%%%%%%%%%%%%%%%%%%%%%%%%%%%%%%
%%%%%%%%%%%%%%%%%%%%%%%%%%%%%%%%%

We show in \reffig{fig:ID_gammaline}(b) the breakdown of the reconstructed branching ratios for the points appearing 
in \reffig{fig:ID_gammaline}(a). One can see that $\textrm{Br}(\chi\chi\rightarrow \gamma\gamma)$ in particular 
can be reconstructed to very good accuracy. 

\section{Summary and conclusions\label{sec:summary}}
 
In this work we addressed the issue of WIMP mass and cross section reconstruction in case a positive
measurement of a DM signal is made in either an underground direct detection
experiment or in a diffuse gamma radiation experiment, or both. For the former we considered
the expected sensitivity of XENON-1T, SuperCDMS-Snolab, and DarkSide-G2, while
for the latter we considered projected 15 yr data from Fermi-LAT
observation of 46 dSphs and projected CTA sensitivity to a signal from
the Galactic Center assuming the default observation time of
500~hours. For each we built a likelihood function assuming realistic
astrophysical uncertainties.

We considered eight WIMP cases (benchmark points) spanning a wide range
of mass from 25\gev\ up to 1\tev\ and featuring  
\sigsip\ (direct detection) or \sigv\ (gamma rays) large enough to be
within the discovery reach of the above experiments. For \sigv\ we
further considered not only a ``default'' WIMP annihilation final
state $b\bar{b}$ but also $\tau^+\tau^-$, $W^+W^-$, and $\mu^+\mu^-$. 
For each case we generated MC simulations of WIMP parameters producing
in different experiments a signal within $2\sigma$ of the considered benchmark
point. We did not assume any specific particle physics model or
scenario, and for this reason neither did we impose any collider
limits nor the relic density constraint.

Our general conclusion is that, even in the optimistic cases of 
cross sections large enough to warrant a strong signal detection, 
reconstructing WIMP properties will for the most part be rather challenging. 
As previously shown, direct detection experiments will only be able to reconstruct
WIMP mass below some 100\gev, above which event spectra become basically
degenerate. Gamma-ray energy spectra, on the other hand, are WIMP mass
dependent but face another type of degeneracy related to different
final states at different WIMP mass. This can severely limit the
ability of Fermi-LAT, in the low mass regime, and, to a lesser extent, CTA, in the high mass regime, 
to determine WIMP mass, \sigv, or
annihilation final states, when considered alone.

However, we found that in several interesting cases a remarkable
improvement in WIMP reconstruction can be achieved by combining discovery data from
Fermi-LAT and/or CTA, or by combining gamma-ray observatories 
with direct detection experiments.

For example, at lower WIMP mass (c.f. 25\gev), good mass determination in direct detection
experiments may not only help substantially reduce the range of \sigv\ allowed by
a positive measurement at Fermi-LAT, but also severely restrict WIMP
annihilation final states.
At the other end of the considered WIMP mass range (1\tev), best
prospects come from CTA which will reach its maximum sensitivity
there, while Fermi-LAT's sensitivity will be poorer, but not always
negligible. We showed that in the case of CTA one can engineer a simple extension of the 
Ring Method that provides significant reduction of the systematic uncertainties in the background normalization 
without the need of a full morphological analysis.
We also quantified the improvement in the WIMP property reconstruction associated 
with increasing the allocated exposure over the default value of 500 hours.

While a signal in CTA would allow for a fairly crude
($b\bar{b}$ final state), good ($W^+W^-$) or even very good ($\tau^+\tau^-$ and $\mu^+\mu^-$) WIMP
mass determination, combining it with an additional signal from
Fermi-LAT may help reducing the range of \sigv\ and the allowed branching
fractions to different final states.

The most difficult mass range appears to be the middle range between
some 100\gev\ and a few hundred GeV where direct detection gives a
very poor mass determination, Fermi-LAT starts losing more and more
sensitivity with increasing WIMP mass, while that of CTA has not
yet reached its full strength. However, even in this tricky parameter space region we have shown 
that in case of a concurrent detection in both a direct detection experiment and
Fermi-LAT data, the former can set a lower bound on the WIMP mass and the latter an upper bound that can be 
combined together to give a rough reconstruction of the WIMP mass and scattering cross section.  

In conclusion, while a detection of a DM signal will be a landmark
achievement whose importance will be difficult of overemphasize, the determination of WIMP properties from a DM
signal in one or more types of experiment is likely to be very
challenging. However, a complementary approach using multiple experimental venues 
can, as is often the case, reduce the degeneracy and provide better information on the DM properties.   

%Future will tell to what extent Nature has been kind to
%us in this respect.
 
%%%%%%%%%%%%%%%%%%%%%%%%%%%%%%%%%%%%%%%%%%%%%%%%%%%%%%%%%%%%%%%%%%%%%%%%%%%%%%%%
\acknowledgments
   We would like to thank Andrzej Hryczuk for helpful comments. 
   The work of LR is supported by the Lancaster-Manchester-Sheffield Consortium for Fundamental Physics under STFC Grant No. ST/L000520/1.
   ST is supported in part by the National Science Centre, Poland, under research grant DEC-2014/13/N/ST2/02555 and by the 
   Polish Ministry of Science and Higher Education under research grant 1309/MOB/IV/2015/0.
   The use of the CIS computer cluster at the National Centre for Nuclear Research is gratefully acknowledged.

%%%%%%%%%%%%%%%%%%%%%%%%%%%%%%%%%%%%%%%%%%%%%%%%%%%%%%%%%%%%%%%%%%%%%%%%%%%%%%%%

\bibliographystyle{JHEP}	% (uses file "plain.bst")

\bibliography{BF_13}

\end{document}